\documentclass{jfm}

\usepackage{natbib}
\usepackage{amsmath,amssymb,mathrsfs}
\usepackage{color}
\usepackage{graphicx}
\usepackage{subfigure}
\usepackage{soul}
\usepackage{bbding}
\usepackage[normalem]{ulem}

\ifCUPmtlplainloaded \else
  \checkfont{eurm10}
  \iffontfound
    \IfFileExists{upmath.sty}
      {\typeout{^^JFound AMS Euler Roman fonts on the system,
                   using the 'upmath' package.^^J}%
       \usepackage{upmath}}
      {\typeout{^^JFound AMS Euler Roman fonts on the system, but you
                   dont seem to have the}%
       \typeout{'upmath' package installed. JFM.cls can take advantage
                 of these fonts,^^Jif you use 'upmath' package.^^J}%
       \providecommand\upi{\pi}%
      }
  \else
    \providecommand\upi{\pi}%
  \fi
\fi

\ifCUPmtlplainloaded \else
  \checkfont{msam10}
  \iffontfound
    \IfFileExists{amssymb.sty}
      {\typeout{^^JFound AMS Symbol fonts on the system, using the
                'amssymb' package.^^J}%
       \usepackage{amssymb}%
         \let\leq=\leqslant
         \let\geq=\geqslant
      }{}
  \fi
\fi

\ifCUPmtlplainloaded \else
  \IfFileExists{amsbsy.sty}
    {\typeout{^^JFound the 'amsbsy' package on the system, using it.^^J}%
     \usepackage{amsbsy}}
    {\providecommand\boldsymbol[1]{\mbox{\boldmath $##1$}}}
\fi

\newcommand\Ca{\mbox{\textit{Bo}}}  % Bond number
\newcommand\We{\mbox{\textit{We}}}  % Weber number
\newcommand{\ms}[1]{\mathcal{#1}}
\newcommand{\mr}[1]{\mathrm{#1}}
\newcommand{\mc}[1]{\mathscr{#1}}

\def\C{\Ca}
\def\bea{\begin{eqnarray}}
\def\eea{\end{eqnarray}}

\newsavebox{\astrutbox}
\sbox{\astrutbox}{\rule[-5pt]{0pt}{20pt}}

\newcommand{\zh}{\boldsymbol}

\def\bea{\begin{eqnarray}}
\def\eea{\end{eqnarray}}
\def\We{W\!e}

\def\ri{\mathrm{i}}
\def\bu{\zh{u}}
\def\bx{\zh{x}}
\def\mbf{\zh{f}}
\def\dd{\mbox{d}}
\def\bn{\zh{n}}

\def\stress{\mbox{\boldmath{$\sigma$}}}

\newcommand\nc{\newcommand}
\nc{\vect}[1]{\mbox{\boldmath $#1$}}
\nc{\pt}{\partial_t}
\nc{\px}{\partial_x}

\newcommand{\mylab}[1]{\label{#1}}
\title[Pulse dynamics on electrified falling films]{Two-dimensional pulse dynamics and the formation of bound states on electrified falling films}
\author[M. G. Blyth and D. Tseluiko and T.-S. Lin and S. Kalliadasis]%
{M. G. Blyth$^1$, \ns
D. Tseluiko$^2$, \ns
T.-S. Lin$^3$, \ns
and S. Kalliadasis$^4$
}

\affiliation{
$^1$School of Mathematics, University of East Anglia, Norwich, NR4 7TJ, UK\\[\affilskip]
$^2$Department of Mathematical Sciences, Loughborough University,
Loughborough,~LE11~3TU,~UK\\[\affilskip]
$^3$Department of Applied Mathematics, National Chiao Tung University,
Hsinchu 30010, Taiwan\\[\affilskip]
$^4$Department of Chemical Engineering, Imperial College London, London, SW7 2AZ, UK
}

\pubyear{2014}
\volume{000}
\pagerange{000--000}
\date{?; revised ?; accepted ?. - To be entered by editorial office}
\begin{document}

\maketitle
%%%%%%%%%%%%%%%%%%%%%%%%%%%%%%%%%%%%%%%%%
\begin{abstract}
The flow of an electrified liquid film down an inclined plane wall is investigated with the focus on coherent structures in the form of travelling waves on the film surface, in particular, single-hump solitary pulses and their interactions. The flow structures are analysed first using a long-wave model, {\color{blue} which is valid in the presence of weak inertia}, and second using the Stokes equations. For obtuse angles, gravity is destablising and solitary pulses exist even in the absence of an electric
field. For acute angles, spatially non-uniform solutions exist only beyond a critical value of the electric field strength; moreover, solitary-pulse solutions are present only at sufficiently high supercritical electric field strengths. The electric field increases the amplitude of the pulses, can generate recirculation zones in the humps, and alters the far-field decay of the pulse tails from exponential to algebraic with a significant impact on pulse interactions. A weak-interaction theory predicts an infinite sequence of bound-state solutions for non-electrified flow, and a finite set for electrified flow. The existence of single-hump pulse solutions and two-pulse bound states is confirmed for the Stokes equations via boundary-element computations. In addition, the electric field is shown to trigger a switch from absolute instability to convective instability, thereby regularising the dynamics, and this is confirmed by time-dependent simulations of the long-wave model.
\end{abstract}

%%%%%%%%%%%%%%%%%%%%%%%%%%%%%%%%%%%%%%%%%
\section{Introduction\mylab{sec:intro}}
%%%%%%%%%%%%%%%%%%%%%%%%%%%%%%%%%%%%%%%%%

Flowing liquid films are central to a wide variety of industrial applications and engineering systems. Spatial heterogeneity in the film thickness, which
arises as a natural feature of the dynamics, may be exploited to enhance the intended purpose of the film, for example in
promoting heat transfer \citep[e.g.][]{park2003three}.

Electric fields have been proposed as a viable control mechanism in liquid film flows. Examples include the creation
of microscale surface patterning in the absence of inertia and shear \citep{schaeffer2000electrically},
and their suggested use in cylindrical space radiators \citep{kim1994}.
In this study, we discuss the use
of an electric field as a means of controlling the dynamics,
most notably the formation of solitary and bound-state pulses, in the flow of a liquid film down an inclined plate.
A flat film is known to become unstable to long-wave perturbations when the Reynolds number exceeds a critical value that depends on the angle of inclination~\citep{benjamin1957wave, yih}; for a vertical plate or one inclined at an obtuse angle
to the horizontal, the film is unstable at any Reynolds number. For an acute angle of inclination, the
flow can be destabilised even at zero Reynolds number by increasing the intensity of the electric field.
The surface profile of an unstable film exhibits a rich variety of spatial and
temporal structures, ranging from almost harmonic waves at the flow inlet to
highly nonlinear wave patterns sufficiently far downstream.
Recent reviews of falling film dynamics can be found in 
\cite{craster2009dynamics}, \cite{kalliadasis2011falling} and \cite{ruyer2014dynamics}.

There are numerous theoretical studies of film flow, but those of particular relevance to the present work and which incorporate the effect of a normal electric field include \cite{gonzales1996}, and \cite{Tseluiko2006}.
Here, the full governing equations were simplified using a long-wave asymptotic analysis to derive a fully nonlinear Benney-type equation for the scaled interface location~\citep{benny1966}, which
includes a non-local term encapsulating the effect of the electric field.
\cite{Tseluiko2006} obtained an even simpler, weakly-nonlinear equation of Kuramoto--Sivashinsky (KS) type with a non-local electric field term.
\cite{Tseluiko2006} carried out an extensive
numerical investigation of the electrified KS equation and found complex
dynamical behaviour. \cite{kawahara1983formation} (for the non-electrified case) and
\cite{tseluiko2010dynamics} (for the electrified case) found the inclusion of dispersion to have a regularising
effect; in particular solutions evolve into weakly interacting pulses. The
regularising effect of dispersion for the non-electrified gKS equation was
analysed in detailed by~\cite{Chang1995} and~\cite{Tseluiko2010a}. Recently \cite{wray_etal_2017} have derived a fully nonlinear model for an electrified falling film which is valid at finite Reynolds number using a weighted residuals approach.
We note that most previous studies have focused on acute inclination angles. Studies of the obtuse inclination angle case are limited in number, but include the recent work by \cite{rohlfs2017hydrodynamic}, who adopted a weighted-residuals approach, as well using direct solutions of the
Navier--Stokes equations, to study pulses on an inverted film in the absence
of an electric field.

Many experiments have been designed to elucidate the behaviour of falling films, quite a number of which have focused on the effects of imposing forced perturbations of a specified frequency at the inlet~\cite[e.g.][]{liu1994, nosoko2004evolution}.
Solitary waves and saturated almost periodic waves
appear close to the inlet at low and high frequencies respectively. Further
downstream, the flow is dominated by two-dimensional solitary-wave pulses and
their mutual interactions~\citep{liu1994, vlachogiannis2001}.
After a stage of weak interaction with each other, the two-dimensional solitary
pulses experience a secondary instability and develop into three-dimensional localised
structures, the highly non-trivial interaction of which eventually gives rise to
interfacial turbulence~\citep{Demekhin2010}.

It is clear, then, that a study of solitary pulses must lie at the heart
of any endeavour to understand the dynamics of a falling film. In this work, we focus on two-dimensional pulses and, in particular, we offer a synthesis of long-wave model calculations and predictions based on fully
nonlinear solutions of the Stokes equations.
We consider unstable films
for both acute and obtuse plate inclination angles, in each case examining steady (in a travelling frame of
reference) solitary pulse solutions and bound-state solutions. Such solutions are themselves inherently unstable: the absolute/convective nature of this is examined using an adaptation of the classical Huerre--Monkewitz approach. Finally, we carry out
numerical time-dependent simulations and demonstrate for the first time that an absolutely unstable flow can
be regularised by increasing the intensity of the electric field, revealing a train of isolated pulses propagating
along the film surface. 

Although our study is restricted to two-dimensional dynamics, it is important to recognise the potentiality for transverse instabilities. Recently \cite{tomlin2017three} studied the spatiotemporal dynamics of electrified falling films using a multidimensional non-local Kuramoto--Sivashinsky model equation. They found that purely transverse modes for a flat film can become unstable for obtuse inclination angles or when the electric Weber number exceeds a certain threshold value. Despite this finding, two-dimensional studies are nevertheless of importance since two-dimensional states are visited during three-dimensional time evolution \cite[e.g.][]{kalliadasis2011falling} or may act as a preferred state in a controlled system \cite[e.g.][]{gomes2017stabilizing}. We also note that the transverse instability of fully-nonlinear electrified two-dimensional travelling-waves has yet to be assessed.

The layout of the paper is as follows. The problem formulation is introduced in \S~\ref{sec:form}.
In \S~\ref{sec:long} we introduce the non-local long-wave model.
We use numerical continuation techniques to explore travelling waves,
analyse the far-field decay of the long-wave pulses and develop a weak interaction theory for multiple
pulses.
In \S~\ref{sec:stokes} we present fully nonlinear pulse solutions computed using the boundary-element method, including single-hump pulses and two-pulse bound states.
In \S~\ref{sec:absolute} we discuss the instability of a single-pulse solution and examine its absolute or convective nature.
Finally, conclusions and discussion of our results are offered in
\S~\ref{sec:conclusions}.

%%%%%%%%%%%%%%%%%%%%%%%%%%%%%%%%%%%%%%%%%
\section{Problem formulation}\label{sec:form}
%%%%%%%%%%%%%%%%%%%%%%%%%%%%%%%%%%%%%%

\begin{figure}
\centering
{\includegraphics[width=2.0in]{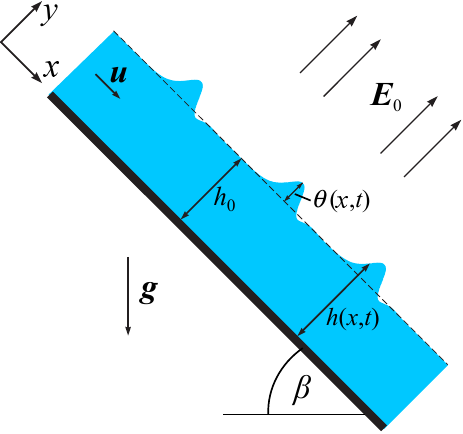}} \vspace{-0.2cm}
\caption{(Color online) Schematic representation of a viscous liquid film flow down an inclined plane wall. {\color{black}
Throughout the study, the waves travel in the positive $x$-direction.}
}
\mylab{fig1}
\vspace{-0.2cm}
\end{figure}

We consider a viscous liquid film that flows down a plane wall inclined at
angle $\beta$ to the horizontal. The film is exposed to an electric field
which acts in the direction normal to the wall and which is uniform with
strength $E_0$ at infinity, as is shown in figure~\ref{fig1}. The fluid is
taken to be either a perfect conductor or a perfect dielectric with relative
permittivity~$\varepsilon_p$. We use Cartesian coordinates $(x,y)$ with $x$
and $y$ measuring distance along the wall and normal to it (pointing into the
liquid), respectively.
The dimensionless Navier--Stokes and continuity equations are
\begin{eqnarray}\label{NS}
Re(\bu_t + \bu \cdot \nabla \bu ) = - \nabla p  + \nabla^2 \bu + \zh{G}, \qquad \nabla \cdot \bu = 0,
\end{eqnarray}
where $\zh{G} = 2(\sin\beta, -\cos \beta)$ is the dimensionless gravity, $\bu=(u,v)$ is the film velocity and $p$ is the film pressure; without loss of generality, we assume that the pressure in the air is zero.
Distances have been made dimensionless using the undisturbed film thickness $h_0$ as the length
scale, and $2\mu/\rho g h_0$ has been used as the time scale, where
$\mu$ and $\rho$ are, respectively, the dynamic viscosity and density of the liquid, and $g$ is gravity. We
have used %$\mu U_0/h_0$
$\rho g h_0/2$ as the pressure scale and, below, we will use $E_0h_0$ as the scale for the electric potential.
The Reynolds number is defined to be
$Re= \rho h_0U_0/\mu$, where $U_0=\rho g h_0^2/2\mu$ is the surface speed of a flat Nusselt film of
thickness $h_0$ flowing down a vertical wall.

The no-slip condition at the wall is
\begin{equation}
\bu = \zh{0} \quad\text{at}\quad y=0,
\end{equation}
and the kinematic condition at the free surface $y=h(x,t)$ is
\begin{eqnarray}
\label{banana}
v=h_t + uh_x.
\end{eqnarray}
The tangential and normal dynamic stress conditions at the free surface are
\begin{eqnarray}
\label{banana}
\zh{t}\cdot \zh{\sigma} \cdot \zh{n} = 0,
\qquad
\zh{n}\cdot \zh{\sigma} \cdot \zh{n} = -\frac{\kappa}{\Ca} + 2\We\, \zh{n}\cdot (\zh{M}_2-\zh{M}_1)\cdot \zh{n}
\end{eqnarray}
where $\zh{t}$ and $\zh{n}$ are the unit tangent and normal vectors at the free surface, respectively, the latter pointing into the liquid, $\kappa$ is the free surface curvature taken to be positive when the surface is concave downwards, and $\zh{\sigma}=-p\zh{I} + (\nabla \bu + \nabla \bu^T)/2$ is the Newtonian stress tensor in the liquid. The Maxwell stress tensor is defined to be
%Here $\varphi(x,y)$ is the electric potential in the air defined so that the electric field $\zh{E}=-\nabla \phi$,
$\zh{M}_j = \zh{E}_j\zh{E}_j - \zh{I}|\zh{E}_j|^2/2$, where $\zh{E}_j=-\nabla \varphi_j$ is the electric field
with electric potential $\varphi_j$ in region $j=1$ (the film) and $j=2$ (the air).
The Bond number $\Ca$ and electric Weber number $\We$ are given by
\begin{equation}\label{cawe}
\Ca = \frac{\rho g h^2_0}{2\gamma}, \quad %= \frac{\mu U_0}{\gamma}, \quad
\We = \left(1-\frac{1}{\varepsilon_p}\right)^2 \frac{\varepsilon_a E^2_0}{\rho g h_0}.
\end{equation}
where $\gamma$ is the surface tension coefficient, and $\varepsilon_p = \varepsilon_1/\varepsilon_2$, where
$\varepsilon_j$ is the electric permittivity of the film ($j=1$) and the air ($j=2$).

In the film and in the air above the film, the electric potentials satisfy Laplace's equation,
$\nabla^2 \varphi_j=0$,
and at the film surface it satisfies the boundary conditions,
\bea
\varphi_1 = \varphi_2, \qquad \varepsilon_p \zh{n}\cdot \nabla \varphi_1 = \zh{n}\cdot \nabla \varphi_2.
\eea
At the wall $\varphi_1=0$, and far above the film the electric field is uniform so that $\nabla \varphi_2 \rightarrow -\zh{j}$ as $y\rightarrow\infty$, where $\zh{j}$ is a unit vector pointing in the $y$-direction.

\vspace{0.125in}
%%%%%%%%%%%%%%%%%%%%%%%%%%%%%%%%%%%%%%%%%
\section{Long-wave model}\label{sec:long}
%%%%%%%%%%%%%%%%%%%%%%%%%%%%%%%%%%%%%%%%%

Assuming that the interfacial deformation wavelength
$\lambda$ is long compared to the undisturbed film thickness $h_0$, so that
the thin film parameter $\delta \equiv h_0/\lambda\ll 1$, \color{black} we can derive the
model equation \color{black} %takes the form
\begin{equation}
h_t + \left[h^3 \left(\frac{2\sin\beta}{3} - \frac{2\cos\beta}{3} h_x +\frac{1}{3\Ca} h_{xxx}
+ \frac{2}{3} \We\, \ms{H}[h_{xx}] \right)
\right]_x=0.
\mylab{eq:TFE1}
\end{equation}
%The long-wave evolution equation for the film thickness, $h(x,t)$, is derived
%in~\citet{Tseluiko2006} where it is assumed that the interfacial deformation
%wavelengths are long compared to the undisturbed thickness, see also Appendix~\ref{appendix:longwave} for the derivation of the equation and the discussion of its range of validity. The dimensionless equation is
%given as, without inertia effects,
Here $\mathcal{H}$ is the
Hilbert transform operator defined by
$\mathcal{H}=\tfrac{1}{\upi}\text{p.v.}\int_{-\infty}^\infty{f(\tilde
x)}/{(x-\tilde x)}d\tilde x$, where p.v. denotes the principal value. For any
finite value of $\varepsilon_p$, the film behaves as a perfect dielectric.
Mathematically, the perfect-conductor case is recovered in the limit $\varepsilon_p \to \infty$.  The equation is valid provided
that $\Ca=O(\delta^2)$, $\We=O(\delta^{-1})$, and then either both $\cot
\beta=O(1)$ and $Re=O(\delta)$ or both $\cot \beta=O(\delta^{-1})$ and
$Re=O(1)$. It is
derived by a systematic asymptotic procedure in the limit $\delta\to 0$ as
described in \cite{kalliadasis2011falling}  \cite[see
also][]{Tseluiko2006,tseluiko2013stability}. Note that for $\cot
\beta=O(\delta^{-1})$ the equation is a leading-order approximation, whereas
for $\cot \beta=O(1)$ it combines both leading-order and first-order
contributions.

If it is additionally assumed that $h(x,t) = 1 + \theta(x,t)$, where
$\theta=O(\delta)$, and a Galilean transformation is made to a moving frame
of reference, namely $x\mapsto x + 2(\sin \beta)t$, we obtain the non-local KS
equation \cite[see also][]{lin2015coherent}, \bea \label{KSeq} \theta_t + 4
(\sin \beta) \theta\theta_x - \frac{2}{3}(\cos \beta) \,\theta_{xx} +
\frac{2}{3}\We \mathcal{H}\left[\theta_{xxx}\right] +
\frac{1}{3\Ca}\theta_{xxxx} = 0. \eea

%{\color{red}\bf (DT: Can we shorten the discussion above? $B$ and $W$ and not used anyway...)}
%{\color{magenta}\bf (TL: I have moved the details to the appendix.)}

%%%%%%%%%%%%%%%%%%%%%%%%%%%%%%%%%%%%%%%%%
%\subsection{Steady-state solutions: Solitary pulses}\label{sec:LW_steady}
\subsection{Travelling-wave solutions}\label{sec:LW_steady}
%%%%%%%%%%%%%%%%%%%%%%%%%%%%%%%%%%%%%%%%%

%For certain parameter values, the long-time evolution of solutions to
%(\ref{eq:TFE1}) is described by a superposition of weakly interacting
%single-hump pulses, each of which resembles an infinite-domain single-hump
%solitary pulse. It is therefore important to analyse such solitary pulses.
%These are travelling-wave solutions which are
%localised in space and which propagate at a constant speed $c^*$.

To analyse travelling-wave solutions, which often characterise falling film flows,
we introduce in (\ref{eq:TFE1}) a moving-frame coordinate via the mapping $x\mapsto x+c^*t$,  where $c^*$ is the wave speed, %$x\mapsto x-c^*t$,
and look for stationary solutions $h^*$ in this frame.
%, which yields
%\begin{equation}
%-c^* h^*_{x} + \left[h^{*3} \left( \frac{2\sin \beta}{3} - \frac{2\cos\beta}{3}  h^*_{x} +\frac{1}{3\Ca} h^*_{xxx}
%+ \frac{2}{3} \We \ms{H}[h^*_{xx}] \right)
%\right]_x=0,
%\mylab{eq:TFE2}
%\end{equation}
%Travelling-wave solutions may be sought in a similar way
%using the non-local KS equation \eqref{KSeq}. These are expected to coincide
%with the long-wave solutions for small amplitude waves found near criticality
%when the film becomes unstable.
For solitary pulses, we will demand that the free-surface
height approaches the Nusselt flat-film solution away from the centre of the
pulse, i.e. $h^*\rightarrow 1$ as $x\to \pm\infty$. Solutions can be obtained using continuation techniques.
{\color{black} The numerical
computations have spectral accuracy and were done in Matlab.} We consider
solutions on a periodic domain, $x\in[-L,L]$, and can obtain solitary-pulse
solutions in the limit $L\rightarrow\infty$ (when it is possible to continue
solution branches to infinitely large $L$). We note that we need to
additionally impose conditions breaking the translational and `volume'
symmetries. {\color{black} (The `volume' symmetry is associated with the
fact that, by slightly changing the volume $\int^L_{-L} h^*\,\mathrm{d}x$, we
obtain a different solution.) To break the translational symmetry, we impose
the condition $h^*_x=0$ at $x=0$, and to break the `volume' symmetry, we
impose the condition $h^*=1$ at $x=L$.

To start
the continuation, we use a small-amplitude nearly sinusoidal solution of a
nearly cut-off wavelength that can be found by a standard linear stability
analysis. For acute inclination angles, gravity is stabilising, and the flat-film
solution is linearly unstable only when the electric field is sufficiently
strong, and in particular in can be shown that this occurs when $\We>\We_c=\sqrt{2\cos\beta/\Ca}$. The range of unstable wavenumbers is
$|k|\in(\kappa_-,\kappa_{+})$, where
\bea
\kappa_{\pm}=\Ca\,\We\pm\sqrt{\Ca^2\We^2-2\Ca\cos\beta}.
\mylab{eq:kappaplusminus}
\eea
For obtuse inclination angles, i.e.
$\beta\in(\upi/2,\upi)$, $\cos \beta<0$,  the flat-film solution is
linearly unstable due to the effect of gravity. The range of
unstable wavenumbers is $|k|\in(0,\kappa_{+})$.

%%%%%%%%%%%%%%%%%
\begin{figure}
\centering
    \begin{minipage}{1.1\textwidth}
        \centering
\hspace{-1.35cm}
\includegraphics[width=1.9in]{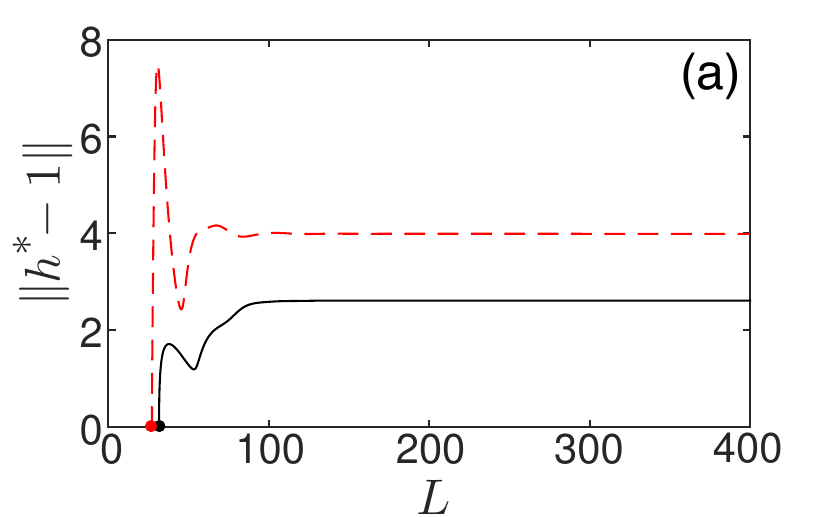}
\hspace{-0.5cm}
%\subfigure[]{\includegraphics[width=2.1in]{speed_bet_0_95pi_Ca_0_005}}
\includegraphics[width=1.9in]{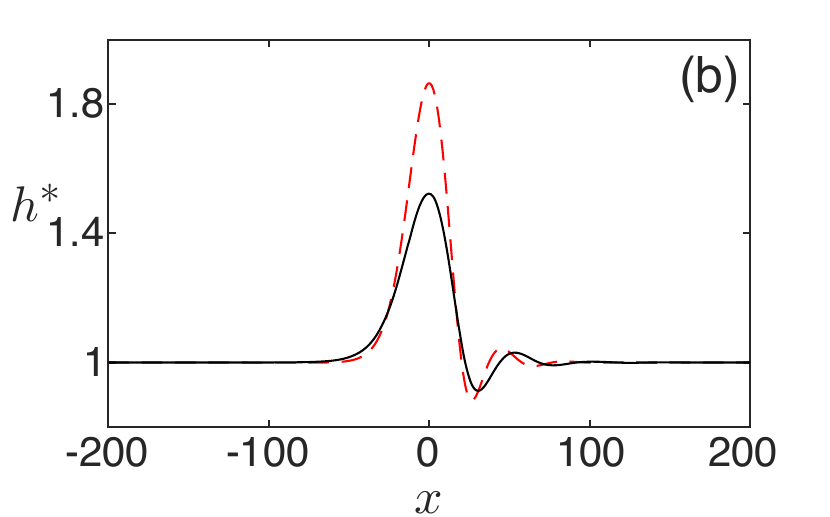}
\hspace{-0.5cm}
\includegraphics[width=1.9in]{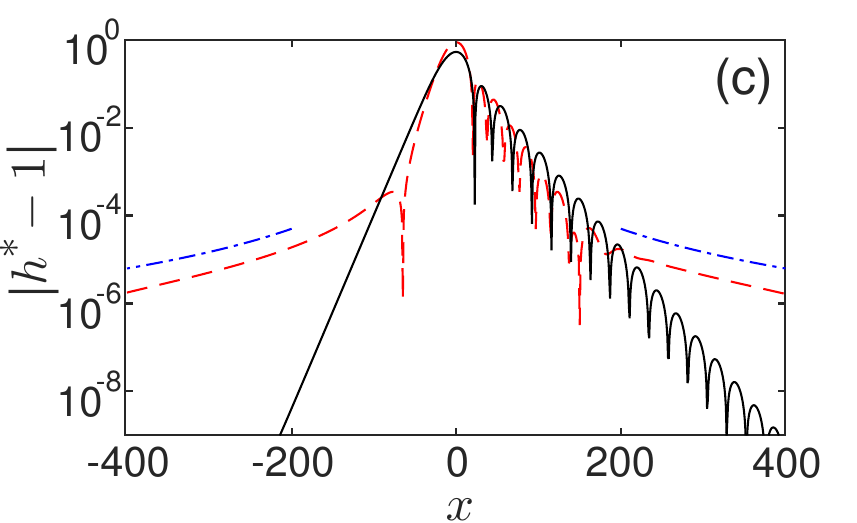}
    \end{minipage}%
    \vspace{-0.1cm}
\caption{(Color online) $\beta=0.95\upi$, $\Ca=0.005$. Solid lines are for $\We=0$ and dashed lines are for $\We=3$: (a)
Bifurcation diagram for the norm $\| h^* -1 \|$ against domain size $L$. The %(black and red)
circles indicate the starting points for the continuations;
(b) Pulse profiles for large $L$; (c) Shifted
profiles $h^*-1$ for large $L$ shown on a logarithmic scale. Also included are dot-dashed lines showing inverse cube decay in $|x|$ for comparison purposes.
}
\mylab{fig:bet_0_95pi_Ca_0_005}
\vspace{-0.2cm}
\end{figure}
%%%%%%%%%%%%

Figure~\ref{fig:bet_0_95pi_Ca_0_005} shows the results of numerical
continuation for an obtuse inclination angle, $\beta=0.95\upi$, when
$\Ca=0.005$ and $\We=0$ and $3$. A Stuart-Landau analysis suggests that the
bifurcation at $L=L_+=\upi/\kappa_+$, both for $\We=0$ and for $\We=3$, is
supercritical, and the numerical results corroborate this. Both the norm and
the speed converge to constants as $L$ increases. Also, for $\We=3$ both the
norm and the speed are greater than those for $\We=0$. Both of the travelling
waves are single-hump solitary pulses preceded by capillary ripples of
decaying amplitudes. The left tail of the pulse for $\We=0$ decays
monotonically to $1$ as $L\rightarrow-\infty$. However, for $\We=3$, the left
tail exhibits a depression just upstream of the main hump before
monotonically approaching $1$. Also, the amplitude of the pulse is larger
when the electric field is switched on, while the amplitude of the ripples is
not so significantly affected by the electric field. The semilog plots show
that the electric field dramatically affects the decay rate of the tails,
namely the decay is exponential when $\We=0$, but for $\We=3$ it becomes much
slower in the far-field.

For acute inclination angles, the solutions of branches
starting at the cut-off domain half-sizes, $L_\pm=\upi/\kappa_\pm$ do not
always converge to single-hump solitary pulses. In fact, we find that only
the solutions of the branch starting at $L_-$
converge to a single-hump solitary pulse when the electric Weber number is greater than a certain value %$\We_1$
that is greater than the linear instability threshold value $\We_c$. We
demonstrate this in figures~\ref{fig:bet_0_25pi_Ca_0_01_We_12_5} and
\ref{fig:bet_0_25pi_Ca_0_01_We_13_5} showing the results for $\beta=0.25\upi$
and $\Ca=0.01$ and for $\We=12.5$ and $13.5$, respectively.
Both of these values of the electric Weber number are above the
linear instability threshold, which is $\We_c\approx 11.89$ in this case.
% that for these values of $\beta$ and $\Ca$ is $\We_c\approx 11.89$.
The bifurcations at $L=L_+$ and $L=L_-$ produce
side branches which move toward larger and smaller values of $L$,
respectively, for $\We=12.5$. This is again in agreement with a Stuart-Landau analysis. For $\We=12.5$, the
points corresponding to the $L_+\approx 19.21$ and $L_-\approx 36.32$ are
connected by a single branch of periodic travelling-wave solutions.
%The squares indicate the points corresponding to where the norm takes the value $0.2$.
For
$\We=13.5$, there are two distinct solution branches
emanating from the points corresponding to the $L_+\approx 15.80$ and
$L_-\approx 44.18$, and both branches apparently extend to infinitely large
values of $L$. Both side branches move immediately toward larger
values of $L$ in agreement with a Stuart-Landau analysis.
For both branches, the norm $\|h^*-1\|$ and the speed $c^*$ tend to constant
values as $L$ increases. The wave profile in panel (b)
apparently converges to a single-hump solitary pulse preceded by capillary
ripples of decaying amplitude, as for the case of an obtuse inclination
angle. Also, like for the obtuse inclination angle, the
left tail of the pulse is not monotonically decaying. There is a depression upstream of
the pulse, and only then the tail tends to $1$ monotonically as
$x\rightarrow-\infty$. The wave profile in panel (c) does not correspond to a
single-hump pulse. Instead, it has the shape of a double-hollow negative
solitary pulse in the terminology of Chang and
co-workers~\citep{chang1994,chang2002} and with capillary ripples upstream
and an elevation downstream, and, moreover, it travels at a speed that is
smaller than that for the single-hump pulse.

\begin{figure}
\centering
    \begin{minipage}{1.1\textwidth}
        \centering
\hspace{-1.35cm}
\includegraphics[width=1.9in]{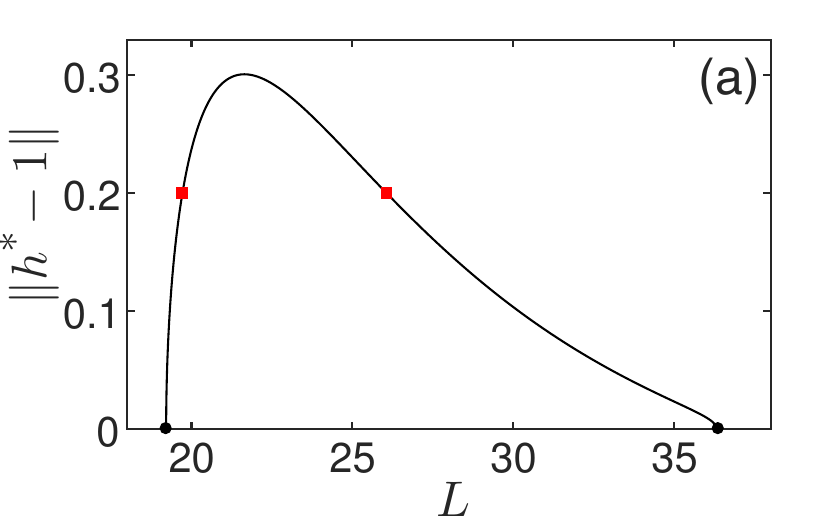}
%\subfigure[]{\includegraphics[width=2.6in]{speed_bet_0_25pi_Ca_0_01_We_12_5_1st_branch}}
\hspace{-0.5cm}
\includegraphics[width=1.9in]{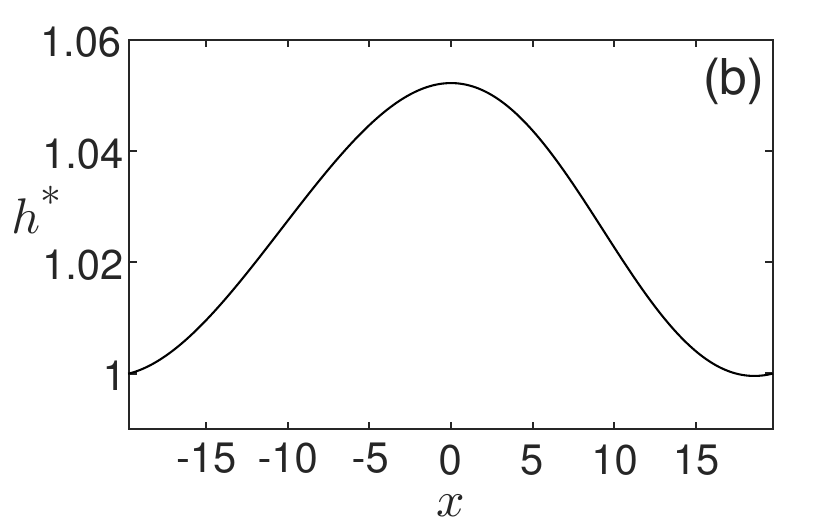}
\hspace{-0.5cm}
\includegraphics[width=1.9in]{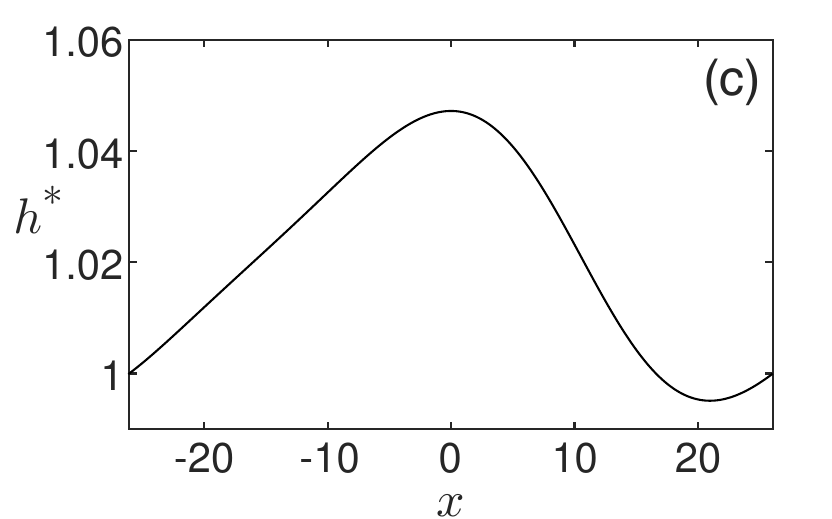}
    \end{minipage}%
\vspace{-0.1cm}
\caption{(Color online) $\beta=0.25\upi$, $\Ca=0.01$, $\We=12.5$
(a) Bifurcation diagram for $\| h^* -1 \|$ against domain size $L$.
%The bifurcation points are at $L_+\approx 19.21$ and $L_-\approx 36.32$.
(b, c) Wave profiles with norm $\| h^*-1\| = 0.2$, indicated by the left and right squares in (a), respectively.
%The corresponding wave speeds are given by the left and right squares in (b).
}
\mylab{fig:bet_0_25pi_Ca_0_01_We_12_5}
\vspace{-0.2cm}
\end{figure}

\begin{figure}
\centering
    \begin{minipage}{1.1\textwidth}
        \centering
\hspace{-1.35cm}
\includegraphics[width=1.85in]{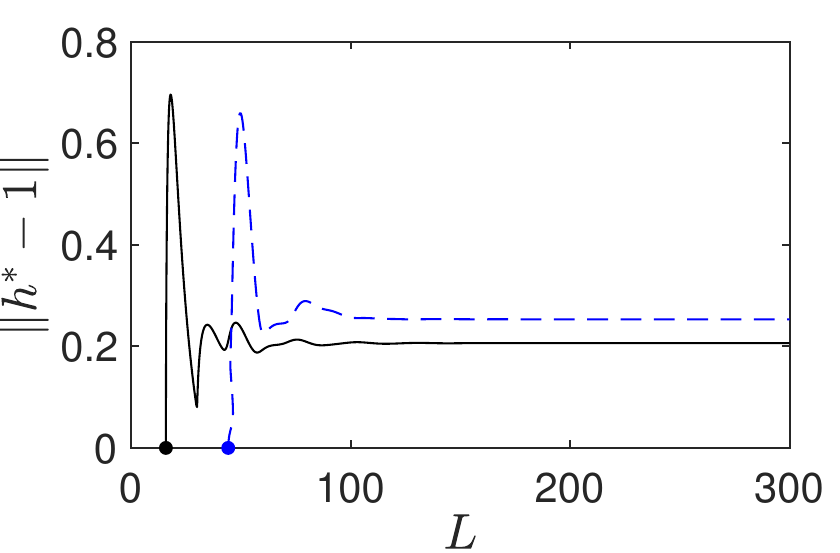}
\hspace{-0.35cm}
\includegraphics[width=1.85in]{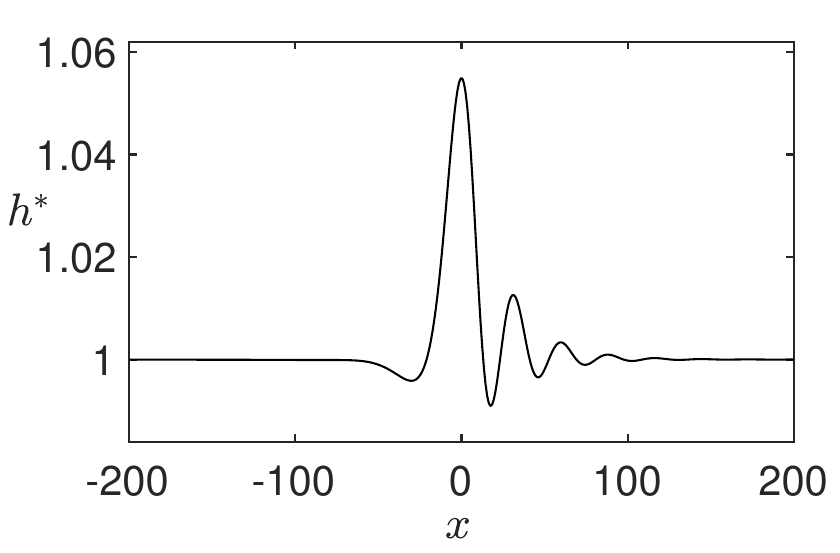}
\hspace{-0.35cm}
\includegraphics[width=1.85in]{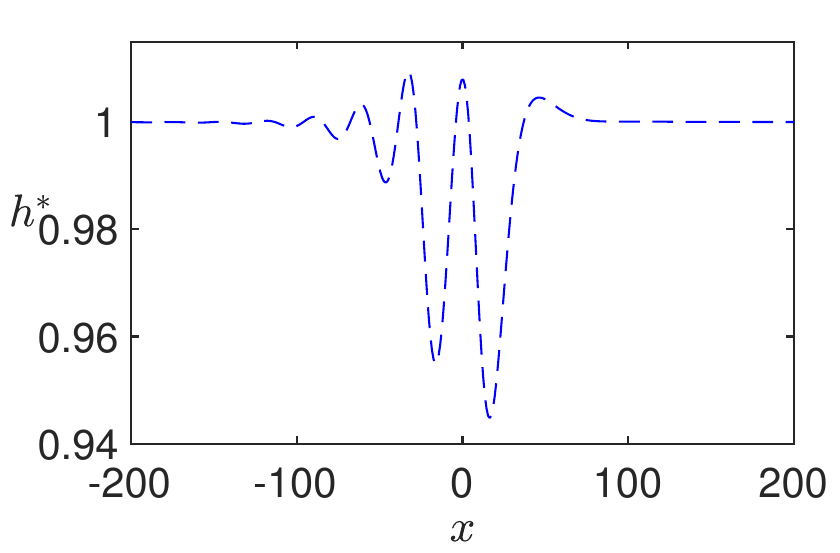}
    \end{minipage}%
\vspace{-0.1cm}
\caption{(Color online) $\beta=0.25\upi$, $\Ca=0.01$, $\We=13.5$. (a) Bifurcation diagram for $\| h^* -1 \|$ against domain size $L$. The solid and dashed lines show bifurcation branches emerging from $L_+=15.80$ and $L_-=44.18$, respectively.  (b, c) Pulse profiles for large $L$ on the branches emerging from $L_+$ and $L_-$, respectively.}
\mylab{fig:bet_0_25pi_Ca_0_01_We_13_5}
\vspace{-0.2cm}
\end{figure}

\begin{figure}
\centering

    \begin{minipage}{1.1\textwidth}
        \centering
\hspace{-1.35cm}
\includegraphics[width=2.0in]{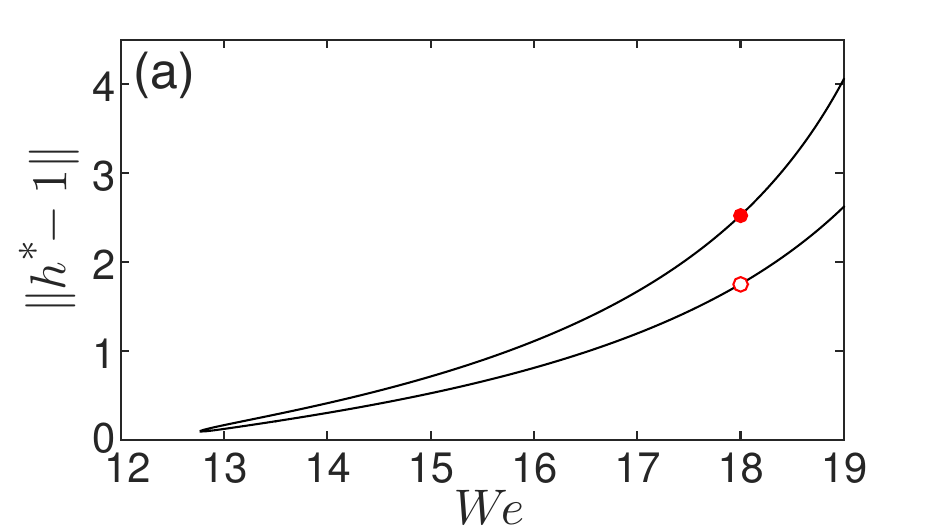}
\hspace{-0.5cm}
\includegraphics[width=1.85in]{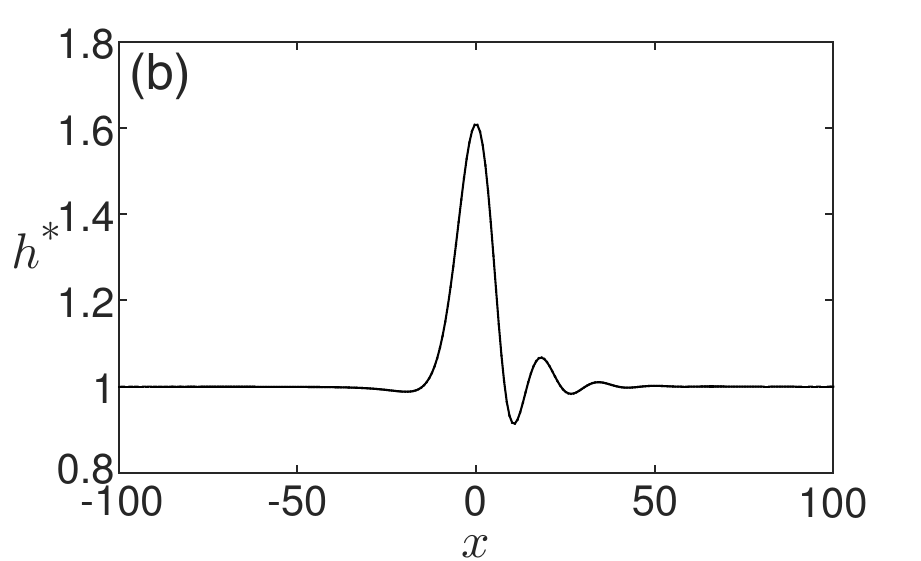}
\hspace{-0.5cm}
\includegraphics[width=1.85in]{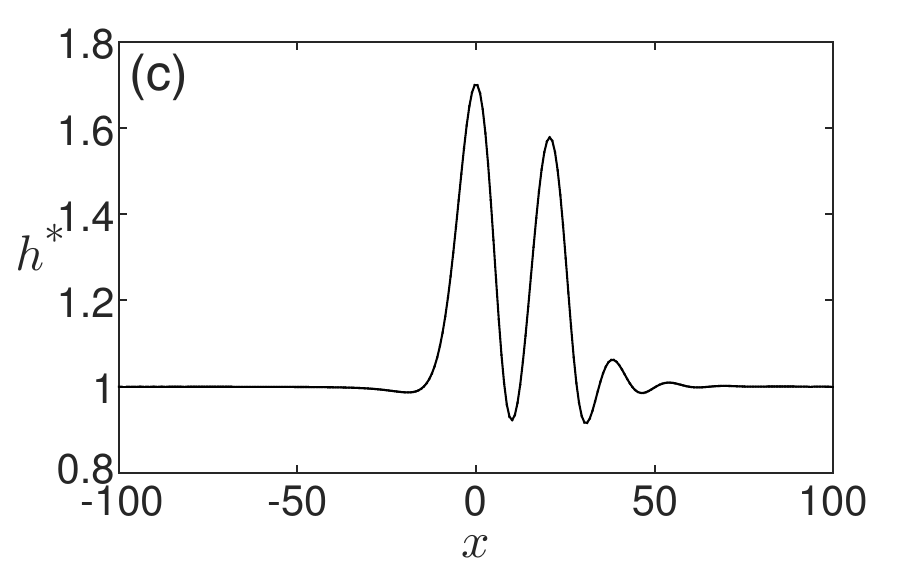}
    \end{minipage}%
        \vspace{-0.1cm}
\caption{(Color online) $\beta=0.25\upi$, $\Ca=0.01$. (a) Solution branch showing the norm $\| h^* -1 \|$ against electric Weber number $\We$.
(b, c) Pulse profiles at $\We=18$, indicated by the
empty and filled circle symbols respectively in panel (a).
%lower and upper dots in panel (a) respectively and the upper and lower dot in panel (b) respectively.
The wave speeds are $c^*= 2.22$ (b), $2.15$ (c).}
\mylab{fig:bet_0_25pi_Ca_0_01}
\vspace{-0.2cm}
\end{figure}

%Branch reconnections must take place as the electric
%Weber number varies between the case shown in figure
%\ref{fig:bet_0_25pi_Ca_0_01_We_12_5} and that shown in figure
%\ref{fig:bet_0_25pi_Ca_0_01_We_13_5} to explain the topological change in the
%solution curves.
%However, as the main focus of the present study is on the single-hump pulse
%solutions, we do not present the full bifurcation structure.

Finally, to investigate in more detail the influence of an electric field on
single-hump solitary pulses, we consider such a pulse for $\beta=0.25\upi$,
$\Ca=0.01$ on the domain $[-300,300]$, and perform continuation in the
electric Weber number, $\We$. The results are shown in
figure~\ref{fig:bet_0_25pi_Ca_0_01}.  The branches have turning points at
$\We=\We^\dagger\approx12.77$, which indicates that single-hump solitary
pulses do not exist for $\We<\We^\dagger$.
The lower part of the branch in panel (a) of figure~\ref{fig:bet_0_25pi_Ca_0_01} corresponds to single-hump
pulses. The upper branch in panel (a) corresponds to
double-hump pulses. Both for single- and double-hump pulses the norm and the speed
monotonically increase as $\We$ increases.

A straightforward far-field analysis for the non-electrified case reveals that if $c^*>2\sin \beta$ we obtain monotonic decay on the upstream side of the pulse and non-monotonic,
oscillatory decay on the downstream side of the tails. Referring back to the scales we have used to non-dimensionalise the
problem, this inequality corresponds to the physical pulse speed being greater than the speed of small amplitude linear long
waves, which themselves propagate at twice the surface speed of a flat film on an inclined plane, namely $2U_0\sin \beta$
\cite[e.g.][]{benjamin1957wave}.
If instead $c^*<2\sin \beta$, we will have oscillatory decay on the upstream side of a
pulse and monotonic decay on the downstream side.

With an electric field present,  the decay becomes algebraic \citep[see also][]{lin2015coherent}. Assuming that the pulse has a non-zero `mass', i.e. $\int_{-\infty}^\infty (h^*-1)dx\neq 0$, it can be shown that
$(h^*-1)\propto 1/x^3$ as $x\rightarrow\pm\infty$.

%%%%%%%%%%%%%%%%%%%%%%%%%%%%%%%%%%%%%%%%%
\subsection{Weak-interaction theory for solitary pulses}\label{sec:weakinter}
%%%%%%%%%%%%%%%%%%%%%%%%%%%%%%%%%%%%%%%%%

We generalise the earlier work of \cite{lin2015coherent} to the case of the general one-dimensional evolution equation
with translational symmetry, $\partial_t u = \partial_x(G[u])$, for some $G$ such that $G[h^*-\alpha]=0$, where $h^*\to \alpha$ as $|x|\to \infty$.   The work of \cite{lin2015coherent} built on earlier studies by \cite{elphick1990patterns}, \cite{elphick1991interacting}, \cite{balmforth1995solitary}, \cite{chang2002} for the case of local equations. Certain technical
details in these studies were re-examined by \cite{Pradas2011} and \cite{Tseluiko2012}.

We consider (\ref{eq:TFE1}) in a frame moving with the speed $c^*$ of a pulse solution. We represent a solution of (\ref{eq:TFE1}) as a superposition of $N$ well-separated quasi-stationary pulses located at $x_1(t)<\cdots<x_N(t)$ and a small overlap function $\hat u$. Suppressing the details in the interest of brevity, we obtain the
dynamical system describing the locations of the pulses \citep[see][]{lin2015coherent}
\begin{equation}
\dot x_k =-\Pi_k\left[ \px\left(G\left[\sum_{|x_i-x_k|<b(\epsilon)} u_i\right] \right)\right],\qquad k=1,\,\ldots,\, N,
\label{eq:dyn_system_locations_c}
\end{equation}
where $b(\epsilon)\gg \epsilon^{-1/p}$ is the range over which the interactions must be taken into account.
Here, $\Pi_k$ is a projection operator defined by $\Pi_k[f]=\int_{-\infty}^\infty f\psi_k\,\mathrm{d}x$, where $\psi_k$ is the zero adjoint eigenfunction for the operator $\px\circ\mc{D}_{u_k}G$ normalised so that $\int_{-\infty}^\infty u_{kx}\psi_k\,\mathrm{d}x=1$. Note that $\mc{D}_{u_k}G$ denotes the Fr\'echet derivative of operator $G$ at $u_k$.
%, that is
%\begin{eqnarray}
%\mc{D}_{u_k}G[f]\equiv c^*f-3(1+u_k)^2\left(\frac{2\sin \beta}{3} - \frac{2\cos\beta}{3}  u_{k\,x} +\frac{1}{3\Ca} u_{k\,xxx} + \frac{2}{3} \We \ms{H}[u_{k\,xx}]\right)f\qquad\nonumber\\
%-(1+u_k)^3\left(-\frac{2\cos\beta}{3} f_{x} +\frac{1}{3\Ca} f_{xxx}  + \frac{2}{3} \We \ms{H}[f_{xx}]\right)\!.\quad\qquad\end{eqnarray}
We note that in \eqref{eq:dyn_system_locations_c}, for the case of an
electrified film, it is necessary to include long-range interactions over
more than just the immediately neighbouring pulses as a direct result of
the algebraic decay of the pulse tails.

Let us consider a two-pulse system, $N=2$. From (\ref{eq:dyn_system_locations_c}), we may derive the
evolution equation for the pulse separation
\begin{equation}\mylab{eq:twopulse}
\dot{l} = \tilde P(l), \qquad \tilde P(l)\equiv \int_{-\infty}^\infty\px(G[  u^*(x+l)  -u^*(x-l)])\psi(x)\,\mathrm{d}x.
\end{equation}
Depending on the initial condition,
the two pulses may attract ($\tilde{P}(l) < 0$) or repel ($\tilde{P}(l) > 0$)
each other. The pulses can also form bound states, the separation distances
for which are zeros of $\tilde{P}(l)$. Figures~\ref{fig:ptilde}(a) and
\ref{fig:ptilde}(b) show $\tilde{P}(l)$ for $\beta=0.95\upi$ and $\Ca=0.005$,
and for $\We=0$ and $2.25$, respectively.
Apparently, there exist infinitely many bound states for $\We=0$, while there
exist only eight possible bound states for $\We=2.25$. To understand this, we
note that for $\We=0$, it can be shown that as $l\to \infty$, $\tilde P$ decays either monotonically or in an oscillatory manner. If the decay is monotonic, there are zero or a finite number of positive solutions to $\tilde P(l)=0$ implying
zero or a finite number of two-pulse bound states
%\bea
%P(l)
%\propto \mbox{e}^{-\lambda_1 l} \quad \mbox{and} \quad P(-l) \propto
%\mbox{e}^{\lambda_2 l},
%\eea
%\bea
%\tilde P(l) \propto \mbox{e}^{-\lambda_1 l} \quad \mbox{or} \quad \tilde P(l) \propto
%\mbox{e}^{-\lambda_2 l},
%\eea
%depending on the relative sizes of $\mbox{Re}(\lambda_1)$ and $\mbox{Re}(\lambda_2)$; here
%$\lambda_1$ is the real root of the cubic
%\eqref{cubic} and $\lambda_2$ is the complex root of the same equation.
If the decay is oscillatory, there is a countably infinite
number of solutions to $\tilde P(l)=0$, and, hence, an infinite number of
two-pulse bound states.

We note that in the case of no electric field, the travelling-wave form of \eqref{eq:TFE1} is local
and can be integrated once to yield a three-dimensional dynamical system.
Then, a single-pulse solution corresponds to a homoclinic orbit in the
three-dimensional phase space for this system. In such a case, the same
conclusion on the number of bound states in existence can be reached using
Shil'nikov's theorem \cite[see, for example,][]{glendinning1984local}. This
gives a criterion for the number of subsidiary homoclinic orbits which
correspond to two-pulse bound states, which exactly coincides with that found via the
monotonic/oscillatory decay argument given above.  However, when
$\We>0$, the travelling-wave form of \eqref{eq:TFE1} is non-local and the Shil'nikov-type approach is not
applicable. Instead, the weak-interaction theory developed here can be used to
analyse bound states. Then, it can be shown that $\tilde{P}(l)$ decays
monotonically algebraically to zero as $l \rightarrow \infty$, and hence
there exist only a finite number of two-pulse bound states.

\begin{figure}
\centering
\includegraphics[width=2.0in]{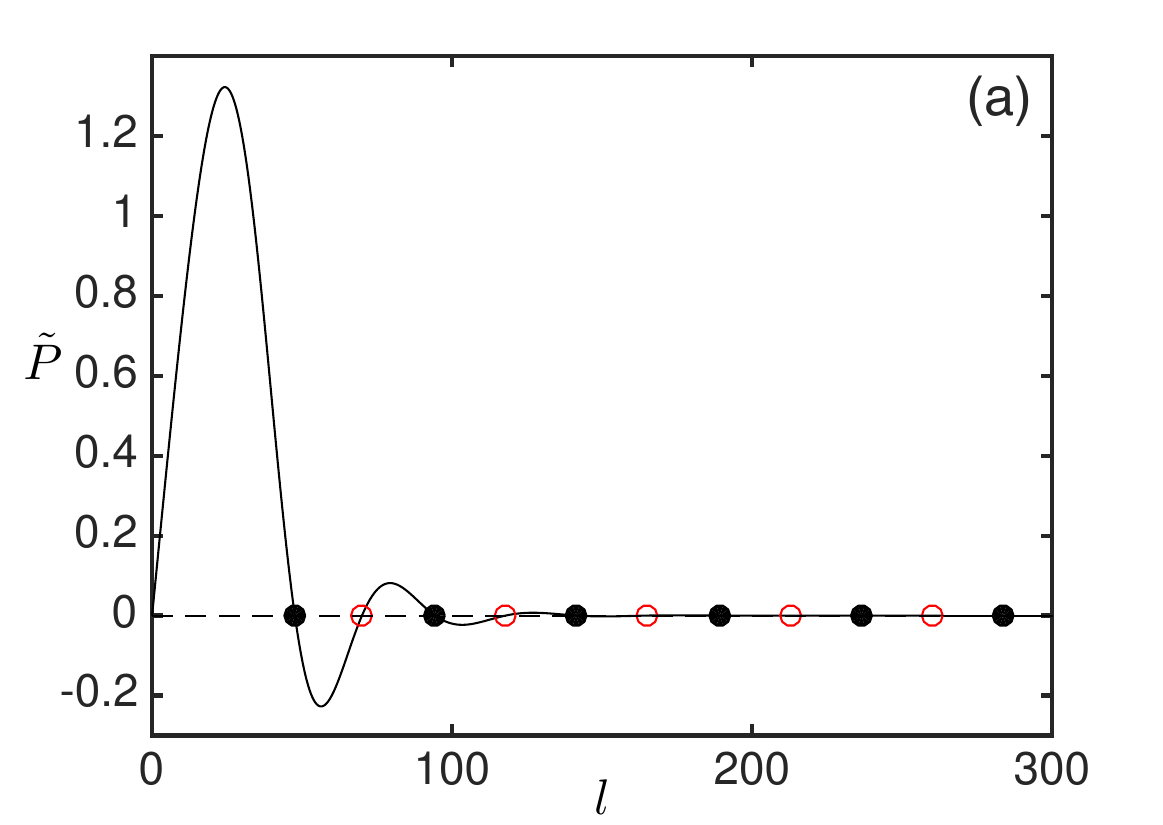}
\includegraphics[width=2.0in]{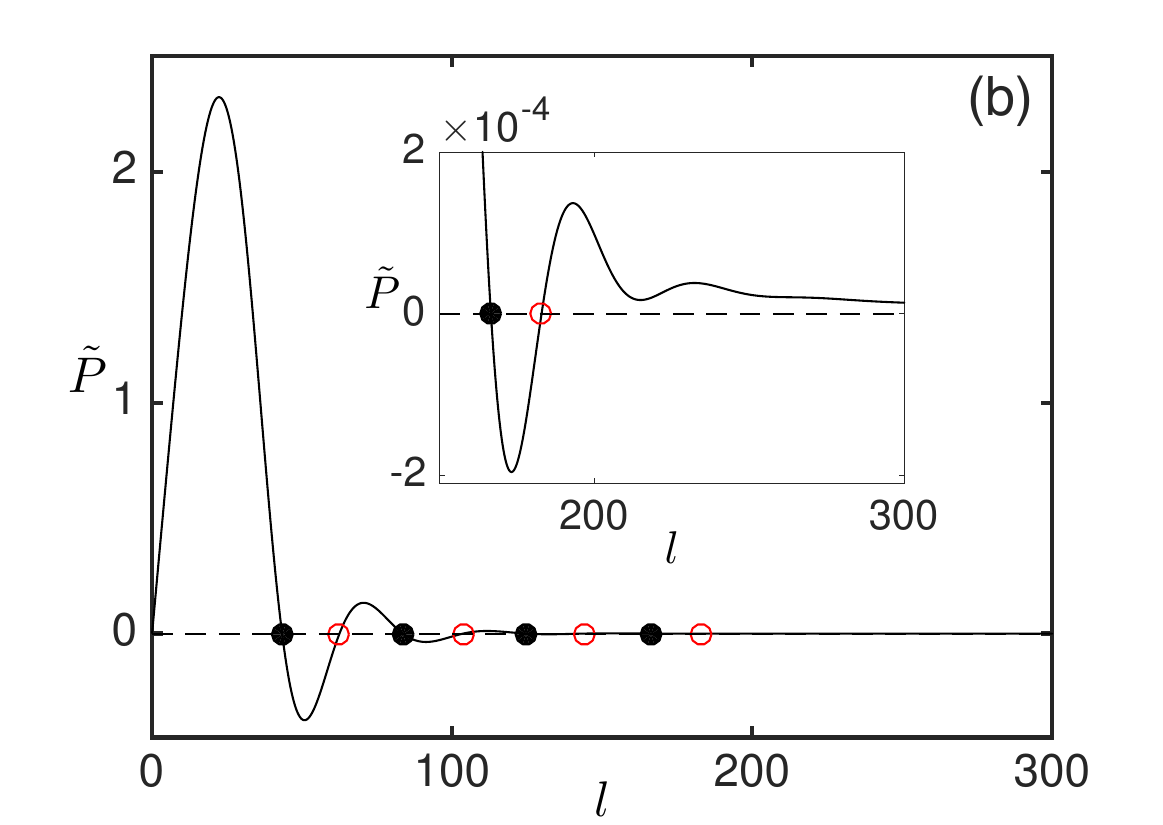}
\caption{(Color online) $\tilde P(l)$ against $l$ for $\beta=0.95\upi$ and $\Ca=0.005$, and for (a) $\We=0$ and (b) $\We=2.25$. Bound states are marked as filled circles (stable) and empty circles (unstable). The inset in panel (b) shows the behaviour for large $l$, which becomes monotonic.}
\label{fig:ptilde}
\vspace{-0.2cm}
\end{figure}

To validate the theoretically predicted pulse dynamics (\ref{eq:twopulse}), we numerically simulated the dynamics of a two-pulse system using a superposition of two pulses as an initial condition for various parameter values and initial separation distances. We found excellent agreement between the simulations and the theory. Similarly good agreement was found for simulations with three and four pulses.
For a physical explanation of pulse attraction and repulsion, see \cite{Duprat2009}.

%%%%%%%%%%%%%%%%%%%%%%%%%%%%%%%%%%%%%%%%%
\section{Fully nonlinear solutions for Stokes flow}\label{sec:stokes}
%%%%%%%%%%%%%%%%%%%%%%%%%%%%%%%%%%%%%%%%%

We now study fully nonlinear pulse solutions at zero Reynolds number. We discuss the
case of a perfect conductor film, that is $\varepsilon_p\to \infty$. The
extension to the case of a liquid dielectric is straightforward
\cite[e.g.][]{tseluiko2008effect}.

We first reformulate
the problem using the boundary-integral method. % \cite[e.g.][]{pozred92}.
%Thereafter, our computational approach
%to find pulse solutions will be to first seek periodic travelling waves, and then to gradually increase the spatial period until we obtain
%solutions which are localised within a long period with approximately flat tails.
%Assuming a travelling wave solution, we work in a frame of reference moving with the wave.
%We use variables which have been made dimensionless according to the prescription laid out in \S~\ref{sec:long}.
The flow is assumed to be periodic in $x$ with half-period $L$.
%Relative to a stationary laboratory frame, the travelling frame is
%moving at a constant, but as yet unknown, speed $c^*$ in the direction parallel to the wall.
We decompose the steady flow velocity $\bu$ and stress tensor $\stress$ into a basic flow part and a disturbance part,
writing $\bu=\bu^B + \bu^D$ and $\stress = \stress^B + \stress^D$. In the travelling frame, the basic flow details are
\bea \label{basicflow}
\bu^B = (y(2-y)\sin \beta - c^*,0), \quad \stress^B = -p_a\zh{I} + 2(1-y)
\begin{pmatrix}
-\cos \beta  & \sin \beta \\
\sin \beta    & -\cos \beta
\end{pmatrix},
\eea
where $p_a$ is the constant dimensionless ambient pressure above the film, and the remaining symbols have been defined
%$\mu$ is the fluid viscosity, and $h_0$ is the thickness of the undisturbed flat
%film as defined
in \S~\ref{sec:long}. Setting $c^*=0$ in (\ref{basicflow}), we recover the classical Nusselt solution for unidirectional flow down
an inclined plane. %\cite[e.g.][]{batchelor2000introduction}.
We note that according to the decomposition described, the
disturbance velocity $\bu^D$ vanishes at the wall, $y=0$.

Following the boundary-integral formalism, we obtain the integral equation for the
disturbance velocity and disturbance traction
$\mbf^D\equiv \stress^D\cdot \bn$ at a point
$\bx_0=(x_0,y_0)$ located on the free surface,
\bea \label{bie}
\hspace{-0.4cm}2\upi u_j^D(\bx_0) = -\int_F G_{ij}(\bx,\bx_0)f_j^D(\bx)\,\dd l(\bx) + \mathrm{p.v.} \int_F u_i^D(\bx)T_{ijk}(\bx,\bx_0)n_k(\bx)\,\dd l(\bx),\,\,
\eea
where p.v. denotes the principal value, $F$ denotes one period of the free surface, $l$ is arc length along $F$, $\bn$ is the unit normal at the free surface
pointing into the fluid, $G_{ij}(\bx,\bx_0)$ is the periodic Green's function for Stokes flow, which vanishes
when $\bx$ is located on the wall at $y=0$, and $T_{ijk}(\bx,\bx_0)$ is the corresponding stress tensor. Closed form expressions
for $\zh{G}$ and $\zh{T}$ can be found in \cite{pozrikidis2002practical}. The kinematic condition at the free surface in the travelling frame requires that
\bea \label{kincon}
\bu^D\cdot \bn = -\bu^B\cdot \bn.
\eea

Following the boundary-integral method for Laplace's equation \cite[e.g.][]{pozrikidis2002practical}, we obtain the integral equation for the electric potential,
\begin{equation}\label{eq:Electric_BI_1}
\begin{array}{rl}
\displaystyle\frac{1}{2}\varphi_2(\zh{x}_0) = % \int_{F}\phi(\zh{x}) \zh{n}(\zh{x})\cdot\nabla G(\zh{x},\,\zh{x}_0)\:\mathrm{d}l(\zh{x}) \displaystyle +
-\int_{F} G(\zh{x},\zh{x}_0) \bn(\bx) \cdot \nabla \varphi_2 (\zh{x})\:\mathrm{d}l(\zh{x}) - y_0 + B,
\end{array}
\end{equation}
where $B$ is an {\it a priori} unknown constant, and $G$ is the singly-periodic upward-biased Green's function with half-period $L$, given by \cite[see][p. 261]{pozrikidis2002practical}
\bea
G(\zh{x},\zh{x}_0) = -\frac{1}{4\upi}\log \left[ 2 \left (\cosh[k(y-y_0)] - \cos[k(x-x_0)] \right )\right ] - \frac{y-y_0}{4L}.
\eea
By integrating Laplace's equation for $\varphi$ over one period of the semi-infinite region above $F$, we obtain the integral condition
\bea \label{intcon}
%\int_F \bn\cdot\nabla \phi \:\dd l(\bx) = E_0 L,
\int_F \bn\cdot\nabla \varphi_2 \:\dd l(\bx) = 2L,
\eea
which is to be satisfied along with (\ref{eq:Electric_BI_1}).

The problem for the electric field and the problem for the fluid flow are coupled together via the dynamic stress boundary condition at the free surface,
\begin{eqnarray} \label{dynstress}
%\mbf^D = -\mbf^B -(\kappa/\Ca  + p_a)\zh{n} + 2\We \zh{M}\bcdot\zh{n},
\mbf^D = -\mbf^B -\frac{\kappa}{\Ca}\zh{n} + \We |\nabla \varphi |^2\zh{n}.
\end{eqnarray}

To complete the formulation, we specify two more conditions akin to those
imposed for the long-wave model. %First, we remove the translational invariance of any travelling-wave solution by stipulating that the wave maximum lies at the midpoint of the computational domain $[-L,L]$, which requires that $\mbox{d}h^*/\mbox{d} x = 0$ at $x=0$, where $h^*(x)$ is the travelling-wave profile. For the remaining condition, one option is to fix the volume of fluid in each computational period. Since we wish to compute solitary-pulse solutions, it is more convenient to fix the height of the film at $x=0$ by demanding that $h^*(\pm L)=1$.
First, we remove the translational invariance by imposing the condition $h^*_x=0$ at $x=0$, and to break the `volume' symmetry, we impose the condition $h^*=1$ at $x=L$.
By fixing the location of the wave maximum and the film height in this way, we efficiently compute pulse solutions (where they exist) by increasing $L$.

With these final two conditions, the coupled integral formulation
comprising (\ref{bie}), (\ref{eq:Electric_BI_1}) and (\ref{intcon}), together with the kinematic
condition (\ref{kincon}) and the dynamic stress condition (\ref{dynstress}) are solved numerically
using the boundary-element method by discretising one period of the {\it a priori} unknown free surface, $F$, with a sequence of
$N$ connected straight elements and treating the unknown variables as constants over the elements. In this way, we derive a set
of nonlinear algebraic equations to be solved using Newton's method for these unknown constant element values, the free-surface profile $h^*$ and the wave speed $c^*$. Full details of the implementation ofthe method are provided by \cite{tseluiko2008effect} and, in the interest of brevity, we do not repeat them here.
%Having made an initial guess for the free-surface shape, the nonlinear algebraic equations are solved using Newton's method to produce the free-surface profile $h^*$, the disturbance velocity $\bu^D$ and disturbance traction $\mbf^D$, and the wave speed speed $c^*$.
%
\begin{figure}
\centering
    \begin{minipage}{1.1\textwidth}
        \centering
\hspace{-1.35cm}
\includegraphics[width=1.9in]{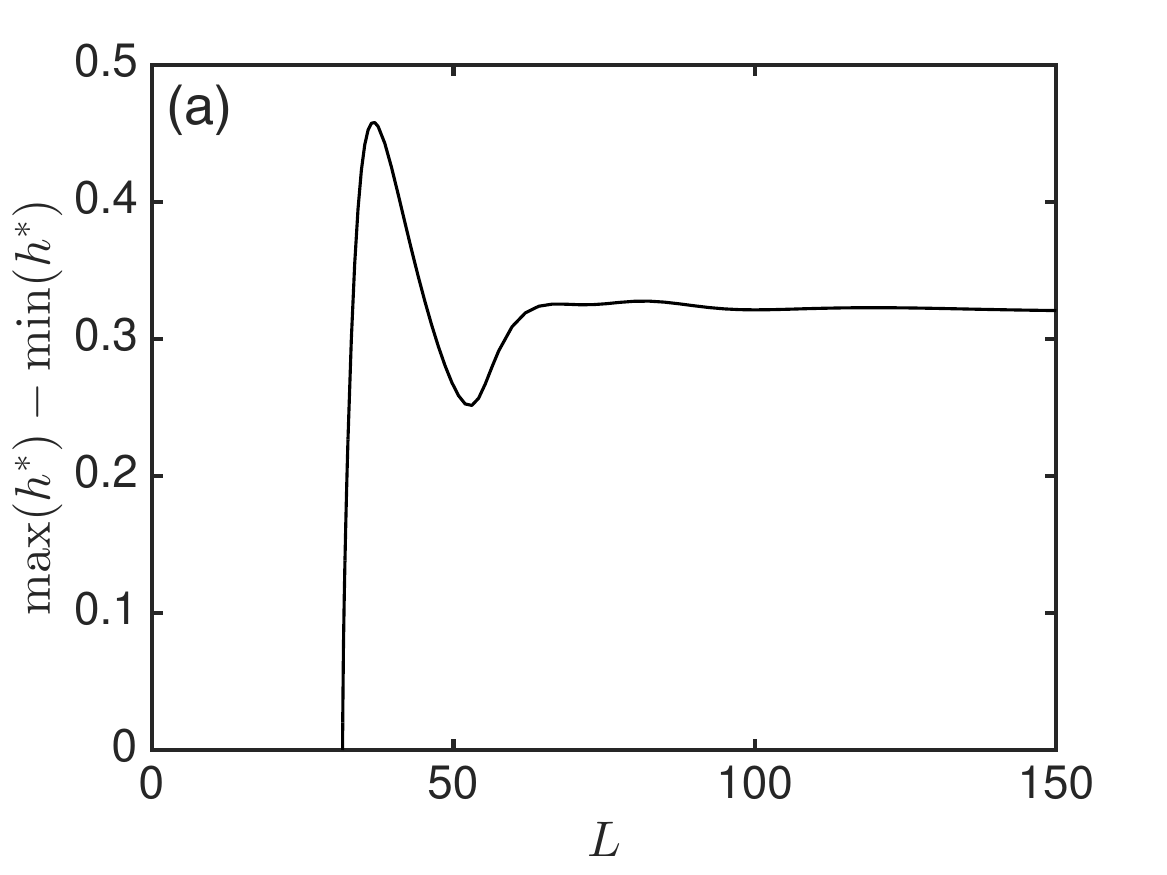}
\hspace{-0.35cm}
\includegraphics[width=1.8in]{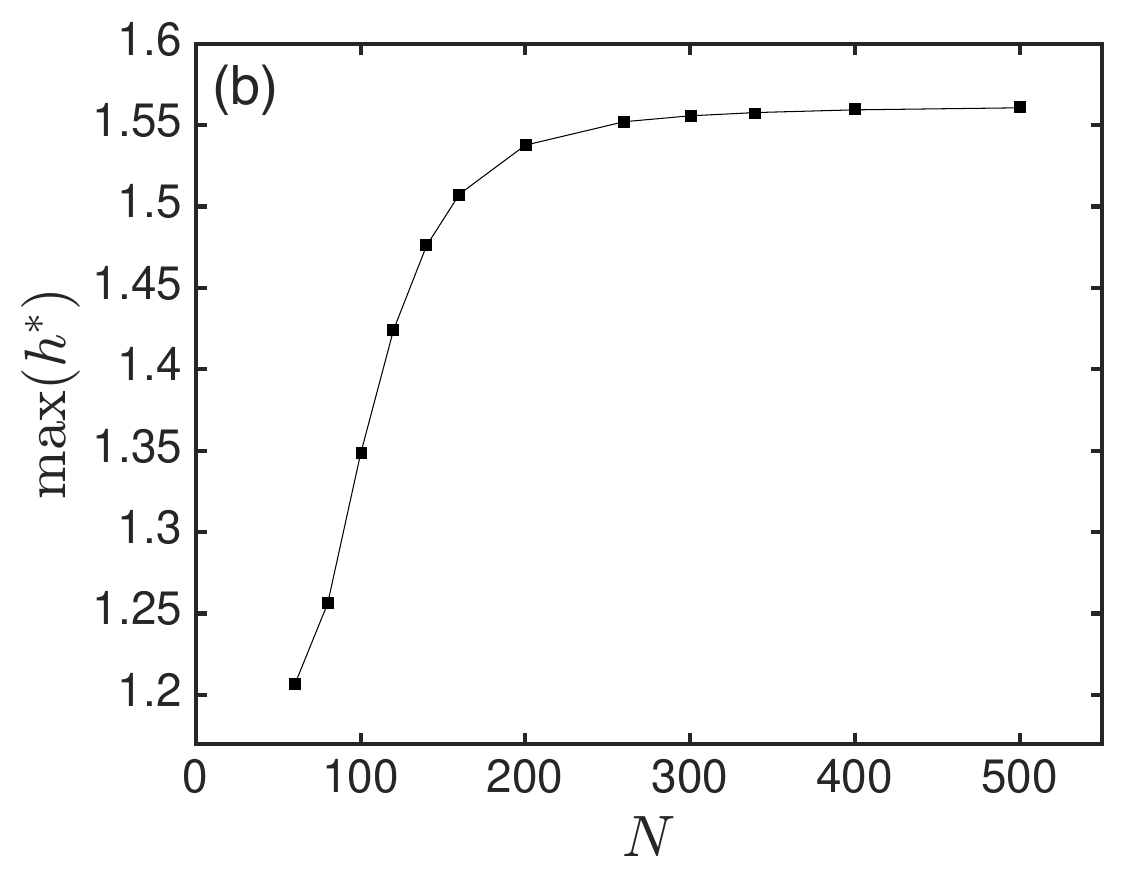}
\hspace{-0.15cm}
\includegraphics[width=1.8in]{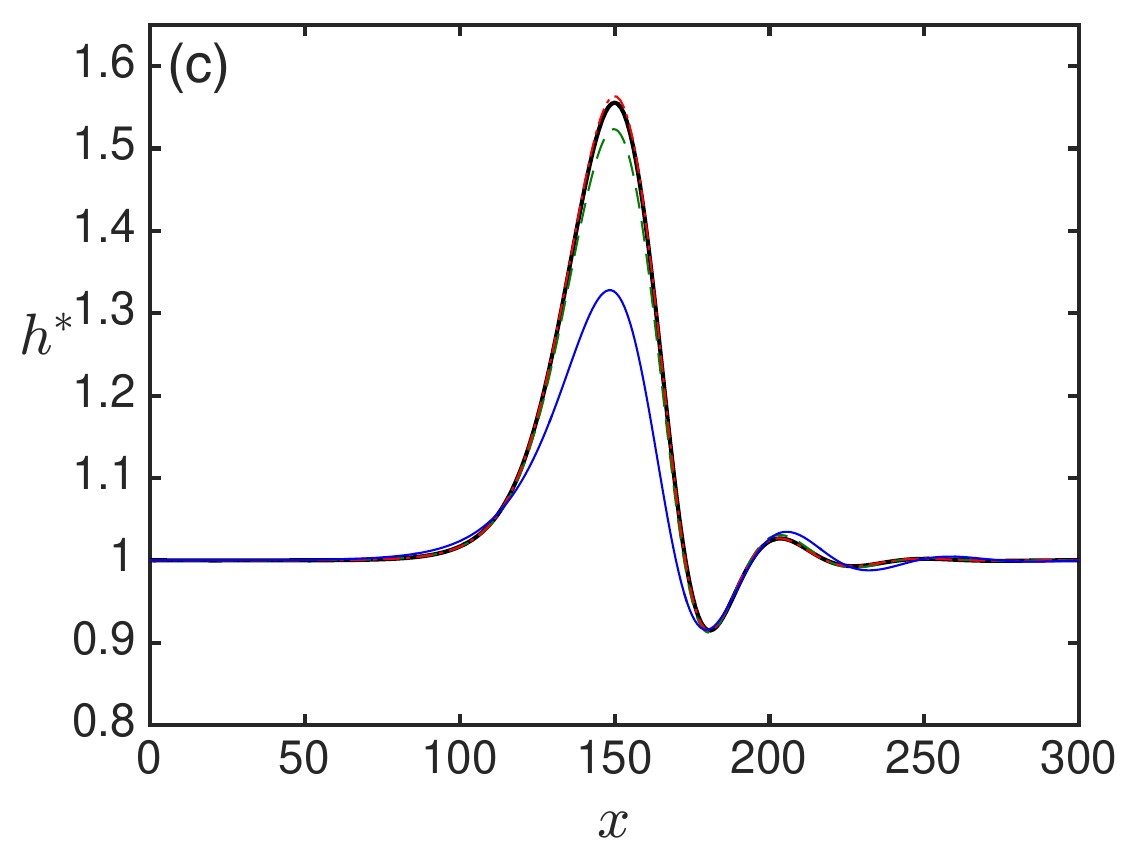}
    \end{minipage}%
    \vspace{-0.1cm}
\caption{(Color online) Stokes flow computations with $\beta=0.95\upi$, $\Ca=0.005$, $\We=0$: (a) Travelling-wave branch for Stokes flow showing the wave height, $\max(h^*)-\min(h^*)$, against $L$, computed using $N=300$ boundary elements. ({\it b}) Numerical convergence for $L=150$ showing $\max(h^*)$ against $N$. ({\it c}) Pulse profile for $L=150$ and $N=300$ (solid line) with the long-wave predictions using \eqref{eq:TFE1} and \eqref{eq:TFE3} (dashed, dot-dashed lines, respectively) and the KS prediction (thin solid line). The wave speed is $c^* = 0.457$ (boundary-element solution) and $c^*=0.449$ (long-wave solution to equation \ref{eq:TFE1}) and $c^*=0.392$ (KS equation \ref{KSeq}).
}
\mylab{fig:branch}
\vspace{-0.2cm}
\end{figure}

A standard normal-mode analysis for small-amplitude periodic
waves \citep[see][]{blyth2008effect} reveals that the cut-off wavenumber coincides with that found from
the long-wave analysis. The properties of the neutral mode are used to construct an initial guess for the Newton iterations and to latch onto the travelling-wave solution branch. This is followed using
continuation in $L$, and in the case where $L$ may be increased indefinitely, we are able to compute solitary-pulse
solutions.
The solution space is determined by the three dimensionless parameters $\Ca$, $We$ and $\beta$ introduced
in \S~\ref{sec:long}.

%
%
%%%%%%%%%%%%%%%%%%%
\begin{figure}
\centering
%\subfigure[]{\includegraphics[width=2.6in]{decay_ca0p75_prof.eps}}
\includegraphics[width=2.0in]{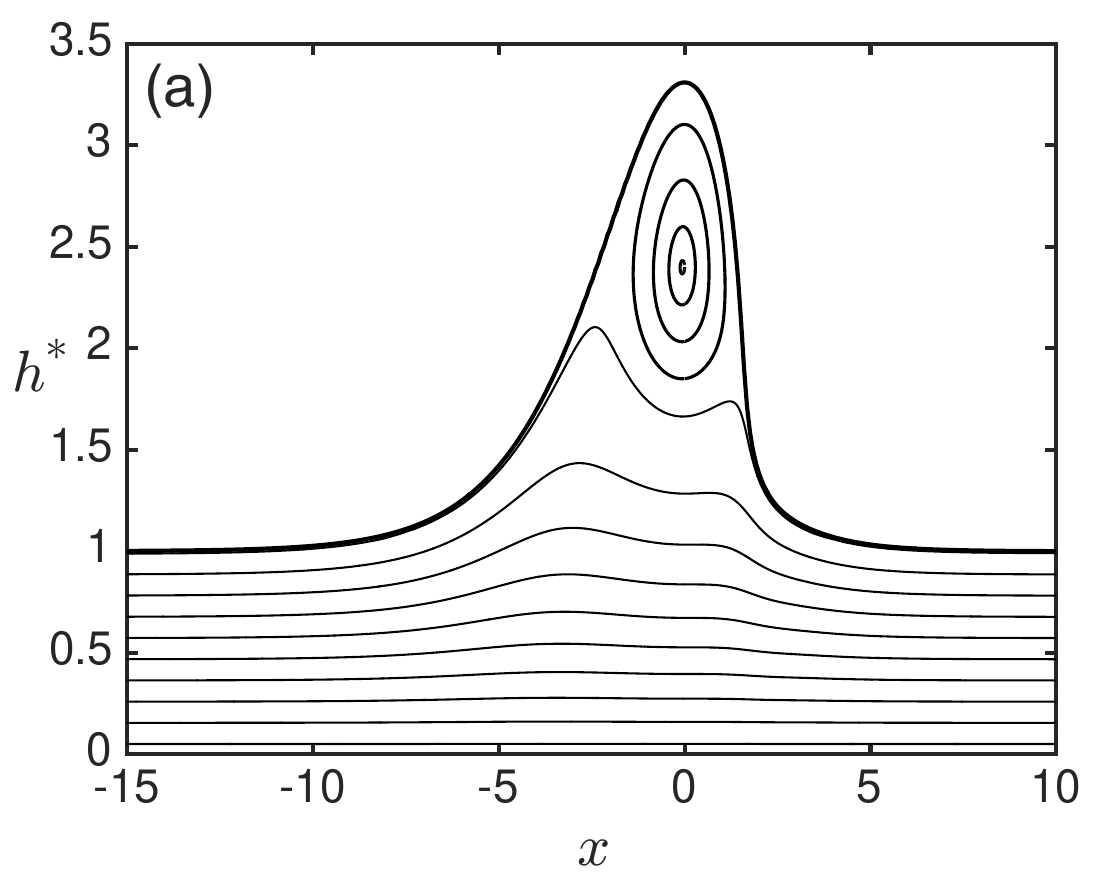}
\hspace{0.75cm}
\includegraphics[width=2.0in]{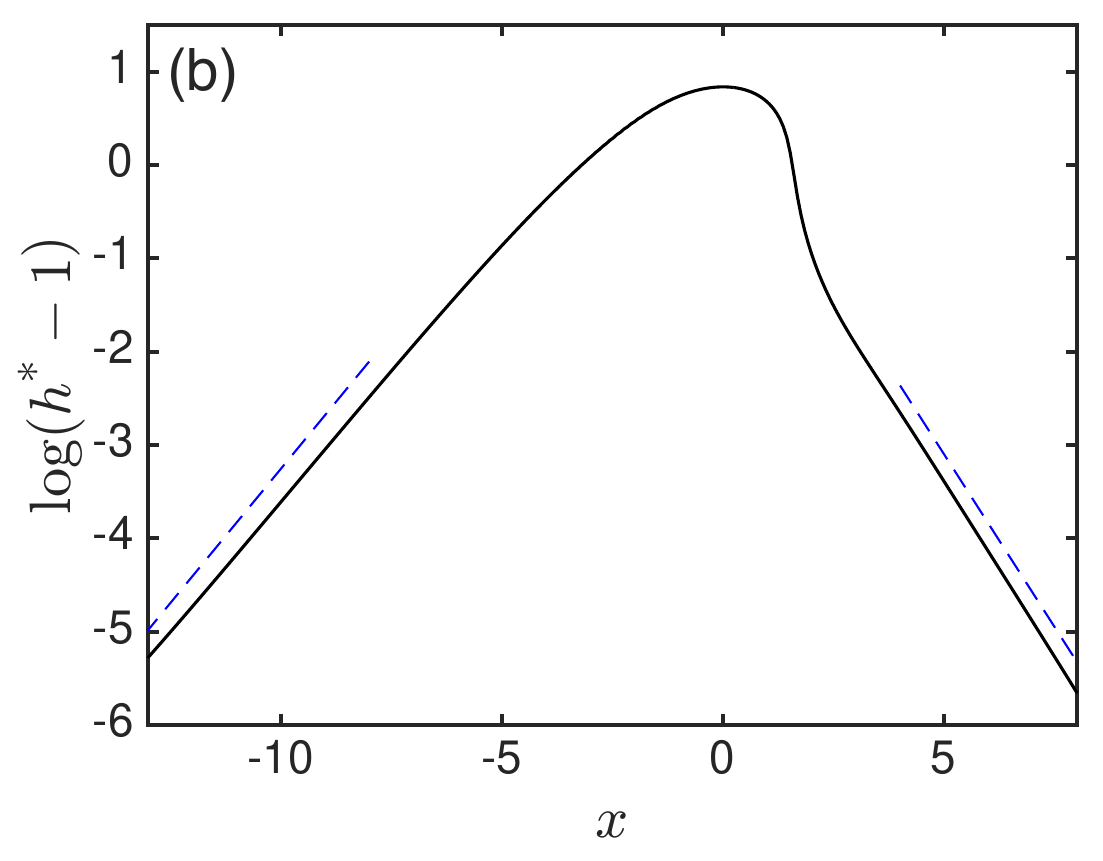}
\vspace{-0.1cm}
\caption{(Color online) Stokes flow calculation for $\beta=0.75\upi$ with $\Ca=0.3$ and $W_e=0$, and with $L=16.0$ and
$N=400$. ({\it a}) Pulse profile and streamlines. ({\it b}) Logarithm of the pulse profile (solid line) with broken lines of gradient
$0.576$ in $x<0$ and $-0.741$ in $x>0$ corresponding to the decay rate predictions (see Appendix \ref{sec:farfield}). The wave speed is $c^*=3.82$.
}
\mylab{fig:ca0p75}
\vspace{-0.2cm}
\end{figure}
%%%%%%%%%%%%%%%%%%%
%
We begin by computing fully nonlinear pulse solutions for no electric field, setting $\We=0$. In figure \ref{fig:branch}(a), we show the results of a sequence of
calculations for increasing domain size $L$. The wave height approaches a constant value, and a pulse solution is
obtained, at around $L \approx 200$. It was found that $N=300$ boundary elements
were sufficient to obtain an accurate wave solution. The numerical convergence is demonstrated
in figure \ref{fig:branch}(b), which shows the variation of the pulse maximum with the number of boundary elements
$N$ at $L=150$. The pulse itself is shown in figure \ref{fig:branch}(c) with a thick solid line. The long-wave
prediction based on \eqref{eq:TFE1} and the weakly-nonlinear prediction based on the KS equation
\eqref{KSeq} are also shown in this figure. The pulse speed is found to be
%$c^* = c/U_0$, where $U_0=\rho g h_0^2/2\mu$ is the surface speed of the equivalent flat vertical Nusselt film, is found to be
%
% NB: U_0 = 1/2 for most of our calculations.
%
$c^*=0.457$ for the boundary-element calculation and $c^*=0.449$
for the long-wave calculation, and so there is reasonable agreement between the two values.
% LW pulse speed in Mark_Dmitri_Tesheng/fri22ndJan2016_boundstate/LW_bound_states/Explain!.

Notably, the performance of the KS equation is rather poor, even at this small value of the Bond number.
Nevertheless, we have confirmed that the long-wave and KS solutions converge as the Bond number is further reduced.
Evidently the long-wave and boundary-element calculations agree very well with a small discrepancy
around the maximum of the pulse. The discrepancy is significantly reduced by including higher-order terms in the long-wave model equation; specifically including second-order terms in the derivation we obtain the extended form of \eqref{eq:TFE1},
\begin{equation}\label{eq:TFE3}
h_t + q_x = 0,
\end{equation}
where $q=q_0+q_1+q_2$, with $q_0=O(1)$, $q_1=O(\delta)$, $q_2=O(\delta^2)$. Specifically,
\begin{equation}
q_0 = \frac{2}{3}(\sin \beta)h^3  - \frac{2}{3}(\cos \beta)h^3 h_x  + \frac{1}{3\C} h^3 h_{xxx}  + \frac{2\We}{3} \,h^3\,\mathcal{H}[h_{xx}],
\end{equation}
and $q_1 = F_1 + \mathcal{E}_1$,
where $F_1$ contains terms all of which vanish when $Re=0$, and
\bea
\mathcal{E}_1 =  \frac{1}{3}\We\,h^3 \left( E_{2x} + 2 h_{x}h_{xx} \right),
\eea
\cite[see][]{tseluiko2010dynamics}
and
\begin{align}
\nonumber
q_2 = &h^3\Bigg(
\sin \beta\left(2 h h_{xx}  + \frac{14}{3}   h_x^2\right)
-2 \cos \beta \Big(
4h h_x h_{xx}  + \frac{3}{5} h^2 h_{xxx}
+ \frac{7}{3}h_x^3
\Big)
\\
&
\hspace{-1.1cm}
+
\frac{1}{\Ca }
\left (
3 h h_x h_{xxxx}
- h_x h_{xx}^{2}
+ \frac{11}{6} h_x^2 h_{xxx}
+ hh_{xx} h_{xxx}
+ \frac{3}{5}h^2 h_{xxxxx}
\right )
\Bigg)
+ F_2 + \mathcal{E}_2,
\end{align}
where $F_2$ contains terms all of which vanish when $Re=0$. The electric-field contribution $\mathcal{E}_2$ must be found by considering the electric field problem at the appropriate order, but is not needed for the present non-electrified case.
Equation \eqref{eq:TFE3} is valid provided that $\Ca=O(\delta^2)$, $\We=O(\delta^{-1})$, $\cot \beta=O(\delta^{-1})$ and $Re=O(1)$.
The pulse solution to this equation is shown in figure \ref{fig:branch}(c) with a dot-dashed line which almost coincides with the thick solid line representing the boundary-element solution.

The pulse profile in figure \ref{fig:branch}(c) decays monotonically on the upstream side and has an oscillatory decay
on the downstream side. Since the pulse speed is greater than the speed of linear long waves, namely $2\sin \beta = 0.313$,
this is consistent with the predictions of the decay rate calculations discussed in Appendix~\ref{sec:farfield}.
These calculations predict that when the Bond number is not small the decay is monotonic both upstream
and downstream of the pulse maximum. Figure \ref{fig:ca0p75}(a) shows the pulse profile for $\Ca=0.3$ and
$\beta=0.75\upi$. The pulse speed is $c^* = 3.82$.
%
% Pulse speed is c = 1.9118456 from BEM file cws.m. So dimless speed is 1.9118/U_0 (and U_0=1/2) here.
%
Evidently the decay is monotonic on both sides of the pulse. For these parameter values the calculation described in Appendix~
\ref{sec:farfield} yields the upstream and downstream decay rates $0.576$ and $-0.741$, respectively, and these show good
agreement with the profile calculated using the boundary-element method, as can be seen in Figure \ref{fig:ca0p75}(b). The
figure also shows streamlines inside the film in a frame of reference travelling at the speed of the pulse. These indicate
the presence of a trapped eddy in the main part of the pulse. Solitary wave eddies have recently been observed experimentally
on a gravity-driven film at non-zero Reynolds number by \cite{reck2015recirculation}.

\begin{figure}
\centering
\includegraphics[width=2.0in]{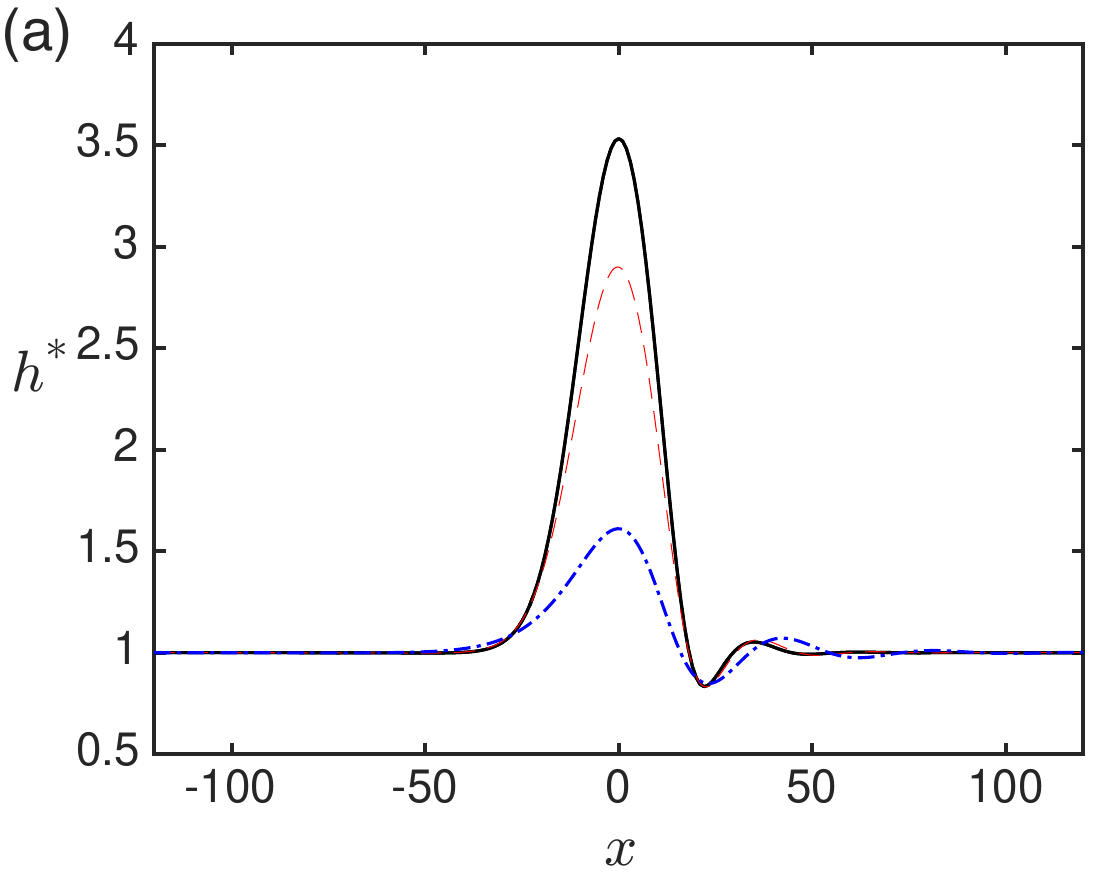}
\hspace{0.75cm}
\includegraphics[width=2.0in]{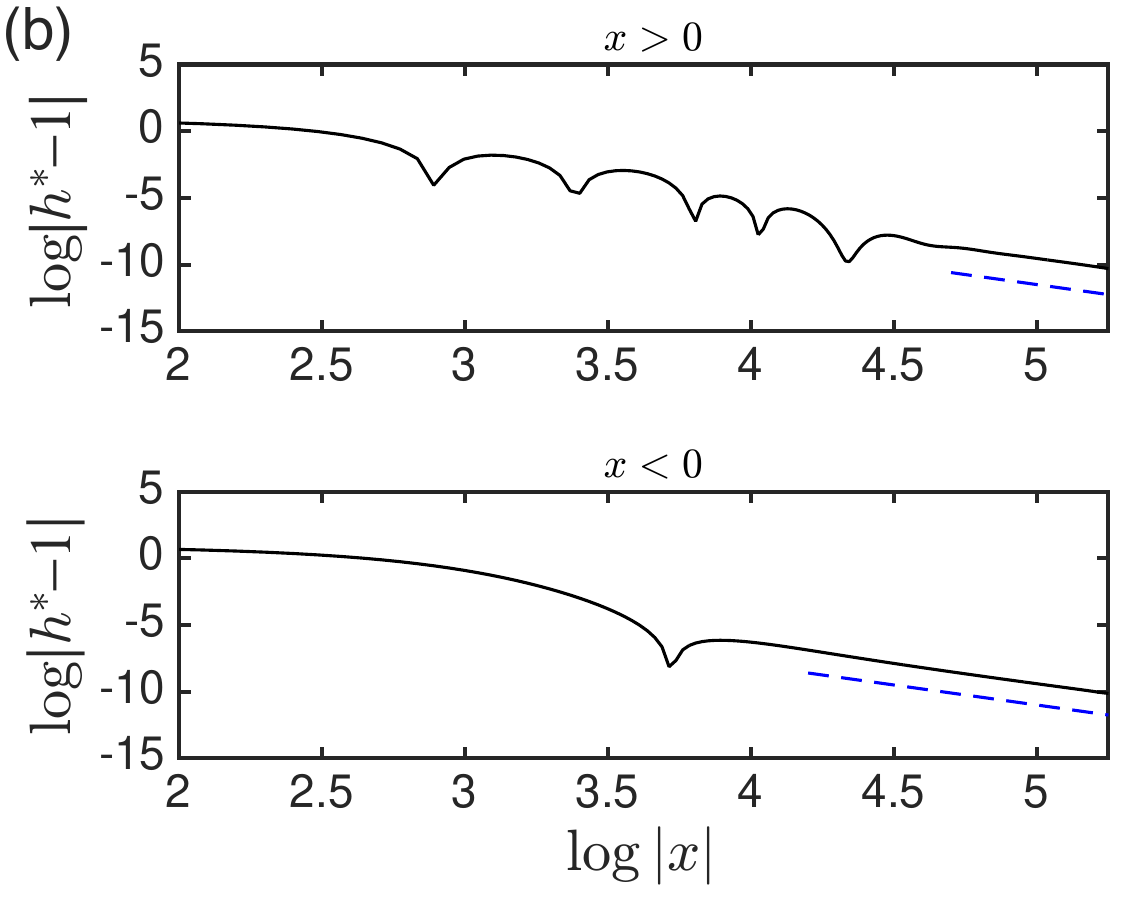}
\vspace{-0.1cm}
\caption{(Color online) Electrified Stokes flow computation for $\beta=0.95\upi$, $\Ca=0.005$, $We=6.25$, and
$L=250$, $N=500$. ({\it a}) Pulse profile (solid line) with the long-wave prediction (dashed line) and the KS prediction (dot-dashed line).
({\it b}) A $\log$-$\log$ plot of the pulse profile in $x>0$ and $x<0$ showing algebraic decay in the far-field with
the expected slope (dashed line) according to the asymptotic theory presented in Appendix~\ref{sec:farfield}.
The wave speed is $c^* = 1.09$ (boundary-element solution) and $c^*=0.88$ (long-wave solution) and $c^*=0.46$ (KS solution).
%{\bf Serafim suggests a possible explanation for the discrepancy is the importance of non-locality in the higher order terms in the LW equation derivation, which are missed out. These may slow the convergence of the series, making higher order terms more important for accuracy.}
}
\mylab{fig:elecalgdecay}
\vspace{-0.2cm}
\end{figure}
An electrified solitary-pulse solution is shown in figure
\ref{fig:elecalgdecay}(a). According to the theory of Appendix~\ref{sec:farfield},
the far-field decay of an electrified pulse is algebraic and so a wide
computational domain and a large number of boundary elements are needed for
an accurate computation. The prediction of the long-wave model equation
\eqref{eq:TFE1} is also shown in the figure with a broken line. Once again, we
see that there is good agreement over most of the pulse profile except near
to the main peak. The visible difference between the Stokes calculation and
the long-wave one at the pulse maximum is exacerbated in the presence of the
electric field (compare the results in figure \ref{fig:branch}c). As for
the non-electrified case studied in figure \ref{fig:branch}(b, c), we would
expect the agreement to improve on using the extended long-wave model
equation \eqref{eq:TFE3}. However, this would require computation of the
corresponding electric field contribution $\mathcal{E}_3$.
%A possible
%explanation for this discrepancy is that the non-locality introduced by the electric field is increasingly poorly captured by the first
%order long-wave model \eqref{eq:TFE1}, and that higher order terms may be needed to furnish a more accurate prediction (see
%\cite{kalliadasis2011falling} section 5.1.2 for a discussion of higher order long-wave Benney-type models).
%{\bf Comment on even worse performance of KS equation in this figure, but note that KS and long-wave converge as $Bo\to 0$}
The decay rate of the pulse tails as $|x|\to \infty$ is investigated in figure~\ref{fig:elecalgdecay}(b).
%To faithfully
%reproduce the expected decay, this calculation was done on a wide domain with $L/h_0=500$ and with a large number of boundary
%elements, $N=500$.
The broken lines shown in the figure indicate the algebraic decay rate
expected from the asymptotic theory of Appendix~\ref{sec:farfield}. The excellent
agreement provides strong evidence of algebraic decay in the far-field and
lends strong credence to the decay rate predicted by the asymptotic theory.
The pulse speed determined from the boundary-element solution is $c^*=1.09$;
this compares with $c^*=0.88$ obtained from the long-wave theory. %, namely the travelling-wave form of equation (\ref{eq:TFE1}).
{\color{black} We note that the pulse travels faster in the
presence of the electric field. %(compare with figure \ref{fig:branch}(c), where $c^*=0.457$).
This trend is in line with the long-wave theory.
% -- see figure \ref{fig:bet_0_25pi_Ca_0_01}(b).}

%
% Pulse speed is c = 0.54699 from BEM file cws.m. So dimless speed is 0.54699/U_0 (and U_0=1/2) here.
%
%
\begin{figure}
\centering
\includegraphics[width=2.0in,height=1.5in]{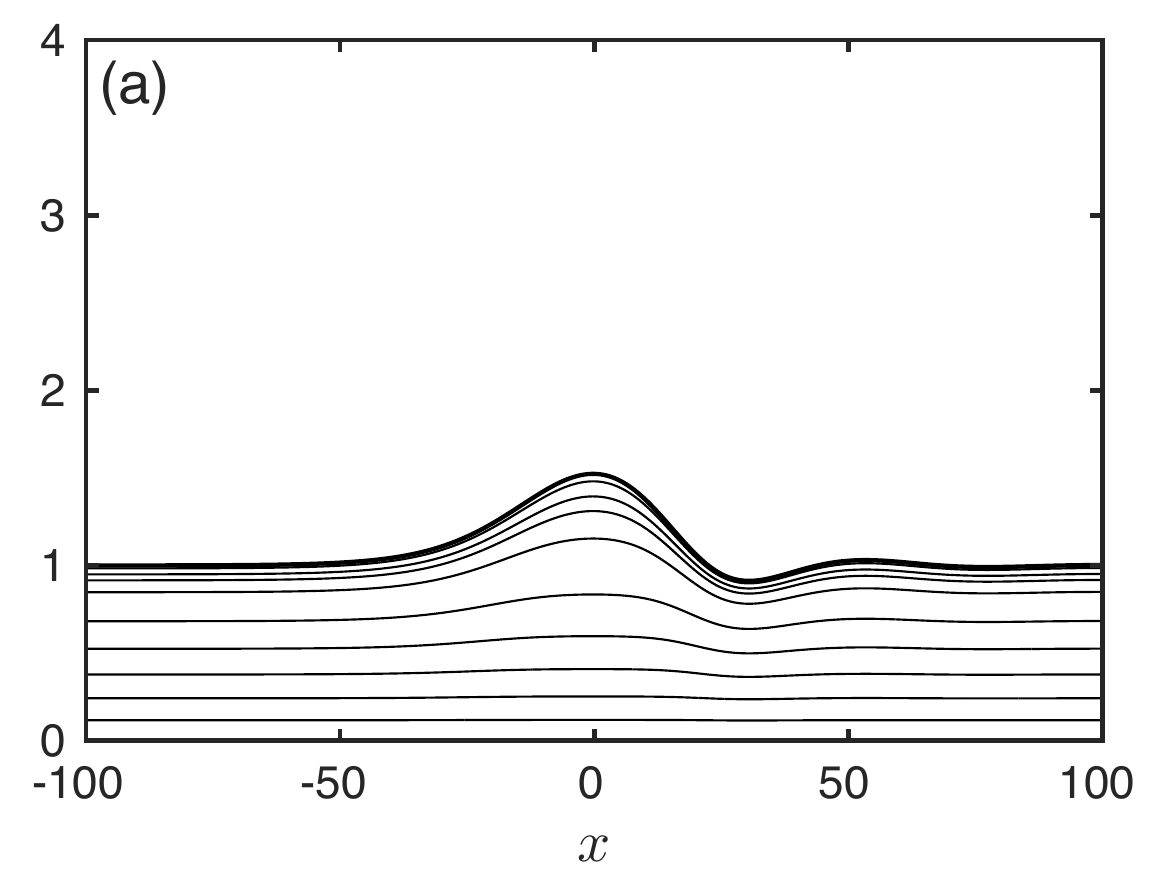}
\hspace{0.75cm}
\includegraphics[width=2.0in,height=1.5in]{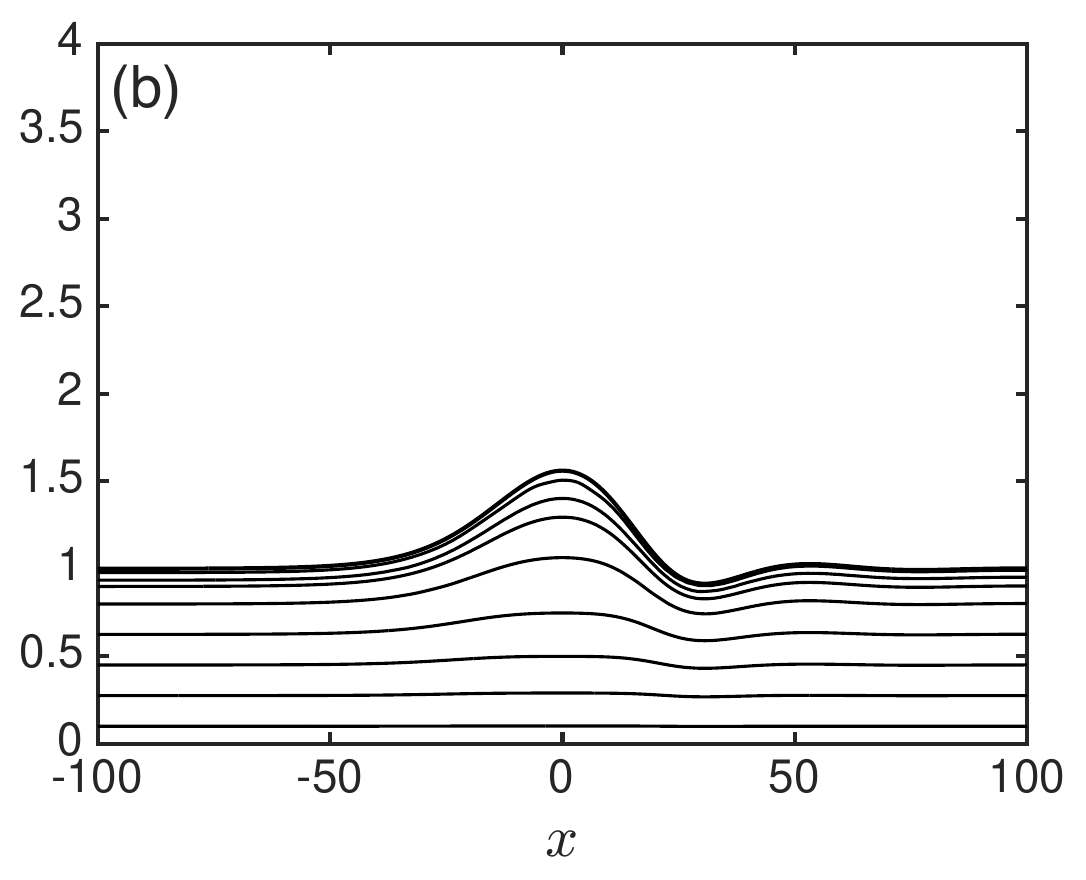}
\includegraphics[width=2.0in,height=1.5in]{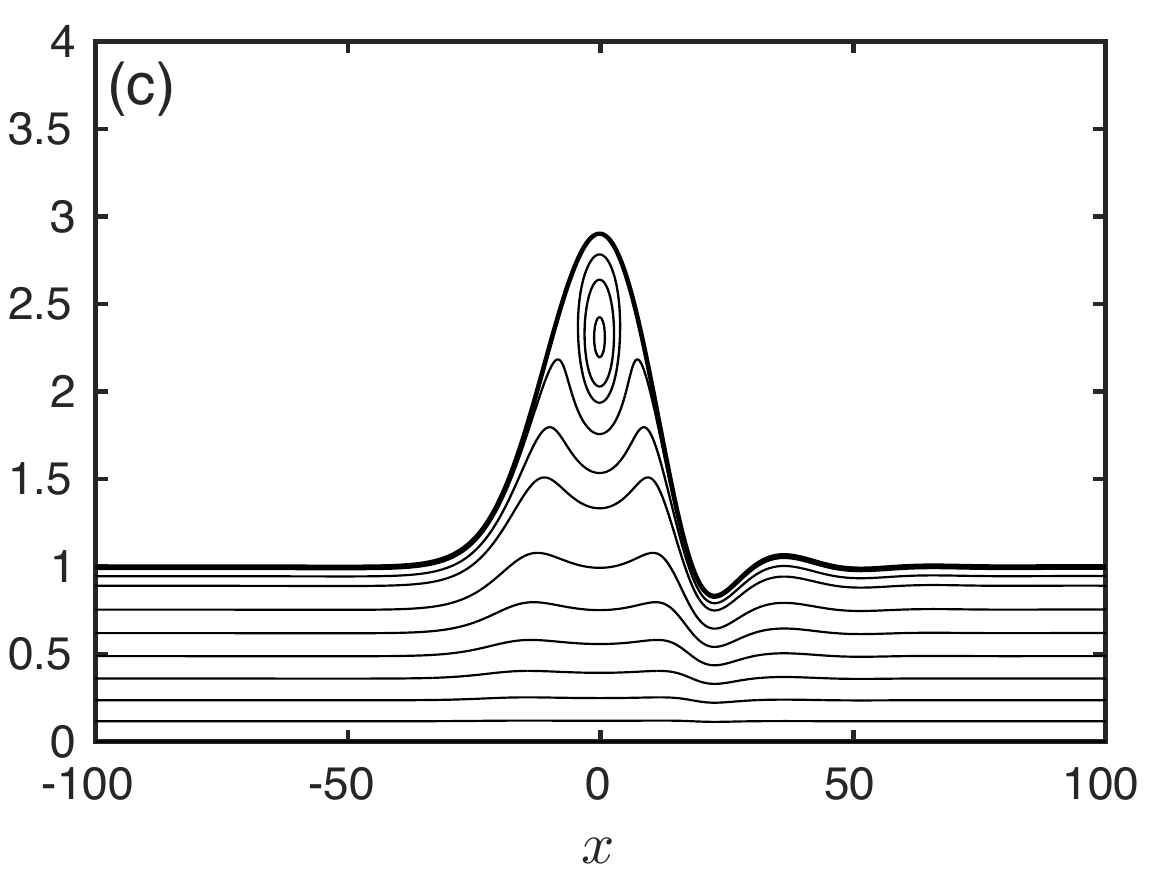}
\hspace{0.75cm}
\includegraphics[width=2.0in,height=1.5in]{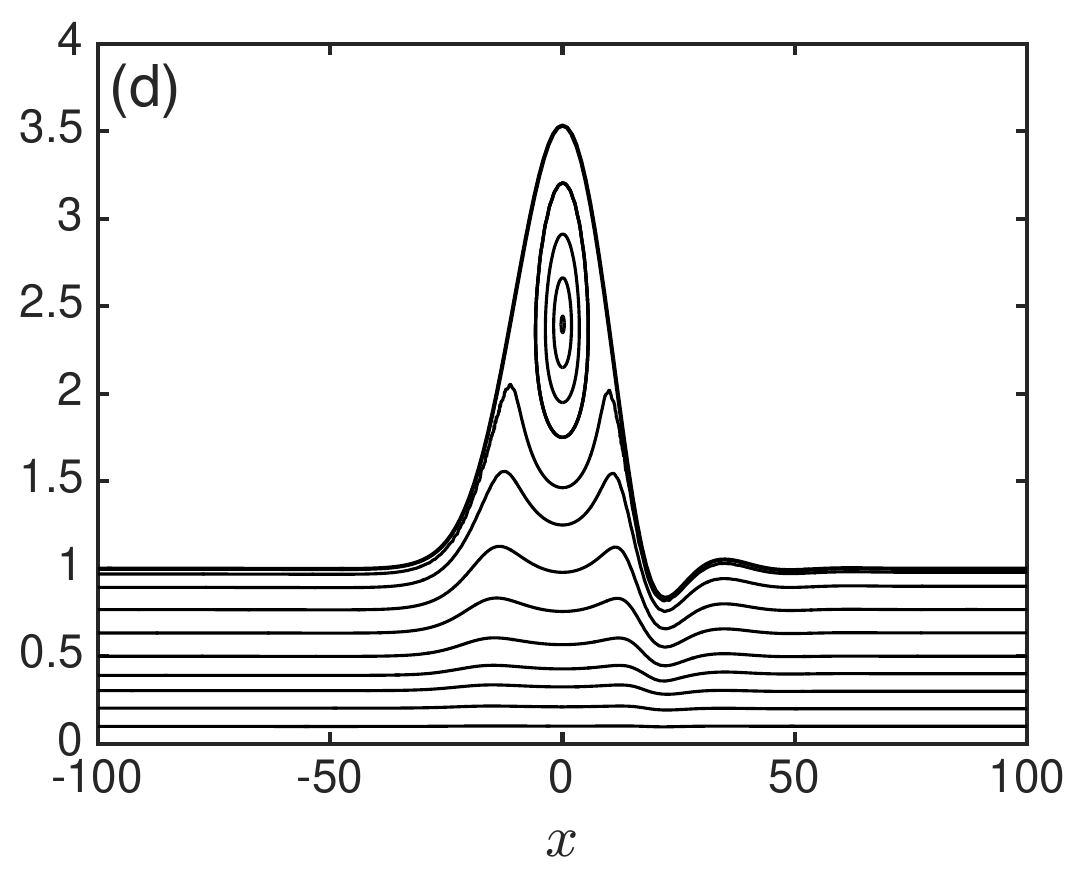}
\vspace{-0.1cm}
\caption{(Color online)
Streamline patterns in a frame moving at the pulse speed for the case $\beta=0.95\upi$ with $\Ca=0.005$: (a, b) $\We=0$, (c, d) $\We=6.25$. We note that (b) and (d) are results for Stokes model while (a) and (c) are for long-wave model. The pulse profiles are shown with a thick line and the streamlines with thin lines.
%The pulse profiles are also in figure \ref{fig:branch}(b) and the electrified pulse shown in figure
%\ref{fig:elecalgdecay}(a) with $\We=6.25$.
}
\mylab{fig:streamlines}
\vspace{-0.2cm}
\end{figure}

Figure \ref{fig:streamlines} depicts the streamline patterns for
$\beta=0.95\upi$ and $\Ca=0.005$ in a frame moving with the pulse for % a non-electrified pulse with
$ \We=0$ (panels a and b) and % an electrified pulse with
$\We=6.25$ (panels c and d). Panels (a) and (c) show results for the
long-wave model and panels (b) and (d) show the results of boundary-element
calculations.
%The stream function for the long-wave model in a frame moving at the pulse speed $c^*$ is given by
%$\Psi(x, y) = \int_0^y u\,\mbox{d} y - c^* y$, where $u$ is given in \eqref{uform}, so that
%\[
%\Psi(x, y) = \left(\sin \beta - (\cos\beta)  h^*_{x} +\frac{1}{2\Ca} h^*_{xxx}
%+ \We \ms{H}[h^*_{xx}]\right)\left(-\frac{y^3}{3} + h^*y^2\right) - c^* y.
%\] \dt{Do we really need $\Psi$ here?}
The two different models give broadly similar results. Notably the electric
field generates an eddy/recirculation zone inside the hump so that a quantity
of fluid is transported along with the pulse.
%This has potential consequences
%for the heat and mass transfer properties of the flow which could be used
%advantageously in industrial applications \cite[see, for
%example,][]{park2003three,kalliadasis2011falling}. We note that if closed
%streamlines exist in the moving frame, a fluid particle is trapped in both
%the moving and the laboratory frame, but the streamlines in the laboratory frame are
%not closed. In the latter frame, a particle simply moves faster and slower than the
%wave crest at the top and bottom halves of the (moving-frame) recirculation
%zone, respectively.

Next, we discuss the computation of bound states for the Stokes equations. We construct
an initial guess made from a superposition of a
converged pulse solution and a duplicate of the same pulse separated by a
nominated distance, $l$, which is taken to coincide with that given by the long-wave
theory of \S~\ref{sec:weakinter}.
%Specifically, we use
%\bea \label{iguess} y = h^*(x) + h^*(x-l) - 1, \eea where $h^*(x)$ is a pulse
%solution, which ensures that the film thickness is unity in the far field.
A good initial guess is required
as the convergence of the Newton iterations is sensitive to the separation
distance. Figure~\ref{fig:bound-state_We} shows four bound-state solutions for the case
$\Ca=0.005$ and $ \beta=0.95\upi$ for $\We=0$ or $2.25$.
%In each case the initial guess \eqref{iguess} was constructed taking $h^*$ to be the pulse
%profile in figure \ref{fig:branch}(c).
For these parameter values and for
the non-electrified case, the long-wave theory predicts the existence of
twelve bound states with separation distances: $47.30$,  $69.74$, $94.13$,
$117.62$, $141.38$, $165.05$, $188.75$, $212.42$, $236.37$, $260.01$,
$283.71$, $307.41$. We computed only a subset of these shown in the figure.
The accuracy of each calculation was confirmed by varying the number of
boundary elements and the size of the computational domain. For each case,
the wave speed $c^*$ predicted by the boundary-element computation and by the
long-wave theory are given in table \ref{table:wspeed}. Evidently, the two
are in good agreement. The wave speed is quite a lot smaller for a
bound-state than for a solitary pulse travelling alone. We also note that for
both the boundary-element and the long-wave calculations, the separation
distances, which in the case of Stokes flow are measured as the distance from
maximum to maximum, decrease as the electric field intensity is increased.
\begin{figure}
\centering
\includegraphics[width=2.1in,height=1.35in]{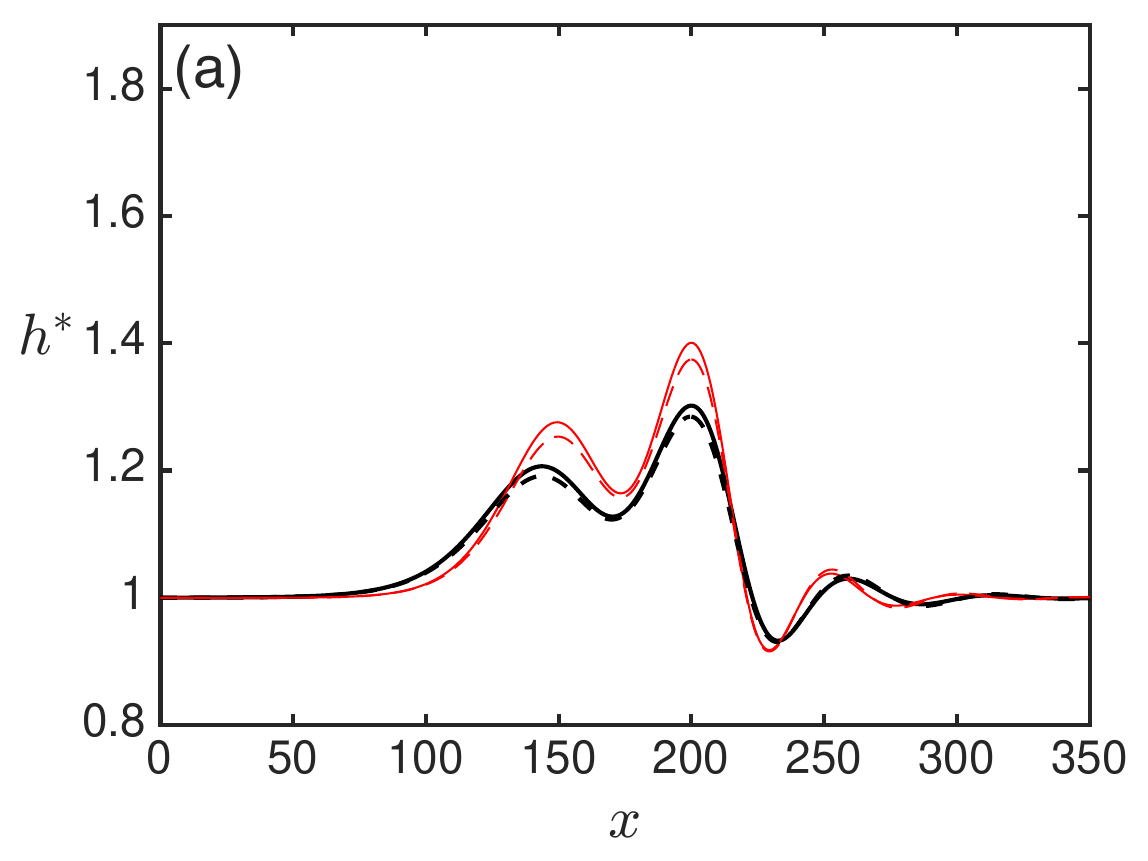}
\hspace{0.75cm}
\includegraphics[width=2.1in,height=1.35in]{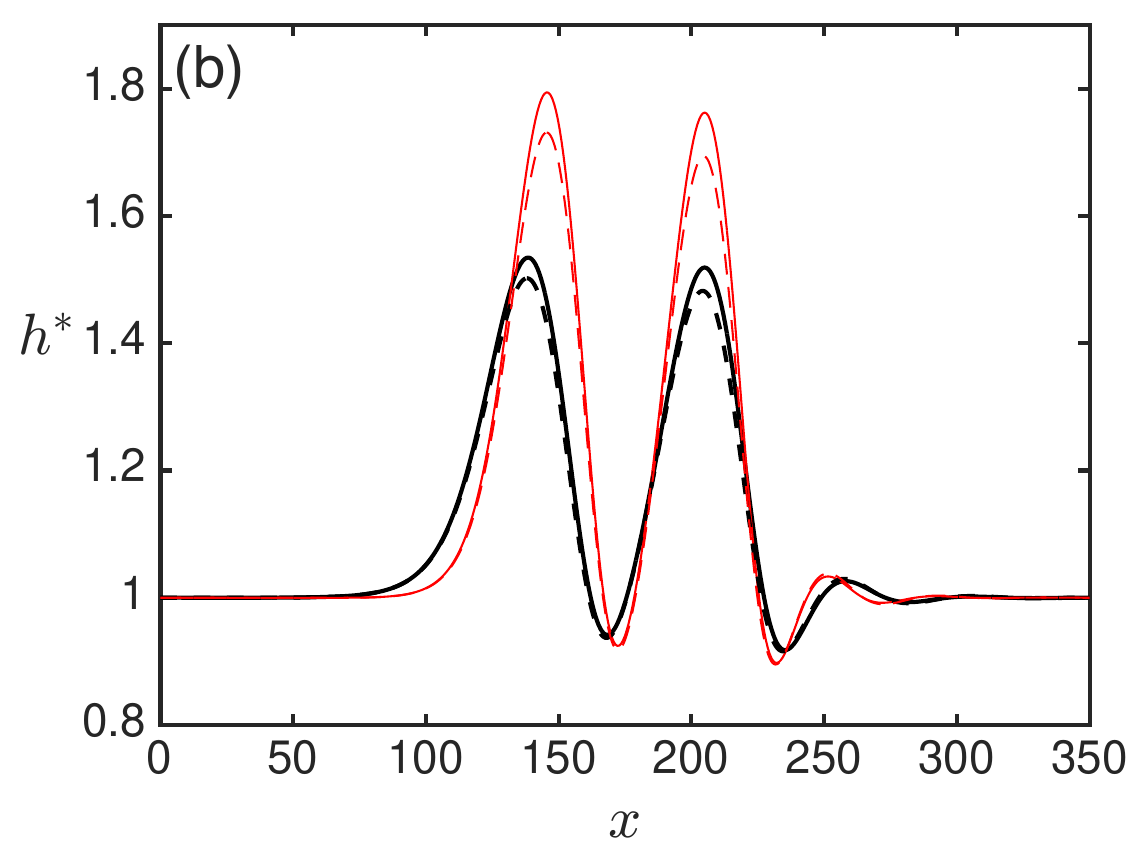}\\
\includegraphics[width=2.1in,height=1.35in]{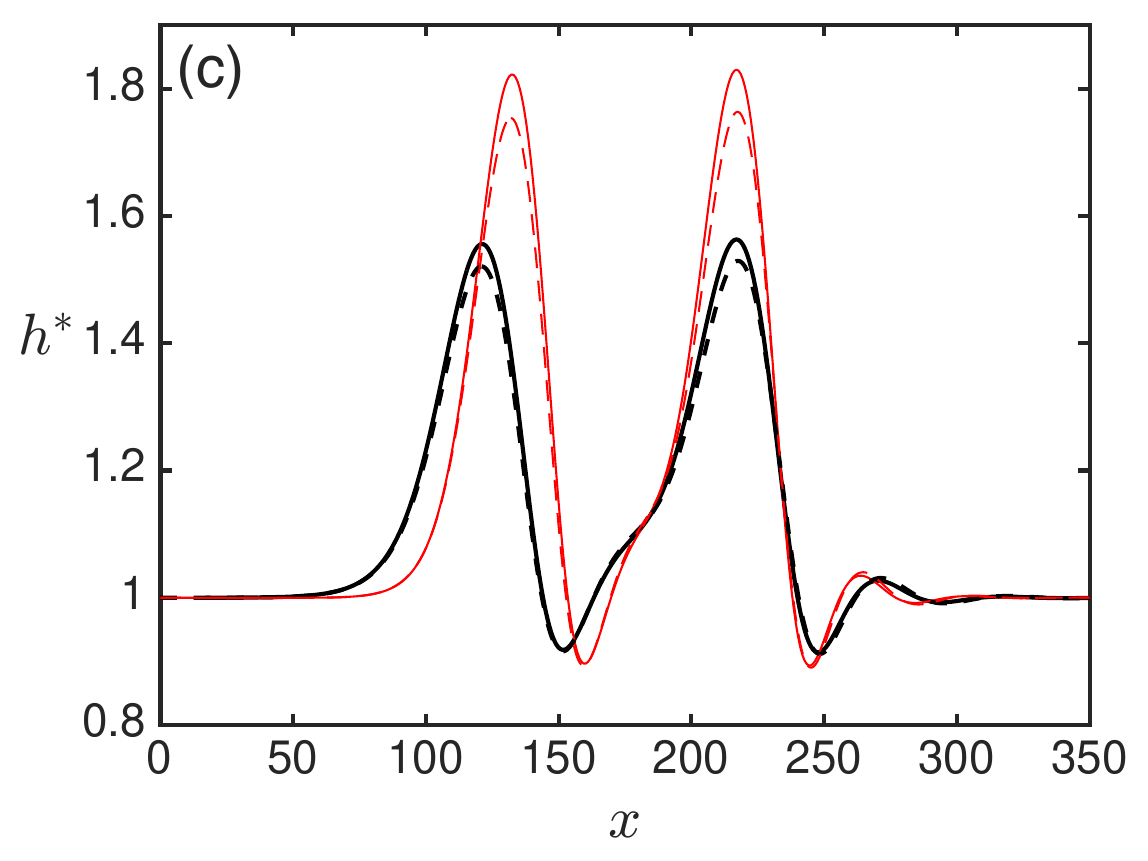}
\hspace{0.75cm}
\includegraphics[width=2.1in,height=1.35in]{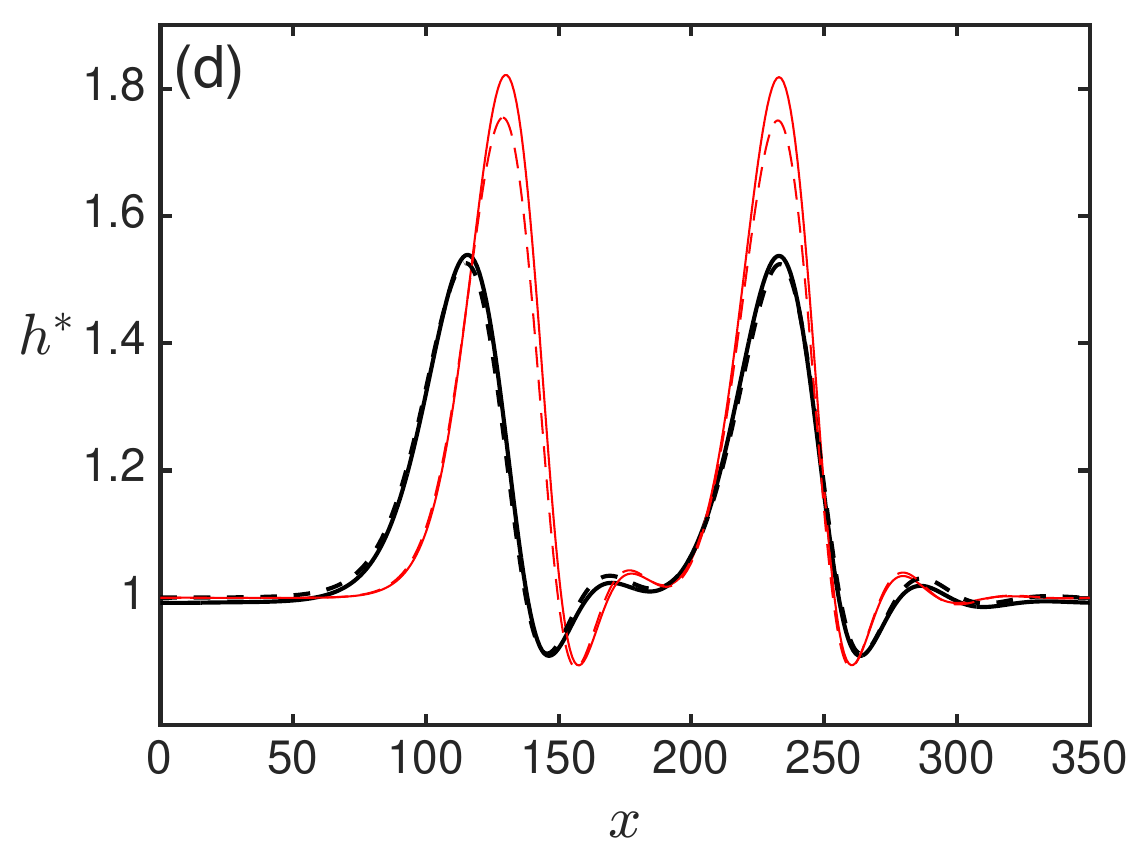}
\vspace{-0.1cm}
\caption{(Color online) Stokes flow bound-state computations for $\beta=0.95\upi$, $\Ca=0.005$ (with $L=175$, $N=500$) for $We=0$ (thick solid lines) and $We=2.25$ (thin solid lines).
The separation distances from maximum to maximum are (with electrified values in brackets):
({\it a}) $56.5$ ($50.5$), ({\it b}) $67.0$ ($60.0$), ({\it c}) $96.0$ ($84.0$), and (d) $117.0$ ($111.0$).
%Boundary element calculations (solid lines) with $L/h_0=350$ and $N=500$
The long-wave predictions for non-electrified (electrified) flows are shown with thick (thin) dashed lines. The long-wave
bound-state separation distances are (with electrified values in brackets): ({\it a}) $47.3$ ($43.4$), ({\it b}) $69.7$ ($62.5$), ({\it c}) $94.1$ ($83.6$), and (d) $117.6$ ($111.2$). The wave speeds are given in table~\ref{table:wspeed}.
}
\mylab{fig:bound-state_We}
\vspace{-0.2cm}
\end{figure}
%
%%
%%%%%%%%%%%
%\begin{table}
%%
%\centering
%{
%\renewcommand{\arraystretch}{1.8}
%\begin{tabular}{ccccccc}
%Figure label &&& $c^*_{BEM}$ &&& $c^*_{LW}$  \\
%\hline
%\ref{fig:bound-state_We}(a) &&&  $0.38$ ($0.40$)  &&& $0.38$ ($0.40$) \\
%\ref{fig:bound-state_We}(b) &&&  $0.46$ ($0.53$)  &&& $0.45$  ($0.51$) \\
%\ref{fig:bound-state_We}(c) &&&  $0.45$  ($0.52$) &&& $0.44$ ($0.51$) \\
%\ref{fig:bound-state_We}(d) &&&  $0.44$ ($0.48$) &&& $0.45$ ($0.47$) \\
%\end{tabular}
%}
%\caption{Wave speeds $c^*$ for the bound states in figure \ref{fig:bound-state_We}. Wave speeds are quoted for the Stokes flow boundary-element  computations ($c^*_{BEM}$) and for the long-wave  model, ($c^*_{LW}$). Values in brackets are for the electrified cases with electric Weber numbers $\We$ as given in the caption to figure  \ref{fig:bound-state_We}.}
%\label{table:wspeed}
%\end{table}
%%%%%%%%%%%%
%%
%
%%%%%%%%%%
\begin{table}
\centering
{
\renewcommand{\arraystretch}{1.5}
\vspace{-0.125in}
\begin{tabular}{cccccccccccccccccc}
Figure label &&& $c^*_{BEM}$ &&& $c^*_{LW}$  &&&& Figure label &&& $c^*_{BEM}$ &&& $c^*_{LW}$  \\
%\hline
\ref{fig:bound-state_We}(a) &&&  $0.38$ ($0.40$)  &&& $0.38$ ($0.40$)
&&&&
\ref{fig:bound-state_We}(c) &&&  $0.45$  ($0.52$) &&& $0.44$ ($0.51$) \\
\ref{fig:bound-state_We}(b) &&&  $0.46$ ($0.53$)  &&& $0.45$  ($0.51$) &&&&
\ref{fig:bound-state_We}(d) &&&  $0.44$ ($0.48$) &&& $0.45$ ($0.47$)
\end{tabular}
}
\caption{Wave speeds for the bound states in figure \ref{fig:bound-state_We} quoted for the Stokes flow boundary-element  computations ($c^*_{BEM}$) and for the long-wave  model, ($c^*_{LW}$). Values in brackets are for the electrified cases with electric Weber numbers $\We$ as given in the caption to figure  \ref{fig:bound-state_We}.}
\label{table:wspeed}
\vspace{-0.2cm}
\end{table}
%%%%%%%%%%%
%

In \S~\ref{sec:LW_steady}, using the long-wave model, we found that
in the case of electrified flow at an acute inclination angle, the travelling-wave
branches which emerge from the two neutral stability points at $L=L_{\pm}$ either connect to form a continuous hoop from
one to the other or else each branch continues independently to large $L$ ultimately producing a solitary-pulse
solution. Figures \ref{fig:case6} and \ref{fig:case7} demonstrate that the same behaviour is observed for Stokes flow. These two figures use the same parameter values as for the long-wave calculations in figures
\ref{fig:bet_0_25pi_Ca_0_01_We_12_5} and \ref{fig:bet_0_25pi_Ca_0_01_We_13_5}. In figure~\ref{fig:case6} ($\We=12.5$), we see that the neutral stability points at $L_+\approx 19.21$ and $L_{-} \approx 36.32$ are connected by a single continuous
travelling-wave branch. Sample profiles along the branch are also shown in the figure. In figure~\ref{fig:case7} ($\We=13.5$), two
independent branches emerge from the neutral points at $L_+\approx 15.80$ and $L_-\approx 44.18$, and continue until pulse solutions are finally attained for large $L$. The pulse profiles are also shown
in the figure.
%%%%%%%%%%%%%%%%%
\begin{figure}
\centering
    \begin{minipage}{1.1\textwidth}
        \centering
\hspace{-1.35cm}
\includegraphics[width=1.80in]{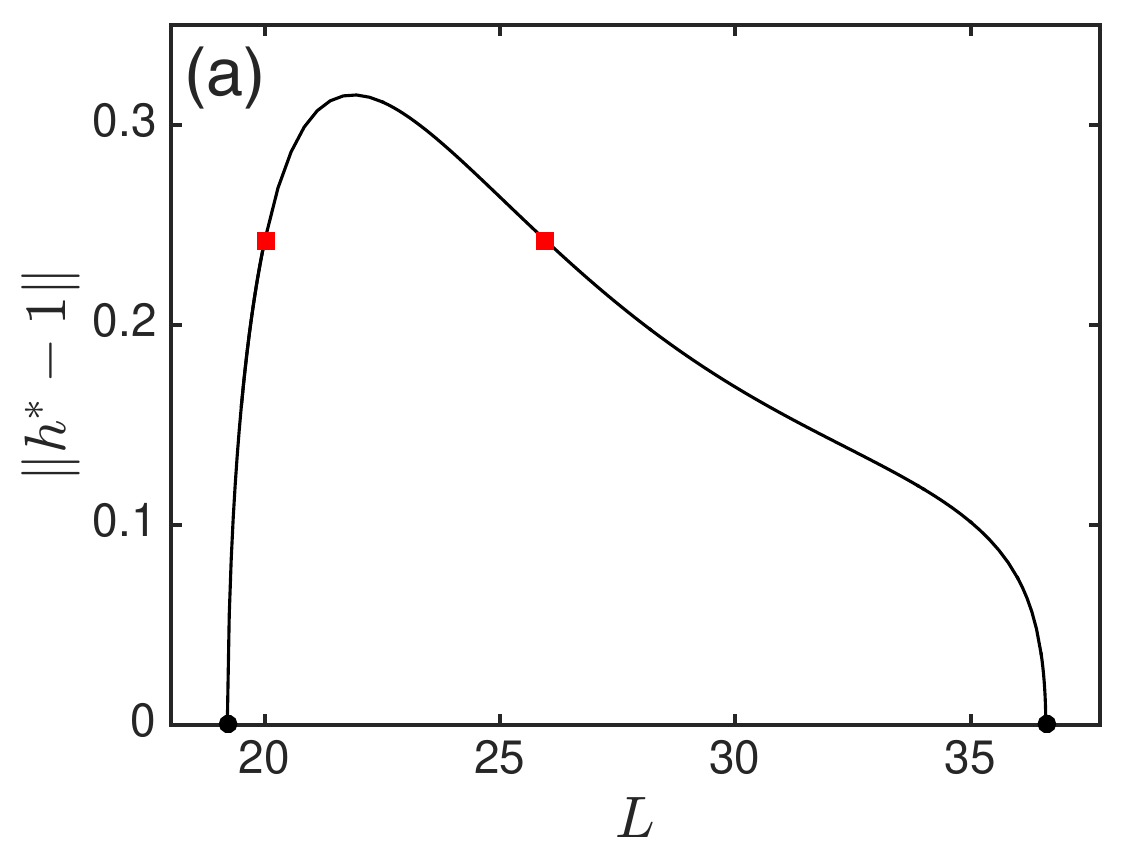}
\hspace{-0.2cm}
\includegraphics[width=1.82in]{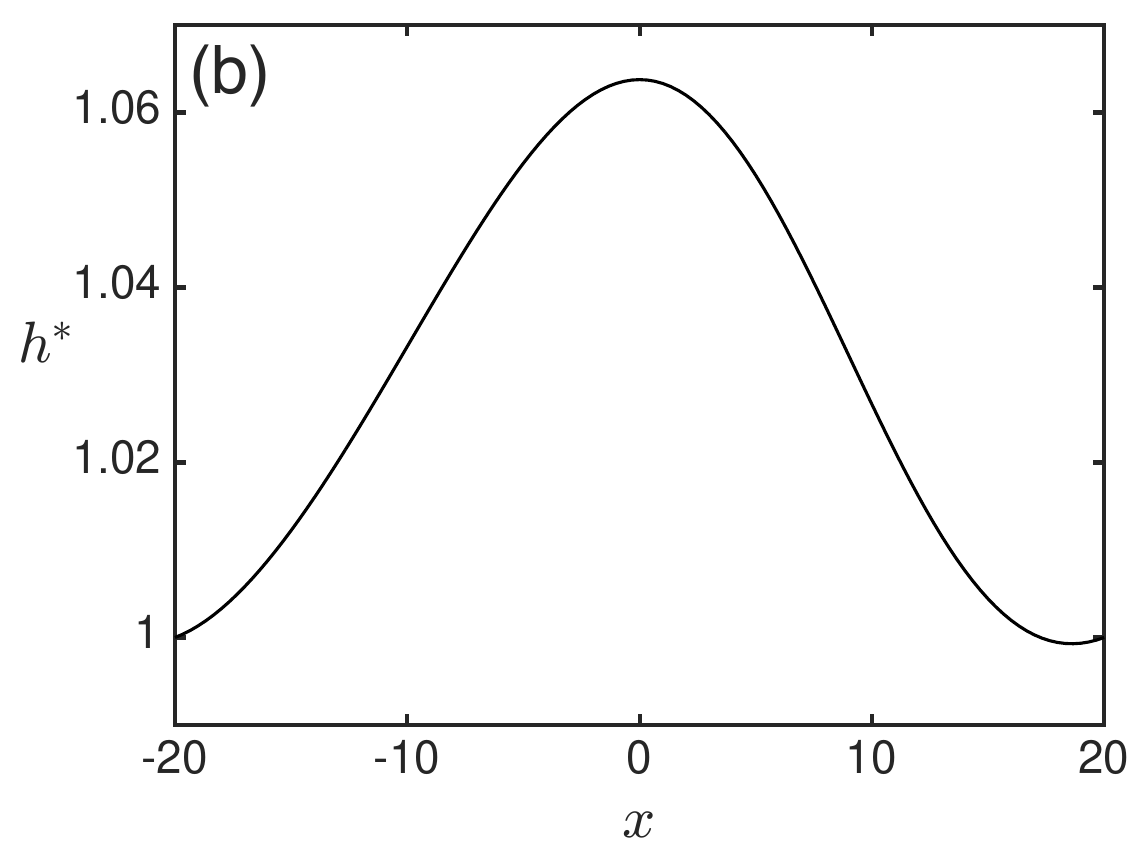}
\hspace{-0.2cm}
\includegraphics[width=1.80in]{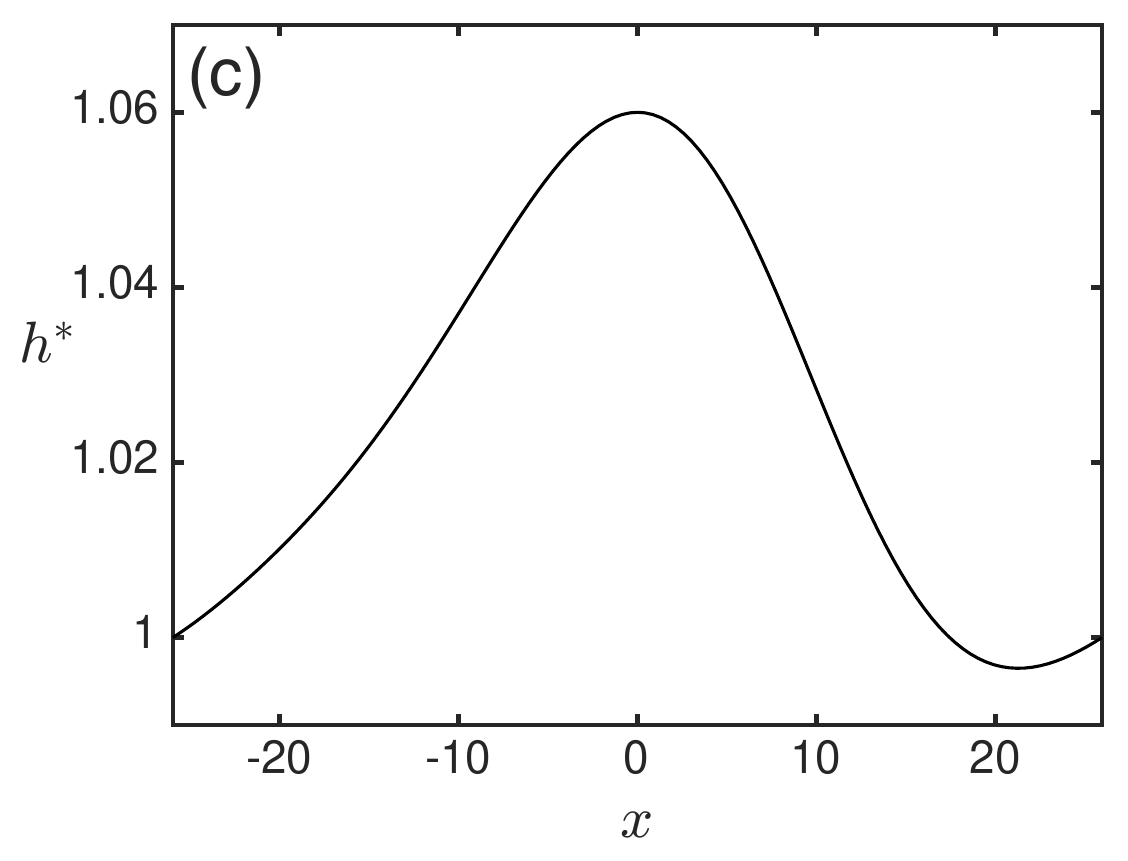}
    \end{minipage}%
    \vspace{-0.1cm}
\caption{(Color online) Stokes flow computations for $\beta=0.25\upi$, $\Ca=0.01$, $\We=12.5$:
(a) Bifurcation diagram for $\| h^* -1 \|$ against $L$.
%The bifurcation points are at $L_+\approx 19.21$ and $L_-\approx 36.32$.
(c, d) Wave profiles with $\| h^*-1\| = 0.2427$, indicated by the left and right squares in (a), respectively.
}
\mylab{fig:case6}
\vspace{-0.2cm}
\end{figure}
%%%%%%%%%%%%

%%%%%%%%%%%%%%%%%
\begin{figure}
\centering
    \begin{minipage}{1.1\textwidth}
        \centering
\hspace{-1.35cm}
\includegraphics[width=1.82in]{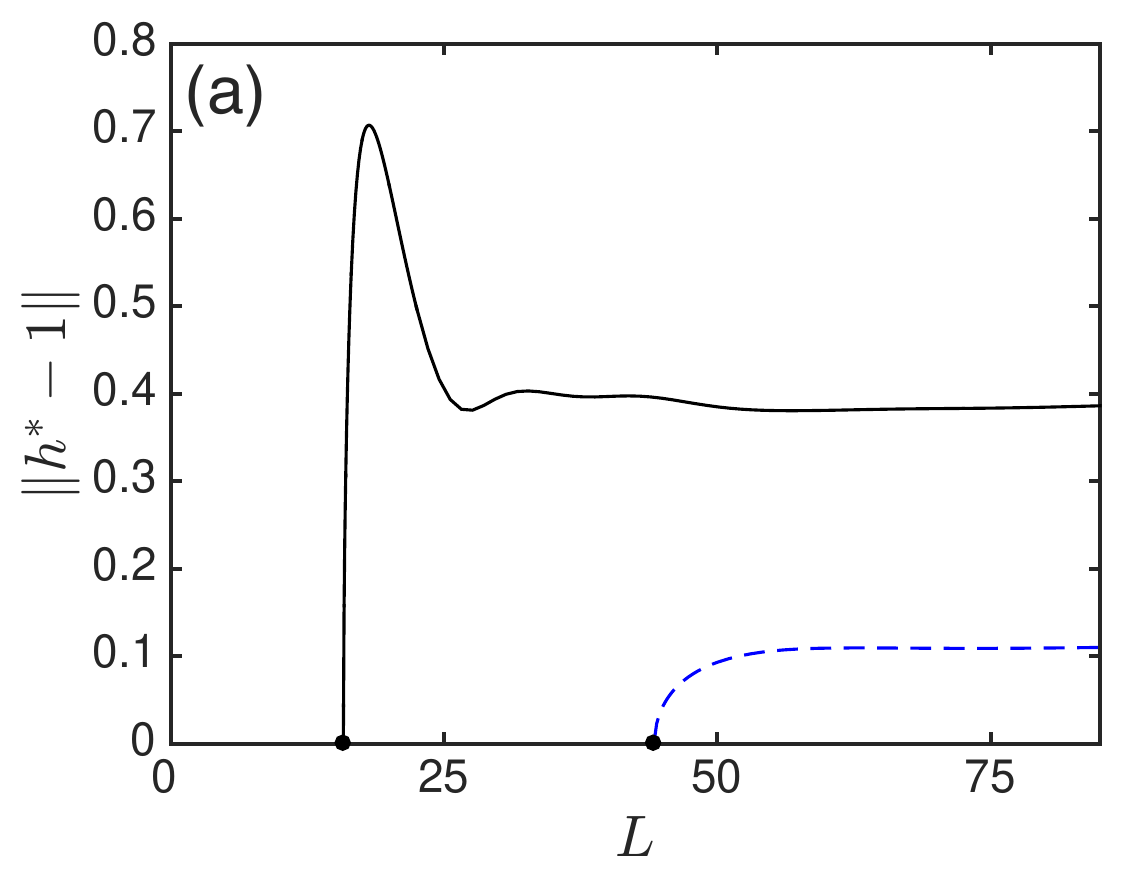}
\hspace{-0.2cm}
\includegraphics[width=1.82in]{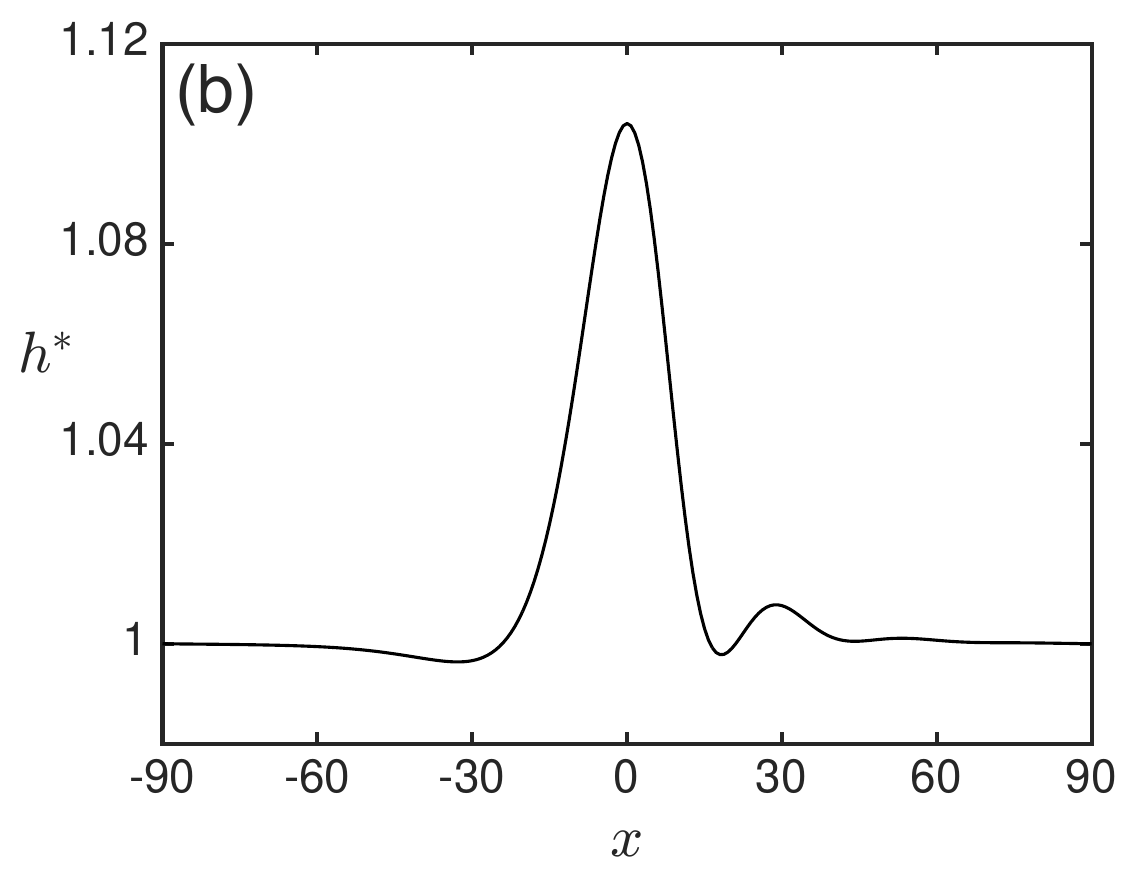}
\hspace{-0.2cm}
\includegraphics[width=1.84in]{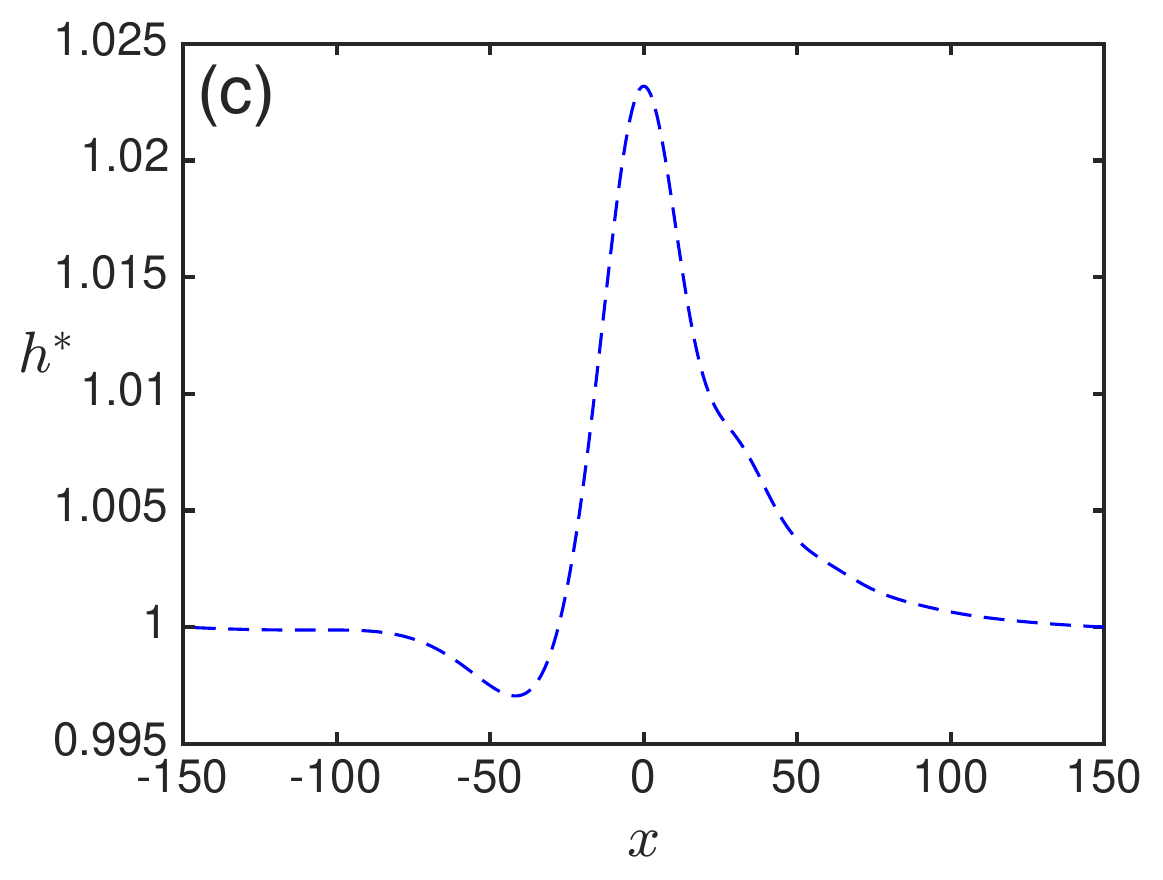}
    \end{minipage}%
    \vspace{-0.1cm}
\caption{(Color online) Stokes flow computations for $\beta=0.25\upi$, $\Ca=0.01$, $\We=13.5$: (a) Bifurcation diagram for the $\| h^* -1 \|$ against $L$ with branches emerging from $L_+\approx 15.80$ (solid line) and $L_-\approx 44.18$ (dashed line). (c, d) Pulse profiles for large $L$ on the branches emerging from $L_+$ and $L_-$, respectively.
}
\mylab{fig:case7}
\vspace{-0.2cm}
\end{figure}
%%%%%%%%%%%%

We present in figure \ref{fig:depression} some
examples of negative solitary pulses. In the non-electrified case shown in
figure \ref{fig:depression}(a), the boundary-element solution is compared
with the prediction for the same wave using the long-wave model. It is
noticeable that, as for the elevation pulse shown in figure
\ref{fig:branch}(c), the greatest discrepancy between the boundary-element and
long-wave calculations is observed at the largest peak. The pulse speed is
$c^*=0.259$ for the Stokes calculation and $c^*=0.257$ for the long-wave
calculation. This is smaller than the speed of linear long waves $2\sin
\beta=0.313$. Evidently the negative pulse is also travelling more slowly
than its elevation pulse counterpart for the same Bond number and inclination
angle shown in figure \ref{fig:branch}(c), whose speed is $c^*=0.457$.

Figure \ref{fig:depression}(b) shows the effect of increasing the electric
field intensity. Similar to the elevation pulses studied above, the electric
field tends to deepen the depression and heighten the amplitude of the
downstream oscillations. The electric field also tends to lower the speed of
the wave (values for the wave speed are quoted in the figure caption).
\begin{figure}
\centering
\includegraphics[width=2.0in]{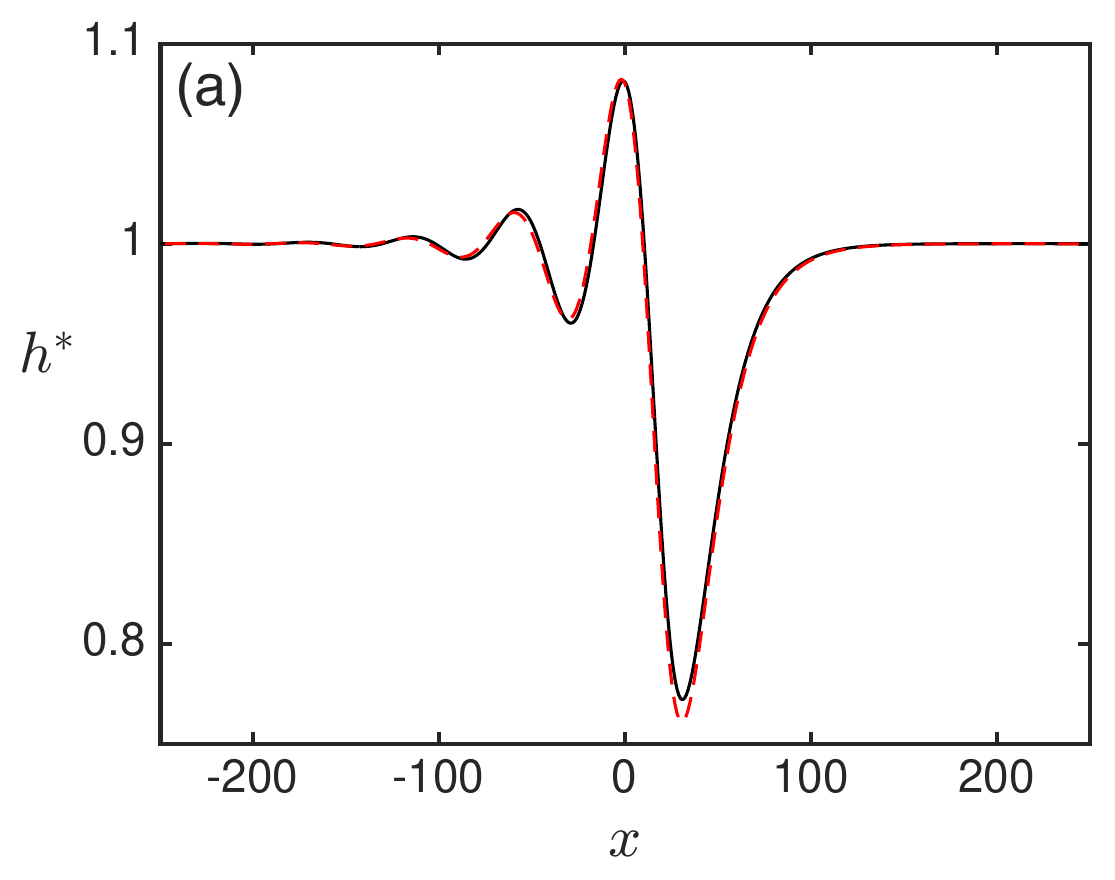}
\hspace{0.75cm}
\includegraphics[width=2.0in]{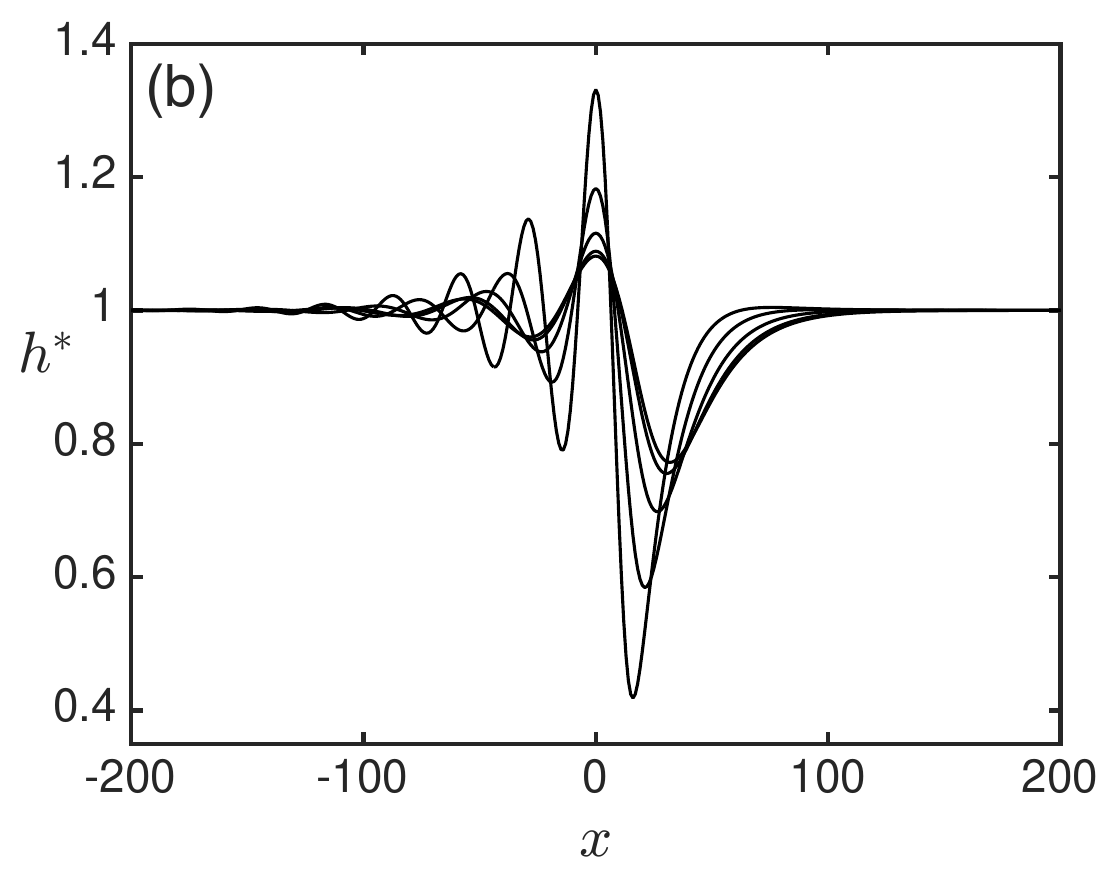}
\vspace{-0.1cm}
\caption{(Color online) Negative pulse profiles for Stokes flow with $\Ca=0.005$, $\beta=0.95\upi$ (with $L=250$, $N=500$). (a) $\We=0$: boundary-element solution
(solid line) and long-wave solution (dashed line). (b) Boundary-element solutions for $We=0$, $1.0$, $4.0$, $9.0$ and $16.0$ with the depth of the main trough increasing with $\We$.
The wave speeds are, respectively, $c^*=0.259$, $0.255$, $0.242$, $0.217$, $0.183$.
}
\mylab{fig:depression}
\vspace{-0.2cm}
\end{figure}
%

%%%%%%%%%%%%%%%%%%%%%%%%%%%%%%%%%%%%%%%%%
\section{Absolute/convective instability of pulse solutions}
\mylab{sec:absolute}
%%%%%%%%%%%%%%%%%%%%%%%%%%%%%%%%%%%%%%%%%

We now discuss the stability of the pulse solutions calculated in the previous sections.
In particular, we classify
pulses as being either convectively unstable or absolutely unstable. Suppose
first that the flow supports pulses which are convectively unstable. Since
such a pulse can tolerate localised disturbances, in time-dependent solutions
we would see the flow ultimately evolve into a state with an array of such
pulses undergoing weak interactions. Suppose instead that the pulses
supported by the flow are absolutely unstable. These can be destroyed by
localised disturbances and consequently we would expect to see highly
irregular dynamics in a general time-dependent simulation. {\color{black} For
further discussion see \cite{Chang1995,chang2002}.}

To study stability, we linearise about a pulse state by imposing a small perturbation in a reference frame which is moving with the pulse at speed $c^*$, writing
\bea \label{hpert}
h(x,t) = h^*(x) + \eta(x,t),
\eea
where $\eta(x,t)$ is a small perturbation localised in space, and $h^*\to 1$ as $|x|\to \infty$.

We consider first long-wave pulses which are solutions of the model equation \eqref{eq:TFE1}. Substituting \eqref{hpert} into (\ref{eq:TFE1}), written in a frame moving at the pulse speed $c^*$, and ignoring nonlinear terms, we obtain
\begin{equation}\mylab{eq:eta1}
\pt \eta=\ms{L}[\eta], \qquad
%\end{equation}
%where the linear operator $\ms{L}$ is given by
%\bea \label{eq:eta11}
\ms{L} \eta = ( a_0 \eta + a_1 \eta_x + a_2\mathcal{H}[\eta_{xx}] + a_3 \eta_{xxx}    )_x
\end{equation}
with
\bea
\nonumber
&&\displaystyle a_0(x) = c^* - 2(\sin\beta)h^{*2} + 2(\cos\beta)h^{*2}h^*_x - \frac{1}{\Ca}h^{*2}h^*_{xxx} - 2\We \: h^{*2}\mathcal{H}[h^*_{xx}],\\
&&\displaystyle a_1(x)=\frac{2\cos\beta}{3}h^{*3}, \qquad a_2(x) = - \frac{2}{3}\We \: h^{*3}, \qquad a_3(x)=-\frac{1}{3\Ca}h^{*3}.
\eea
The solution of (\ref{eq:eta1}) can be written as \cite[e.g.][]{Chang1996,lin2015coherent}
\begin{equation}
\eta(x,t)=
\sum_{i}\mathrm{e}^{\lambda_i t}B_i\phi_i(x)
+ \int_{-\infty}^{\infty}\mr{e}^{\sigma(\kappa)t}B(\kappa)\phi(x,\kappa)\:\mr{d}\kappa,
\mylab{eq:eta2}
\end{equation}
where the summation is over all the isolated eigenvalues $\lambda_i$ (the discrete spectrum) with corresponding
eigenfunctions $\phi_i(x)$, that is the functions belonging to the null space of $\lambda_i I-\ms{L}$,
and $B_i$ are constants.
%
% The calculation of the spectrum is done in:
%
%  Mark_Dmitri-Shared/LW_abs_conv/spectrum/L_spectrum
%
In fact, for the given model, we find numerically that there is only one isolated eigenvalue, which is real and negative. The same happens, for example, for the gKS equation \cite[][]{Chang1996,Tseluiko2012}.
In the second integral, $\sigma(\kappa)$ is the essential spectrum of $\ms{L}$
which we find coincides precisely with the spectrum for a flat film as is the case with the established theory for ordinary differential operators
\cite[][]{edmunds1987spectral,pego1992eigenvalues}.
For a flat film of unit thickness this is given by
\bea \label{eq:dispersion}
\sigma =  (c^*-2\sin \beta)\ri\kappa - \frac{2\cos \beta}{3} \:\kappa^2 + \frac{2\We}{3} \:|\kappa|^3 - \frac{\kappa^4}{3\Ca},
\eea
where $\kappa \in \mathbb{R}$.
In (\ref{eq:eta2}), $\phi(x,\kappa)$ are functions in to the null space of $\sigma I-\ms{L}$,
and (\ref{eq:dispersion}) implies that $\sigma(-\kappa)=\overline{\sigma(\kappa)}$ and also
$\phi(x,-\kappa)=\overline{\phi(x,\kappa)}$, where the overline denotes
complex conjugation. For a real perturbation $\eta$, the coefficients $B(\kappa)$ satisfy
$B(-\kappa)=\overline{B(\kappa)}$.

Since the discrete spectrum contains only $\lambda_1$, and $\lambda_1$ is negative, the
first term in (\ref{eq:eta2}) decays to zero as $t\to \infty$, and the convective/absolute nature of the instability is determined by the second, integral term,
which we label $I_c$. From (\ref{eq:dispersion}), if $\We^2 \Ca \leq 2\cos \beta$ then $\mathrm{Re}(\sigma)\leq 0$ and the flow is spectrally stable. However, if
$\We^2 \Ca > 2\cos \beta$ then $\mathrm{Re}(\sigma)>0$
if $|\kappa| \in (\kappa_1,\kappa_2)$, where
%\bea
%\kappa_{1} &=& \max \left \{0, \We \:\Ca - (\We^2 - 2 (\cos \beta)/\Ca)^{1/2}\right\}, \\
%\kappa_{2} &=& \We \:\Ca + (\We^2 - 2 (\cos \beta)/\Ca)^{1/2},
%\eea
%{\bf MGB: These formulae look to be wrong. Should be
\begin{align} \label{linstabform}
%\kappa_{1} &=& \max \left \{0, \We \:\Ca - (We^2\Ca^2 - 2 \Ca\cos \beta)^{1/2}\right\}, \\
\kappa_{1} &=
\left \{
\begin{array}{lll}
\kappa_{-} &\mbox{if} & \phantom{\upi/}0 < \beta \leq \upi/2\\
0 &  \mbox{if} & \upi/2 < \beta < \upi
\end{array}
\right .,
\qquad \kappa_2 = \kappa_{+},
\end{align}
%All the possible cases for $\beta$ are covered by the range $(0,\upi)$.
where $\kappa_{\pm}$ are given in \eqref{eq:kappaplusminus},
% = \We \:\Ca \pm (\We^2\Ca^2 - 2 \Ca\cos \beta)^{1/2}$,
and the flow is spectrally unstable. When the flow is unstable, we can determine the nature of the instability by
%considering the path of integration taken for $I$. Specifically, if is is
looking to see if it is possible to deform the contour of integration for $I_c$ so that it is entirely contained inside the region of the complex $\kappa$ plane where $\mathrm{Re}
(\sigma)<0$. If this is possible, then the instability is convective and otherwise the instability is absolute.
However, $\sigma(\kappa)$ is not analytic in the complex $\kappa$ plane
because of the $|\kappa|^3$ term in \eqref{eq:dispersion}
and so the classical approach of \cite{huerre1990local} cannot be applied directly. To proceed, we note that we can
rewrite $I_c$ as \cite[see][]{lin2015coherent,Vellingiri2015}
$I_c = 2\,\mathrm{Re}(K(x,t))$,
where $K(x,t)=\int_{0}^{\infty}\mathrm{e}^{\sigma^+(\kappa)t}B(\kappa)\phi(x,\kappa)\:\mathrm{d}\kappa$, where
$\sigma^+(\kappa)$ is given by (\ref{eq:dispersion}) with $|\kappa|^3$ replaced by $\kappa^3$ and is an analytic function in the entire complex $\kappa$
plane. Furthermore, it is sufficient to consider the range of integration where $\mathrm{Re(\sigma^+})>0$ and examine the integral
\begin{equation}
K_1(x,t)=\int_{\kappa_1}^{\kappa_2}\mr{e}^{\sigma^+(\kappa)t}B(\kappa)\phi(x,\kappa)\:\mr{d}\kappa.
\end{equation}
Since the integrand is analytic in the $\kappa$
plane, we may deform the contour of integration into any path $\Gamma$ connecting $\kappa_1$ and $\kappa_2$.
%\begin{equation}
%K_1(x,t)=\int_\Gamma \mr{e}^{\sigma^+(\kappa)t}B(\kappa)\phi(x,\kappa)\:\mr{d}\kappa,
%\end{equation}
%where the path of integration $\Gamma$ starts at $\kappa=\kappa_1$ and ends at $\kappa=\kappa_2$.
The case of convective instability, in which the deformed contour $\Gamma$ lies entirely within a region with $\mbox{Re}\: \sigma^+(\kappa)<0$, is
shown in figure~\ref{fig:convabs}({\it a}). The case of absolute instability, for which the contour $\Gamma$ must pass through a region with $\mbox{Re}\:
\sigma^+(\kappa)>0$, is shown in figure \ref{fig:convabs}({\it c}). The transition from convective to absolute instability under a continuous parameter
change must happen by passing through the situation illustrated in figure \ref{fig:convabs}({\it b}) where the two lines on which
$\mbox{Re}\, \sigma^+
(\kappa)=0$ have pinched together at a saddle point, where $\mbox{d}\sigma^+/\mbox{d}\kappa=0$ (note that since $\sigma^+$ is analytic,
points where $\mbox{d}\sigma^+/\mbox{d}\kappa$ vanishes are necessarily saddle points).
\begin{figure}
\centering
{\includegraphics[width=5.25in]{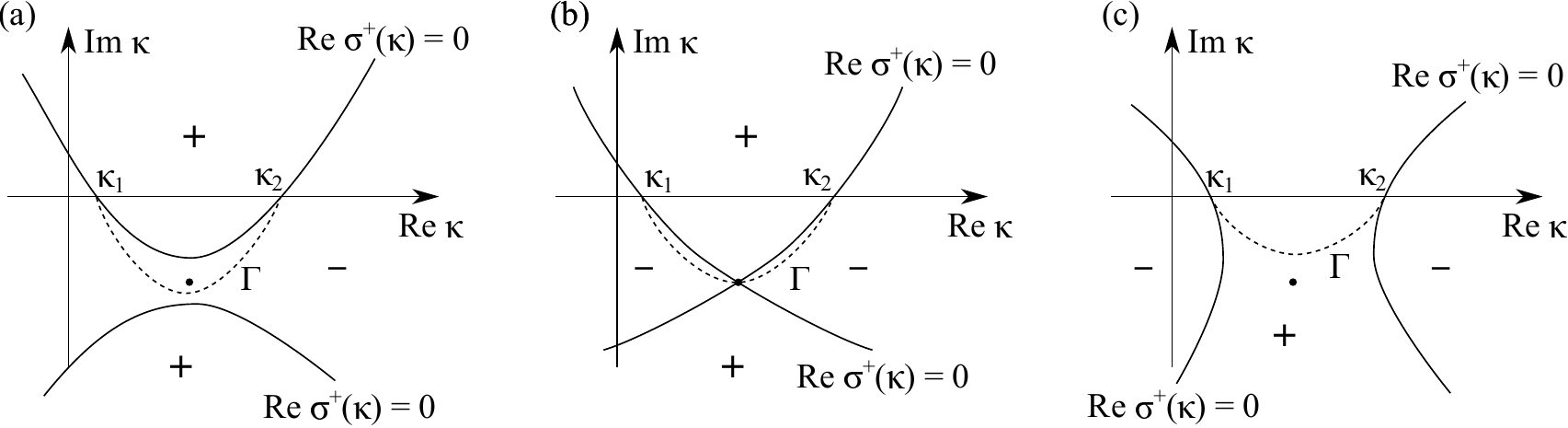}}
\vspace{-0.1cm}
\caption{(Color online) Illustration of the change from ({\it a}) convective instability to ({\it c})
absolute instability. The solid lines shows where $\mbox{Re}\:
\sigma^+(\kappa)=0$, and the broken line shows the integration contour
$\Gamma$. Regions with $\mbox{Re} \:\sigma^+(\kappa)>0$ and $\mbox{Re}\:
\sigma^+(\kappa)<0$ are shown with a $+/-$ respectively. The black dot
indicates a saddle point. } \mylab{fig:convabs}
\vspace{-0.2cm}
\end{figure}

Following the above discussion, to identify the transition from convective to
absolute instability, we seek the parameter values for which $ \mbox{Re}\:
\sigma^+(\kappa)=0$ and $\mbox{d}\sigma^+/\mbox{d}\kappa=0$ at the relevant
saddle point. To achieve this, we first take a pulse solution and solve the
cubic equation $\mbox{d}\sigma^+/\mbox{d}\kappa=0$ to find the relevant saddle point
(by scrutinising the contours) in the $\kappa$ plane. We then adjust $\We$
until $\mbox{Re}\: \sigma^+(\kappa)=0$ is satisfied. We may then continue in
any desired parameter to trace out the transition boundary. Figure
\ref{fig:absconv} shows the boundary between absolute and convective
instability in the $(\Ca,\We)$ plane for a range of values of the inclination
angle $\beta$. Evidently for fixed Bond number $\Ca$, the flow is either
convectively unstable for any value of $\We\geq 0$, or else as $\We$
increases it undergoes a transition from absolute to convective instability,
and we, therefore, expect that sufficiently strong electric field should have
a regularising effect on the dynamics. {\color{black} From the physical point
of view, this can be explained as follows. The electric field has a
destabilising effect on the flow, and results in the generation of
larger-amplitude pulses. The speed of the pulses is also amplified by the
electric field. If the electric field becomes sufficiently strong, the pulses
become sufficiently fast so that they can escape expanding wave packets
generated by localised disturbances. As a result, the pulses become
convectively unstable.}

Our results are consistent with those found by
\cite{lin2015coherent} for the non-local KS equation~\eqref{KSeq}, which
%obtained from \eqref{eq:TFE1} from a weakly nonlinear expansion as described in
%\S~\ref{sec:long}.
supports pulse solutions which are
absolutely unstable for any $\We$ \cite[][]{lin2015coherent}. This is
consistent with the findings shown in figure \ref{fig:absconv}: for any Bond
number there is a transition from absolute to convective instability as the
Weber number is increased, and the smaller the Bond number is the larger the
Weber number at which the transition occurs.

Next, we turn our attention to a travelling pulse that is a solution to the full Stokes equations, as discussed in \S~\ref{sec:stokes}.
%For Stokes flow, the growth rate of small-amplitude disturbances on a flat film is given by (\ref{stokessig}).
For a disturbance $\eta(x,t)$,
which is superimposed onto a pulse solution, as for the long-wave case, the solution to the linearised stability problem may be
written in the form \eqref{eq:eta2}. For the present purposes, we will assume that the discrete spectrum does not contain any
eigenvalues which lie in the right half plane; consequently the first term in \eqref{eq:eta2} decays to zero as $t\to \infty$ and, the
convective/absolute nature of the instability is determined by study of the second term. As for the long-wave case the nature of the instability hinges on the ability to deform the contour of integration of the integral term $I_c$, as discussed in figure
\ref{fig:convabs}.

%Table \ref{table:acon} summarises the results found for the Stokes solutions presented in \S~\ref{sec:stokes}.
All of the elevation boundary-element pulse solutions for Stokes
flow computed in \S~\ref{sec:stokes}, both electrified and non-electrified
(see figures~\ref{fig:branch}c, \ref{fig:ca0p75}a and
\ref{fig:elecalgdecay}a), are found to be convectively unstable. The negative
pulse solutions presented in figure \ref{fig:depression} are all absolutely
unstable.

%%%%%%%%%
\begin{figure}
\centering
\includegraphics[width=2.5in]{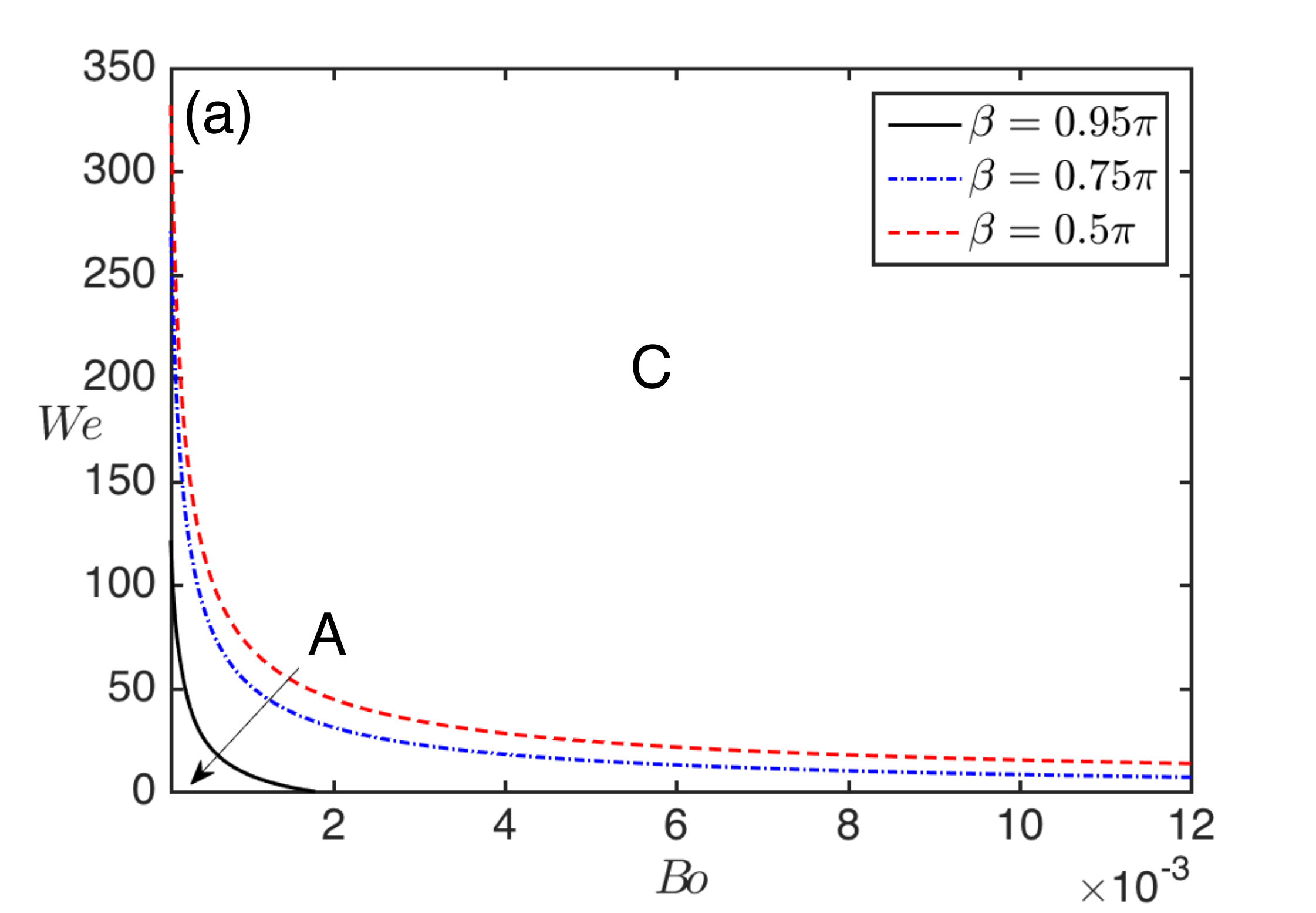}
\includegraphics[width=2.5in]{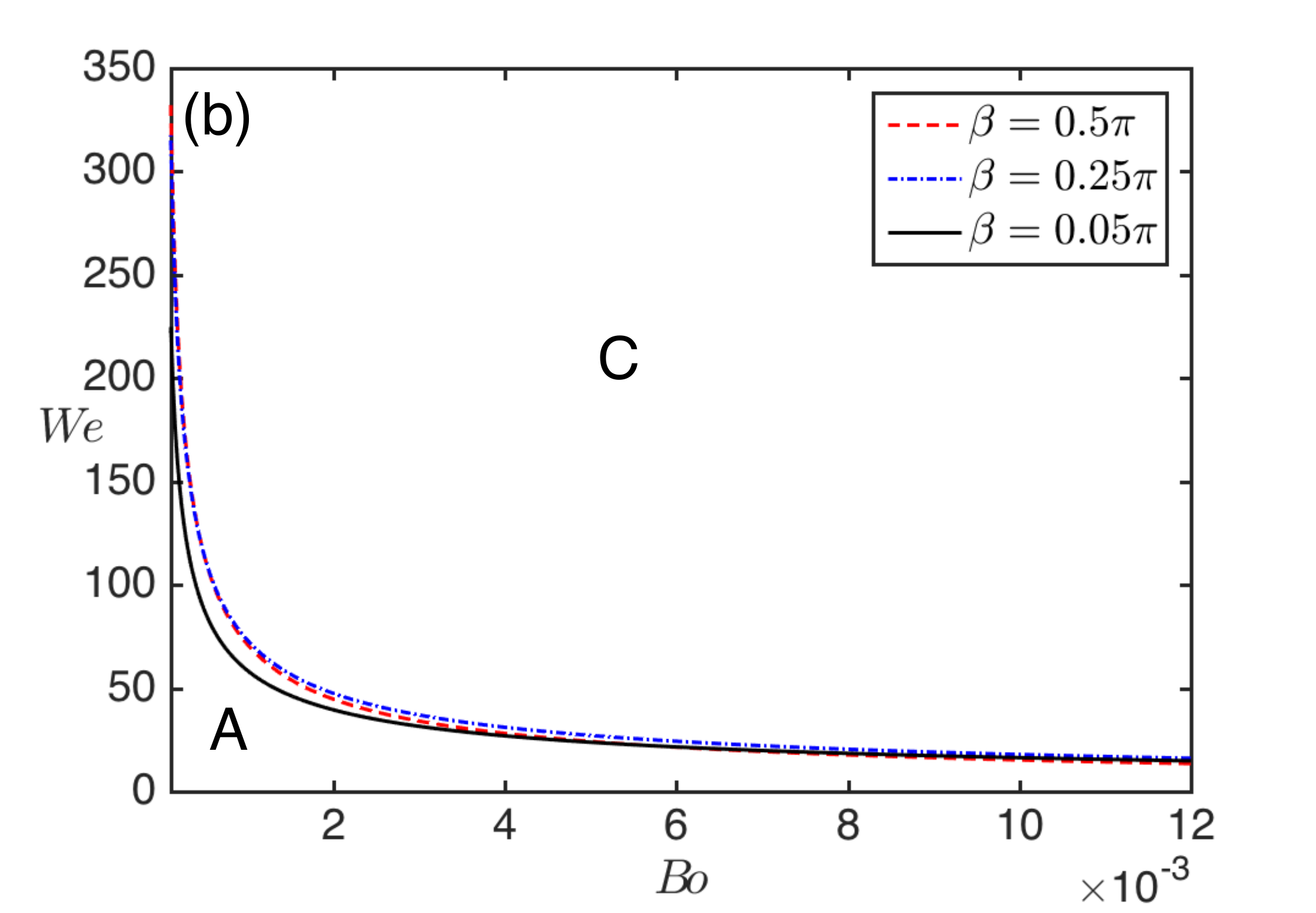}
\vspace{-0.1cm}
\caption{(Color online) Regions of convective (C) and absolute (A) instability in the $(\Ca,We)$ plane for a range of inclination angles for the long-wave model \eqref{eq:eta1}: (a) angles $\beta=0.5\upi,\,0.75\upi,\,0.95\upi$ and (b) angles $\beta=0.05\upi,\,0.25\upi,\,0.5\upi$.}
\mylab{fig:absconv}
\vspace{-0.2cm}
\end{figure}
%%%%%%%%%%

%%%%%%%%%%%%%%%%%%%%%%%%%%%%%%%%%%%
%%%%%%%%%%%%%%%%%%%%%%%%%%%%%%%%%%%
%%%%%%%%%%%%%%%%%%%%%%%%%%%%%%%%%%%

\begin{figure}%[!htb]
    \centering
    \begin{minipage}{.38\textwidth}
        \centering
        \includegraphics[height=0.2\textheight]{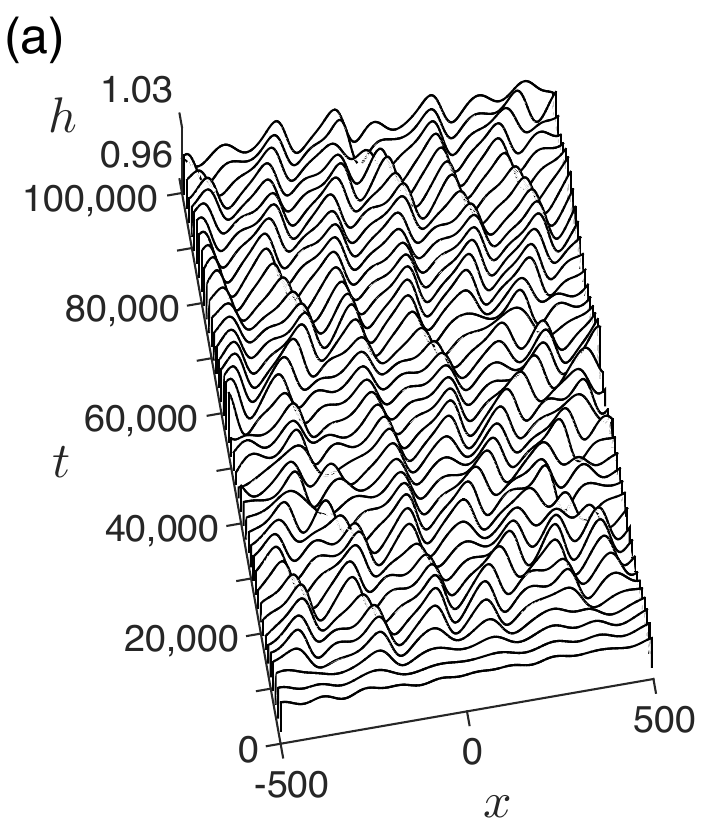}
        \includegraphics[height=0.09\textheight]{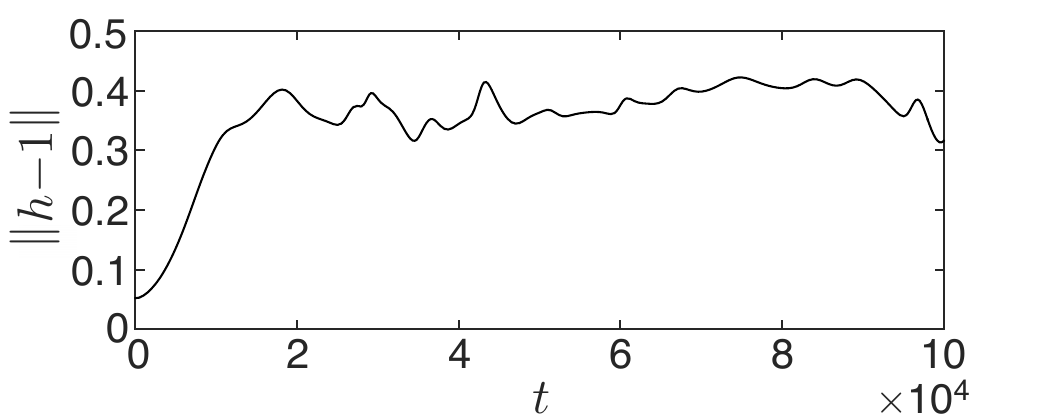}
%        \caption{$dt=0.1$}
%        \label{fig:prob1_6_2}
    \end{minipage}%
    \begin{minipage}{0.38\textwidth}
        \centering
        \includegraphics[height=0.2\textheight]{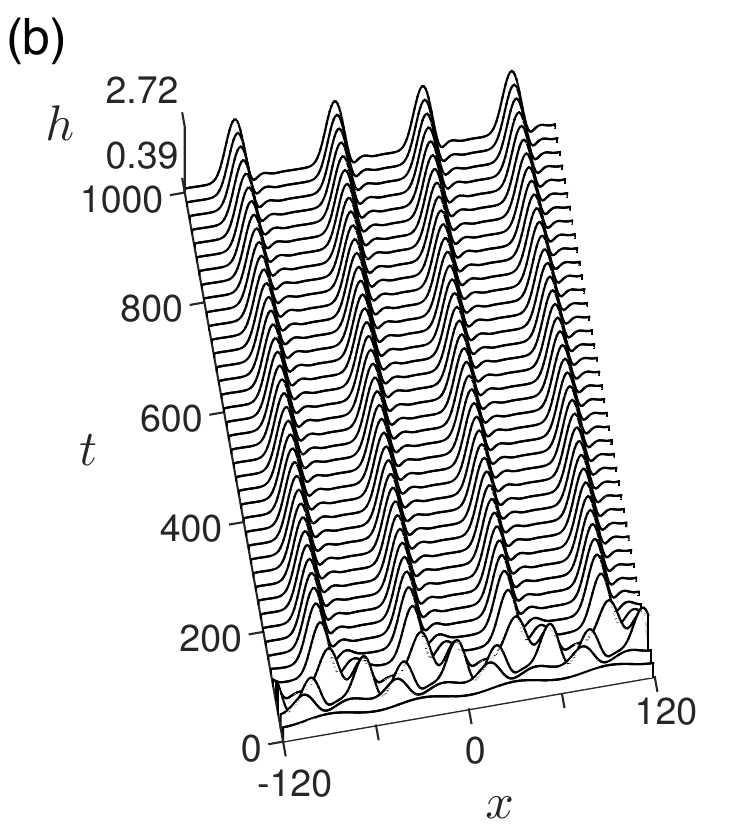}
        \includegraphics[height=0.09\textheight]{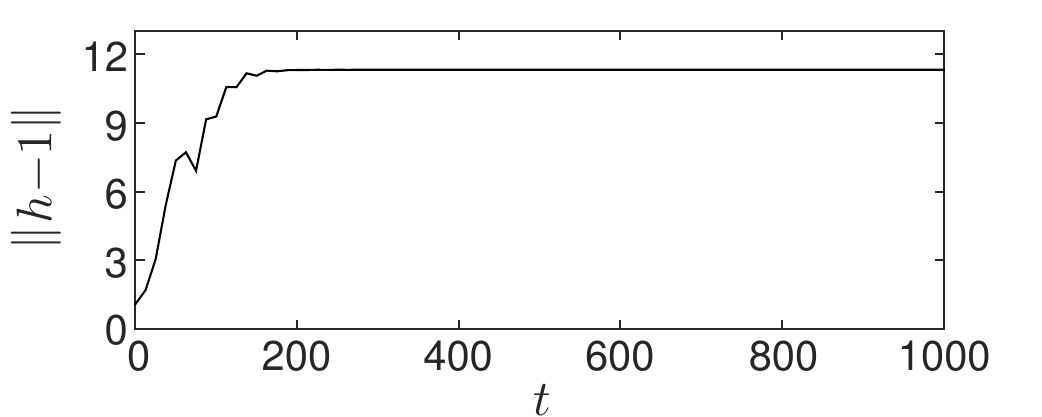}
    \end{minipage}%\vspace{0.6cm}
   \caption{Time-dependent solutions of \eqref{eq:TFE1} for  $\beta=0.75\upi$, $\Ca=0.002$: (a) $\We=0$ and (b) $\We=62$,
    The top panels show the time evolution in a frame moving at speed (a) $1.4$ and (b) $2.22$, and the lower panels show the evolution of the norm $\| h-1\|$.
 %In panel (a), for presentational purposes, the solution is shown in a frame moving at speed $1.4$. Obtuse angle, showing the regime for absolute instability.
 %In panel (b), the solution evolves into an array of weakly interacting pulses.}
   }
    \label{fig:absconv_bet_0_75pi_Ca_0_002}
    \vspace{-0.2cm}
\end{figure}

%\newpage
\begin{figure}%[!htb]
    \centering
    \begin{minipage}{.33\textwidth}
        \centering
        \includegraphics[height=0.2\textheight]{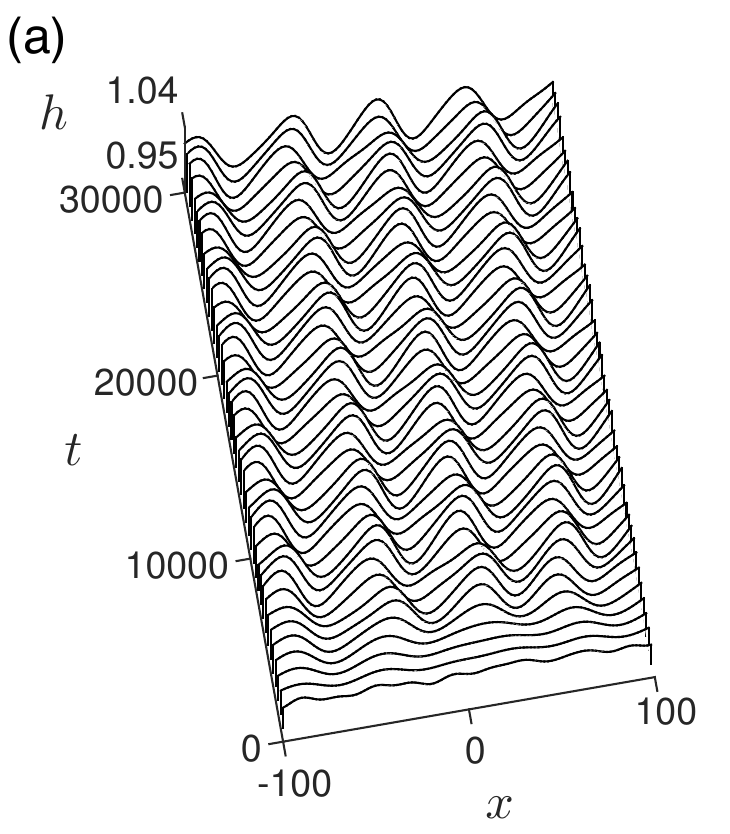}
        \includegraphics[height=0.09\textheight]{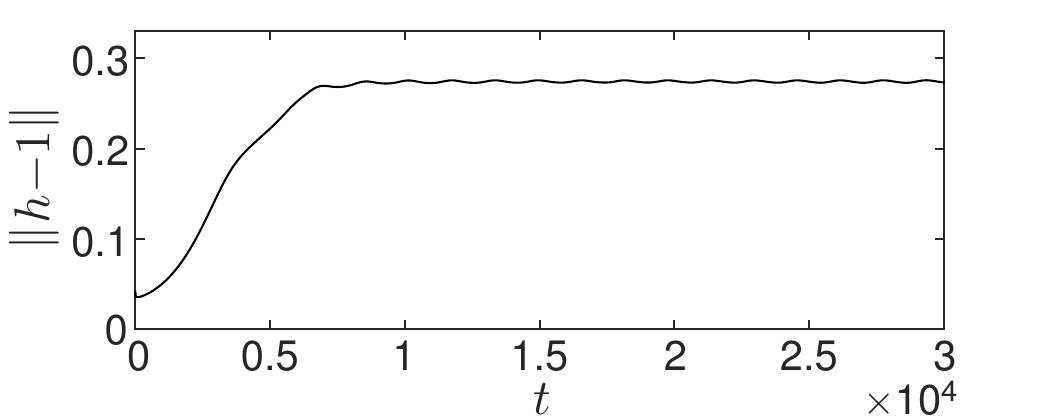}
%        \caption{$dt=0.1$}
%        \label{fig:prob1_6_2}
    \end{minipage}%
    \centering
    \begin{minipage}{.33\textwidth}
        \centering
        \includegraphics[height=0.2\textheight]{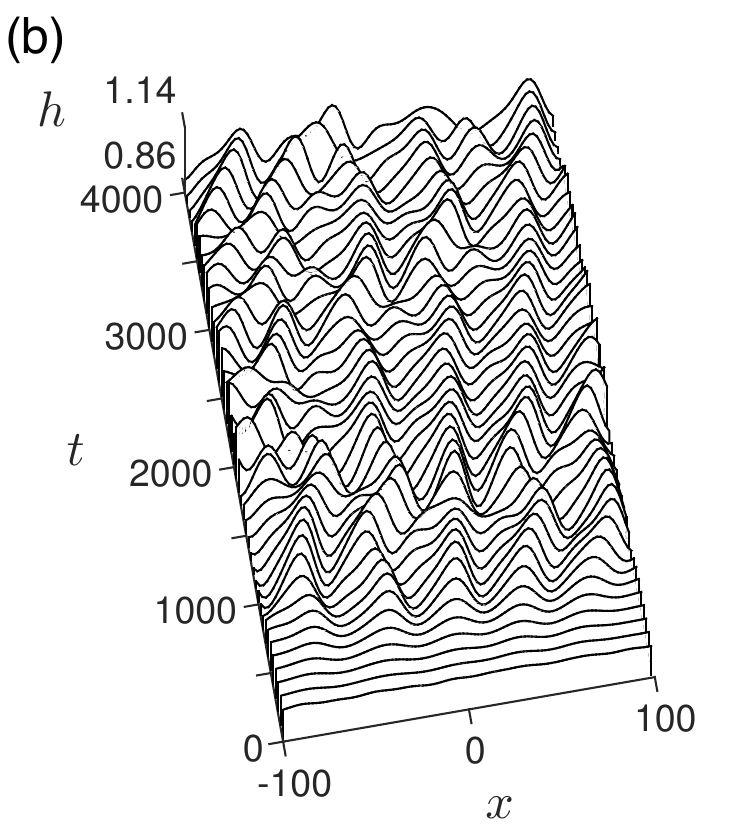}
        \includegraphics[height=0.09\textheight]{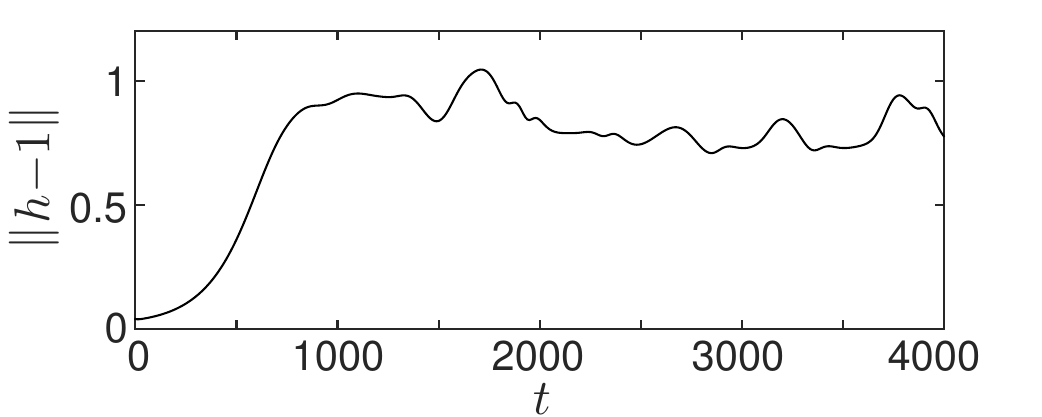}
%        \caption{$dt=0.1$}
%        \label{fig:prob1_6_2}
    \end{minipage}%
    \begin{minipage}{.33\textwidth}
        \centering
        \includegraphics[height=0.2\textheight]{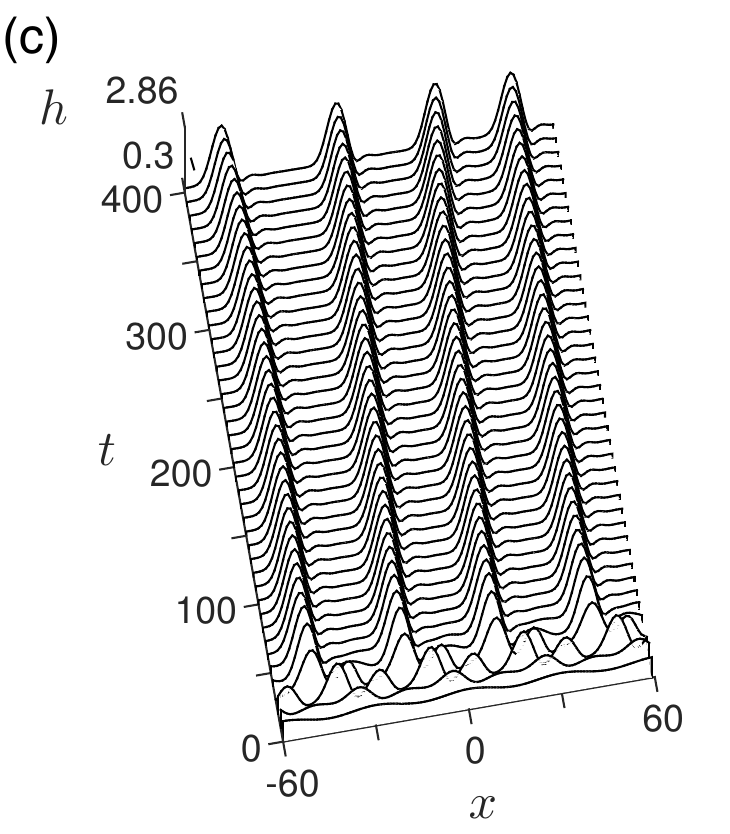}
         \includegraphics[height=0.09\textheight]{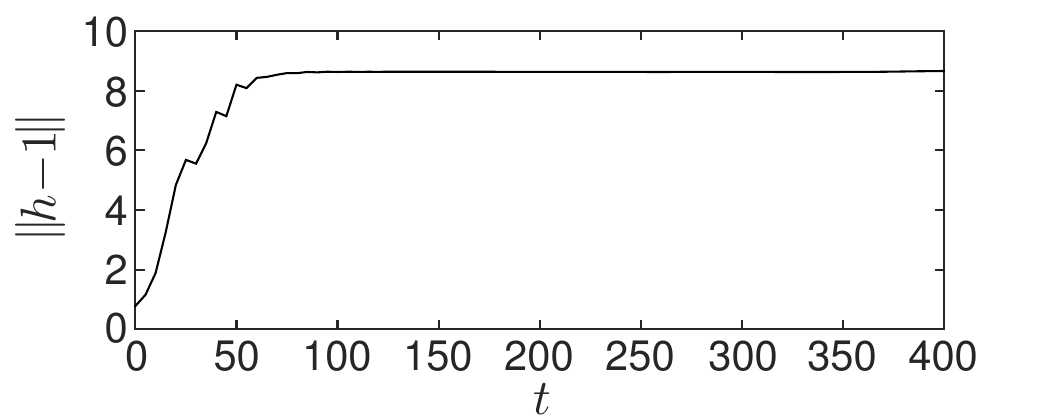}
%        \caption{$dt=0.1$}
%        \label{fig:prob1_6_2}
    \end{minipage}%
    \caption{Time-dependent solutions of \eqref{eq:TFE1} for $\beta=0.25\upi$, $\Ca=0.01$: (a) $\We=12.5$,
    % Acute angle, showing the development of a modulated array of  short-wavelength waves.
    (b) $\We=14.5$,
    %Acute angle, showing the regime for absolute instability. $\beta=0.25\upi$, $\Ca=0.01$,
    (c) $\We=26$.
    %Acute angle, showing the regime for convective instability. The solution evolves into an array of weakly interacting pulses.
     The top panels show the time evolution of the solution in a frame moving at speed (a, b) $1.4$ and (c) $2.08$. The lower panels show the evolution of the norm $\| h-1\|$.
    }
    \label{fig:absconv_bet_0_25pi_Ca_0_01}
    \vspace{-0.2cm}
\end{figure}

\color{black}

%%%%%%%%%%%
%\begin{table}
%%
%\centering
%{
%\renewcommand{\arraystretch}{1.8}
%\begin{tabular}{ccccccc}
%Figure label &&& Absolutely unstable &&& Convectively unstable  \\
%\hline
%\ref{fig:branch}(b) &&&  -- &&& \Checkmark  \\
%\ref{fig:ca0p75}(a) &&&  -- &&& \Checkmark  \\
%\ref{fig:elecalgdecay}(a) &&&  -- &&& \Checkmark  \\
%\ref{fig:depression}(a) &&&  \Checkmark  &&& -- \\
%\ref{fig:depression}(b) [$\We=0-16$] &&&  \Checkmark  &&& --
%\end{tabular}
%}
%\caption{Instability character for the Stokes single pulse solutions in \S~\ref{sec:stokes}. {\bf CONSIDER REMOVING THIS TABLE AND SAYING IN WORDS}}
%\label{table:acon}
%\end{table}
%%%%%%%%%%%%
%%

\subsection{Time-dependent simulations}\label{sec:timedep}

We have confirmed the predictions of the absolute/convective instability analysis of the long-wave model carried out
above by conducting time-dependent simulations
of \eqref{eq:TFE1}. In each case, the initial condition was chosen to be comprised of a superposition of the first few Fourier modes with randomly assigned
amplitudes. Figures~\ref{fig:absconv_bet_0_75pi_Ca_0_002} and \ref{fig:absconv_bet_0_25pi_Ca_0_01} show the results of simulations for the obtuse
inclination angle $\beta=0.75\upi$, in which case gravity is destabilising, and the acute inclination angle $\beta=0.25\upi$, when gravity is stabilising,
respectively. The former set of simulations was conducted at $\Ca=0.002$ while the latter set was done with $\Ca=0.01$. For each of the simulations we show
the time evolution of the film profile in the leftmost panel, and the final profile at the end of the simulation and evolution of the norm $\|h-1\|$ in the top and
bottom rightmost panels, respectively. For the obtuse inclination angle, as can be seen in panel (a) of figure \ref{fig:absconv_bet_0_75pi_Ca_0_002}, non-trivial
dynamics is observed even in the absence of an electric field due to the
destabilising effect of gravity, and the dynamics remains highly irregular throughout the simulation.
This is in agreement with the results shown in figure \ref{fig:absconv}(a) which predicts that pulse solutions are absolutely unstable at this value of the electric Weber number.
However,
the dynamics becomes regularised when an electric field of sufficient intensity is introduced. This is demonstrated in panel (b) of figure
\ref{fig:absconv_bet_0_75pi_Ca_0_002} for the electric Weber number $\We=62$. The film surface evolves relatively quickly
into an array of weakly interacting pulses. Although the norm reaches a plateau
by about $t\approx 200$, the flow remains time-dependent as the pulses
continue to rearrange their relative positions through weak attractions and repulsions.

For an acute inclination angle, a sufficiently strong electric field is
required to produce non-trivial dynamics. For the case shown in figure
\ref{fig:absconv_bet_0_25pi_Ca_0_01}, we see that at relatively low electric
field strength ($\We=12.5$ in panel a), the film surface exhibits a nearly
periodic modulated wave train during the time simulation. Pulses are not
observed until the electric field strength exceeds  a critical value in
agreement with our earlier finding shown in figure
\ref{fig:bet_0_25pi_Ca_0_01}, which predicts that pulses are only present
when $\We>12.77$. Although pulses should theoretically exist at the Weber
number considered in panel (b), namely $14.5$, they are absolutely unstable
according to the analysis of \S~\ref{sec:absolute} (see figure
\ref{fig:absconv}(b); for $ \Ca=0.01$ and $\beta=0.25\pi$ the threshold
between absolute and convective instability is at $\We\approx 17.9$). This
conforms with what is seen in the dynamic simulation which shows highly
irregular behaviour throughout. For a sufficiently large value of the
electric Weber number, a single pulse is convectively unstable, and indeed
pulses are observed during the simulation conducted at the larger electric
Weber number $\We=26$ shown in panel~(c).

{\color{black}

%%%%%%%%%%%%%%%%%%%%%%%%%%%%%%%%%%%%%%%%
\section{Physical context}\mylab{sec:physical}
%%%%%%%%%%%%%%%%%%%%%%%%%%%%%%%%%%%%%%%%

%{\bf We should also add something about the dielectric film case. For water-glycerol solutions (low Reynolds number) we can take some bits from Tseluiko \& Papageorgiou, Phys Rev E {\bf 82} (2010) paper.}

It should be possible to observe the phenomena described in this study in a
physical experiment. In a typical experimental set-up, a liquid film emerges from an inlet and flows down an inclined plane.
We would expect to observe two-dimensional wave phenomena for not too large Reynolds number and sufficiently close to the inlet. However, three-dimensional behaviour is likely further away from the inlet, in particular for hanging films as was recently observed by \cite{charogiannis2016application}.

%and in this section we note
%typical physical parameter values at which the restrictions of our theory should be satisfied.
Previous experimental studies have used films down to just a few microns in thickness. % ({\bf references}).
%Our long-wave model is valid provided that the Bond number $\Ca$ is small and the electric Weber number is large, and %if the Reynolds number is small for $O(1)$ inclination angles, or if the Reynolds number is $O(1)$
%when the plate is almost horizontal.
Assuming an aqueous film of thickness $h_0=100\:\mu\mathrm{m}$ at $20^\mathrm{o}\:\mathrm{C}$,
%, and taking $\rho=10^3\:\mathrm{kg}\cdot\mathrm{m}^{-3}$, $\mu=10^{-3}\:\mathrm{Pa}\cdot\mathrm{s}$, and $\gamma=7\times10^{-2}\:\mathrm{N}\cdot\mathrm{m}^{-1}$,
we find that
$\Ca=7.0\times 10^{-4}$ and $Re=4.9$. Referring to figure
\ref{fig:absconv}(a), the threshold between convective and absolute
instability is at $\We=15.65$ and so in order to observe the regularisation
of the dynamics discussed in section \ref{sec:timedep} we might take
$\We=20$, say, for an experiment with $\cot \beta=0.95\upi$ (an inverted
film). This requires an electric field of strength $E_0=1.487\times
10^6\:\mathrm{V}\cdot\mathrm{m}^{-1}$.
%, which is large but still some way below the critical value for dielectric breakdown of air, namely $3\times10^6\:\mbox{V}\cdot\mbox{m}^{-1}$.
%Let us note here that previous experiments have been conducted with very large electrode voltages of theorder of tens of thousands of volts \cite[e.g.][]{griffing2006}.
Assuming an
electrode potential of $1000\:V$, a gap between the film and the electrode of
about 672$\:\mu\mathrm{m}$ would be required, and this is almost seven times
the film height (our analysis assumes that the electrode is a long way from
the film surface). For the same parameter values, but considering a slightly
inclined plate with $\cot \beta=0.05\pi$, the transition between convective
and absolute instability occurs at $\We=71.45$ (see figure
\ref{fig:absconv}b). Taking $\We=75$ would require an electric field strength
$E_0=2.880\times 10^6\:\mathrm{V}\cdot\mathrm{m}^{-1}$, which is also below
the dielectric breakdown limit (in this case with the same electrode
potential the electrode-film gap is approximately 3.5 times the film height).

}

%
%
%

%%%%%%%%%%%%%%%%%%%%%%%%%%%%%%%%%%%%%%%%%
\section{Conclusions}\mylab{sec:conclusions}
%%%%%%%%%%%%%%%%%%%%%%%%%%%%%%%%%%%%%%%%%

%We have examined the flow of a perfect conductor or perfect dielectric liquid
%film down an inclined planar wall in the presence an electric field that is
%uniform and directed normal to the wall far away from the film. Previous
%studies have shown that for certain parameter values the dynamics is
%characterised by weakly interacting pulses, each of which resembles an
%infinite-domain single-hump solitary pulse.
In this study, we have provided
the first analysis of mutually-interacting pulses on electrified falling films using
%both a weakly nonlinear model and
a quasi-linear long-wave model with a
non-local term representing the effect of the electric field, as well as the
full system of governing equations in the Stokes-flow regime when the effect
of inertia is neglected.

%For an obtuse inclination angle, gravity is destabilising and solitary-pulse solutions exist even in the absence of an
%electric field.
%%The linear stability analysis shows that there is a band of unstable wave numbers extending from zero to a cut-off ...
For an acute inclination angle, if the electric field strength is supercritical but not too strong, solitary-pulse
solutions do not exist, and only spatially-periodic solutions are found. Solitary-pulse solutions only appear  if the electric field strength exceeds a certain supercritical value. Generally speaking, in the
presence of an electric field the pulse solutions have a larger amplitude and may have
recirculation zones in their humps.
%The latter feature means that a quantity of fluid is transported along with the
%pulse thereby promoting the heat and mass transfer properties of the flow, and so might be used to advantage in
%industrial applications.
We have also shown that the tails of a pulse decay algebraically at infinity when an electric
field is present, in contrast to the exponential decay found for the non-electrified case.
%All of the aforementioned qualitative features were also found for fully nonlinear pulse solutions for Stokes
%flow.
%We compared pulse solutions computed on wide
%domains in the Stokes regime with those obtained using the long-wave model and also those found from a weakly
%nonlinear model. As expected, the long-wave results are in better agreement with the fully nonlinear solutions than
%the weakly nonlinear results, but the differences found between the various models is exacerbated in the presence
%of an electric field.
Within its domain of validity the long-wave model is in excellent agreement with the Stokes results over almost all of the
domain except at the pulse hump. Improved agreement was obtained by extending the long-wave model to include
higher-order terms.

We used a weak-interaction theory to study pulse interactions for the
long-wave model. The algebraic decay of the tails in the electrified case
requires long-range interactions to be
taken into account. Our chief interest was in the formation of bound states, for which the pulse separation distance
does not change in time.
%We analysed such states in detail and calculated the corresponding
%separation distances. In the absence of an electric field, the governing
%equation is local and a pulse solution can be viewed as a homoclinic orbit in
%a three-dimensional phase space. Bound states can be constructed as
%subsidiary homoclinic orbits using an approach based on Shil'nikov's theory
%in dynamical systems. However, since the governing equations for electrified
%pulses are non-local, Shil'nikov's approach is not applicable which then
%necessitates weak-interaction theory to obtain bound states.
With no electric field, there are a countably infinite set of two-pulse
bound states, but in the presence of an electric field only a finite number
of such states is present. Moreover, the long-range dynamics becomes
repulsive meaning that if the initial pulse separation distance exceeds a
threshold then bound states will not form and the pulses will instead
indefinitely repel each other. We also obtained two-pulse bound states for
Stokes flow both with and without an electric field and obtained good
agreement with the results for the long-wave model.

The solitary pulses we computed are inherently unstable by virtue of the fact
that the flat film far upstream and far downstream is itself unstable.
Nevertheless determining the absolute or convective nature of the instability
provides important insight into the expected dynamics in a time-dependent
simulation. In general, a sufficiently strong
electric field can switch the nature of the instability from absolute to
convective, and in so doing regularise the dynamics. This possibility was
confirmed by time-dependent simulations for the long-wave model. We also
determined the nature of the instability for the single-hump Stokes pulse solutions. We
found that all of these are convectively
unstable and would therefore be relevant to the time evolution of Stokes flow. Such simulations were not performed here, however, and
are left as a topic for future work.

\vspace{0.2cm}
%%%%%%%%%%%%%%%%%%%%%%%%%%%%%%%%%%%%%%%%%
We acknowledge financial support by the EPSRC under grants EP/J001740/1 and
EP/K041134/1. T.-S. and by the Ministry of Science and Technology of Taiwan
under research grant MOST-103-2115-M-009-015-MY2.

\appendix

%%%%%%%%%%%%%%%%%%%%%%%%%%
\section{Decay rates for a solitary Stokes pulse}\label{sec:farfield}

We assume the existence of a solitary pulse which is propagating at speed $c^*$ in the positive $x$ direction, and work in a frame
of reference fixed in the pulse.
%Using the same physical scales to non-dimensionalise variables that were introduced in
%\S~\ref{sec:long},
The dimensionless momentum and continuity equations are given by
(\ref{NS}) with $Re=0$.
%The no-slip and no-penetration boundary conditions at the wall are
%\begin{equation}
%u=-c^*,\quad v=0\quad\text{at}\quad y=0,
%\end{equation}
%and the kinematic, and tangential and normal stress conditions at $y=h^*(x)$ are
%\begin{eqnarray}
%\label{banana}
%&v=(u-c^*)h^*_x,&\\
%&(1-h_x^{*2})(u_y+v_x)-2h^*_x(u_x-v_y)=0,&\\
%\label{pooch0}
%&\displaystyle -p+\frac{2}{1+h_x^{*2}}[v_y+h_x^{*2} u_x-(v_x+u_y)h^*_x]=\frac{h^*_{xx}}{\Ca(1+h_x^{*2})^{3/2}}+\We(1+h_x^{*2})\varphi_{2y}^2. &
%\end{eqnarray}
The electric field problem in the air above the film was stated in \S\ref{sec:form}. Here we focus on the case of a perfect conductor film ($\varepsilon_p\to \infty$) in which case $\varphi_2$ satisfies Laplace's equation in the air with the stated far-field condition together with the condition that $\varphi_2=0$ on $y=h^*(x)$.

%In the air above the viscous film, the electric potential satisfies Laplace's equation, so that
%$\varphi_{xx}+\varphi_{yy}=0$.
%Since the film is assumed to be a perfect conductor, the electric potential is taken to vanish at the free surface, so that
%$\varphi=0$ at $y=h^*(x)$. Far above the film, the electric field is uniform so that $\varphi_x\rightarrow 0$ and $\varphi_{2y}\rightarrow -1$ as $y\rightarrow\infty$.

For a flat film, ${h^*(x)\equiv 1}$, the base state electric field solution is
$\varphi_0=1-y$. We write $\varphi=\varphi_0+\Phi$, where $\Phi$ is the displacement field from the basic state, and to introduce the change of independent variables
$\xi=x$ and $\zeta=y-h^*(x)$,
so that the free surface is flat with respect to the new variables.
The problem for $\Phi$ is then given by
\begin{eqnarray}
\label{stokesfar}
\Phi_{\xi\xi}-2h^*_\xi\Phi_{\xi\zeta}+(1+h_\xi^{*2})\Phi_{\zeta\zeta}-h^*_{\xi\xi}\Phi_\zeta=0,
\eea
with $ \Phi=f$ at $\zeta=0$ and $\Phi_\xi,\, \Phi_\zeta\rightarrow 0$ as $\zeta\rightarrow\infty$,
where $f\equiv h^*-1$.
Assuming that $\Phi$ and $f$ tend to zero algebraically as $|\xi|\rightarrow\infty$, it is easy to see that
$|2h^*_\xi\Phi_{\xi\zeta}|$, $|h_\xi^{*2}\Phi_{\zeta\zeta}|$ and $|h^*_{\xi\xi}\Phi_{\zeta}|$ are asymptotically smaller than $|\Phi_{\xi\xi}|$ and $|\Phi_{\zeta\zeta}|$ in the far-field. Thus the far-field behaviour of the solution to \eqref{stokesfar} coincides with that of the solution to the problem
\begin{eqnarray}
\widetilde\Phi_{\xi\xi}+\widetilde\Phi_{\zeta\zeta}=0, \qquad
\widetilde\Phi=f\quad \text{at}\quad \zeta=0, \qquad
\widetilde\Phi_\xi,\, \widetilde\Phi_\zeta\rightarrow 0\quad\text{as}\quad \zeta\rightarrow\infty.
\end{eqnarray}
(We obtain the same problem if we assume that as $|\xi|\to \infty$,
$\Phi\to 0$  either algebraically or exponentially, and $f\to 0$ exponentially). Taking a Fourier transform in $\xi$, it is
straightforward to show that the solution in
Fourier space %(when the Fourier transforms are computed with respect to $\xi$)
is given by $\mathcal{F}\big [\,{\widetilde \Phi}\,\big ]=\hat{f}(k,t)\mathrm{e}^{-|k|\zeta}$, where $k$ is the wavenumber in the $\xi$-direction.

%Proceeding, we assume that $\hat{f}$ is smooth for sufficiently small positive values of $k$ and for sufficiently small negative values of $k$, but that it may be non-smooth at $k=0$.
Exploiting the fact that $f$ is real, and using Maclaurin series
expansions for positive and negative values of $k$, we obtain the general expansion for $\mathcal{F}[f]$,
\begin{equation}\label{fexpansion}
\mathcal{F}[f]=d_0+d_1|k|+d_2k^2+\cdots,
\end{equation}
where $d_n=d_n^r+\mathrm{i}\,\mathrm{sgn}(k)d_n^i$,
with $d_n^r$ and $d_n^i$, $n=0,1,2,\,\ldots,$ all real. Hence for small $k$,
\begin{eqnarray}
\nonumber
\mathcal{F}\big[ \:\! \widetilde{\Phi}\:\! \big ]&=&(d_0+d_1|k|+d_2k^2+\cdots)\Bigl(1-\zeta|k|+\frac{\zeta^2}{2}k^2+\cdots\Bigr)
= d_0-(d_0\zeta-d_1)|k|+O(k^2)\\
&=&  d_0^r + \mathrm{i}\mbox{sgn}(k)d_0^i - (d_0^r\zeta-d_1^r)|k| + \mathrm{i}(d_1^i -d_0^i \zeta) k+O(k^2).
\end{eqnarray}
As it will become clear below, we must take $d_0^i=d_1^r=0$; otherwise we
would obtain a contradiction. Then, assuming that $d_0^r\neq 0$,  that is
$\int_{-\infty}^\infty f\,\mathrm{d}x\neq 0$,  we find that the leading
singularity in $\mathcal{F}\big[ \:\! \widetilde{\Phi}\:\! \big ]$ at $k=0$ is $-d_0^r\zeta|k|$. Thus, using
Theorem 19 on p. 52 of \cite{Lighthill1958}, we obtain that $\widetilde \Phi\propto
\zeta/\xi^2$ as $|\xi|\rightarrow\infty$, and this implies in particular that
on $y=h^*$ \bea \label{phifar} \Phi_{xy}\propto x^{-3} \quad \mbox{as} \quad
|x|\rightarrow\infty, \eea which will be needed below. The
integral condition $\int_{-\infty}^\infty f\,\mathrm{d}x\neq 0$ demands a non-zero pulse mass; such is found to occur in general, and so the algebraic
decay \eqref{phifar} is expected.

Next, we turn to the problem in the liquid film. For convenience, we define the stream function $\Psi$ such that $u=\Psi_y$ and $v=-\Psi_x$, and without loss of generality, we replace the no-penetration condition by $\Psi=0$ at $y=0$. The base-state solution for $h^*\equiv 1$ is
\begin{equation}
\Psi_0=\Bigl(y^2-\frac{y^3}{3}\Bigr)\sin\beta-c^*\,y,\quad p_0=2\cos\beta\,(y-1)-\We.
\end{equation}
To describe a pulse solution, we write $\Psi=\Psi_0+\Psi_1$ and $p=p_0+p_1$,
where $\Psi_1$ and $p_1$ represent the deviation from the flat film state.
We are interested in the limit $|x|\rightarrow\infty$, in
which case we assume that $\Psi_1$ and $p_1$ are small, and consider the
linearised form of the problem. On eliminating the pressure, the free-surface
conditions become
%\begin{equation}
%\nabla^4\Psi_1=0,
%\end{equation}
%with the following boundary conditions at $z=0$:
%\begin{equation}
%\Psi_{1z}=0,\quad \Psi=0
%\end{equation}
%and the following boundary conditions are $z=1$:
\begin{eqnarray}
%&&\Psi_{1x}+(1-c)f_x=0,\label{eq:lin1}\\
&&\Psi_{1x}+(\sin \beta-c^*)f_x=0,\label{eq:lin1}\\ % \sin \beta added by MGB 21/3/16
%&&\Psi_{1zz}-\Psi_{1xx}=2f,\label{eq:lin2}\\
&&\Psi_{1yy}-\Psi_{1xx}=2(\sin \beta) f,\label{eq:lin2}\\  % \sin \beta added by MGB 21/3/16
&&\Psi_{1yyy}+3\Psi_{1xxy}=-\frac{1}{\Ca}f_{xxx}+2(\cos\beta)f_x+2\We\, \Phi_{xz}\label{eq:lin3}
\end{eqnarray}
at $y=h^*$. The kinematic condition (\ref{eq:lin1}) implies that if $f\sim\alpha/x^m$ as $|x|\rightarrow\infty$, for some constants $\alpha$
and $m$, then $\Psi_1\sim{a(y)}/{x^m}$ as $|x|\rightarrow\infty$, where $a(y)$ is such that $a(1)=\alpha(c^*-1)$.
%$\Psi_1(z=1)\sim\alpha(c-1)/x^k$ as $|x|\rightarrow\infty$
Assuming that $a'''(1)\neq 0$, the leading-order balance in (\ref{eq:lin3}) implies that
$\Psi_{1yyy}\propto \Phi_{xy}$ as $|x|\rightarrow\infty$. Taking into account (\ref{phifar}), this means that $m=3$ so that
\bea \label{ffar}
f\propto x^{-3}  \quad \mbox{as} \quad |x|\rightarrow\infty.
\eea
This result forces $d_0^i=d_1^r=0$ in \eqref{fexpansion} and highlights the potential contradiction alluded to
above. Hence for an electrified solitary pulse with non-zero mass, $\int_{-\infty}^{\infty} f\,\mbox{d} x\neq 0$, the
far-field decay is algebraic and given by \eqref{ffar}.
We note that this is the same decay behaviour as was found for the long-wave model. % in \eqref{LWdecay_far}.

When $\We=0$ the pulse tails decay exponentially fast in the far-field. To determine the rate of decay we write
$f = f_0\exp (\lambda x)$ as $|x| \to \infty$, where $|f_0|\ll 1$, and aim to calculate the real part of $\lambda$, which itself is
generally complex.  Finally we obtain a nonlinear relation $d(\lambda;\Ca,\beta)=0$ from which $\lambda$ can be extracted
numerically using Newton's method.
%For downstream decay as $x\to \infty$ the relevant decay rates are those with $\mbox{Re}(\lambda)<0$, and for upstream %decay as $x\to -\infty$ the relevant decay rates are those with $\mbox{Re}(\lambda)>0$.
Invoking the Argument Principle, and computing the pertinent contour integral numerically, we can determine the upstream or downstream decay rate with the smallest (in magnitude) real part. Note, however, that to determine the decay rate requires knowledge of the pulse speed $c^*$ and so requires global information about the solution.

\bibliographystyle{jfm}
\bibliography{JFM}

\end{document}